\documentclass[sigconf, nonacm]{acmart}

\newcommand\vldbdoi{XX.XX/XXX.XX}
\newcommand\vldbpages{XXX-XXX}
\newcommand\vldbvolume{14}
\newcommand\vldbissue{1}
\newcommand\vldbyear{2020}
\newcommand\vldbauthors{\authors}
\newcommand\vldbtitle{\shorttitle} 

\usepackage{latexsym}
\usepackage{anyfontsize}
\usepackage{url}

\usepackage{lipsum}

\usepackage{amsthm,amsmath}
\usepackage{environ}
\usepackage[polish, english]{babel}
\usepackage{graphicx} % Required for the inclusion of images
\usepackage[utf8]{inputenc}
\usepackage{amsthm} %http://ctan.org/pkg/amsthm
\makeatletter
\newcommand{\deftheorem}[3]{\@ifundefined{#1}{\newtheorem{#1}{#2}[#3]}{}}
\makeatother
\deftheorem{theorem}{Theorem}{}
\deftheorem{lemma}{Lemma}{}
\deftheorem{corollary}{Corollary}{}
\deftheorem{claim}{Claim}{}

\usepackage{xspace}
\usepackage{mathtools}
\usepackage{thmtools, thm-restate}
\usepackage{multirow}

\usepackage{etoolbox}
\providetoggle{sketches}
\settoggle{sketches}{false}

\providetoggle{extended}
\settoggle{extended}{true}

\usepackage{balance}

\usepackage{etoolbox}
\providetoggle{sketches}
\settoggle{sketches}{false}

\usepackage{tikz}
\usetikzlibrary{arrows,
                calc,
                petri,
                topaths,
                positioning,
                shapes,
                patterns,
                decorations.pathmorphing}
\usepackage{tabularx}
\usepackage{savesym}
\usepackage{newlfont}

\newcommand{\HistoryNameColumn}[1]{\vspace{-1.2cm} #1}
\newcommand{\HistoryDiagramColumn}[1]{\scalebox{0.85}{#1}}%

\newcolumntype{H}{>{\collectcell\HistoryNameColumn} >{\centering\arraybackslash}m{0.25in}  <{\endcollectcell}}
\newcolumntype{D}{>{\collectcell\HistoryDiagramColumn}{l}<{\endcollectcell}}

\newcommand\etal{\emph{et al.}\xspace}

\usepackage{collcell}
\usepackage{comment}

\usepackage{multirow}
\usepackage{centernot}
\usepackage{multicol}
\usepackage{algorithm}
\usepackage{algpseudocode}

\definecolor{yellow-green}{rgb}{0.6, 0.8, 0.2}

\makeatletter
\newcommand{\algcolor}[2]{%
  \hskip-\ALG@thistlm\colorbox{#1}{\parbox{\dimexpr\linewidth-2\fboxsep}{\hskip\ALG@thistlm\relax #2}}%
}

\makeatother

\usepackage{cleveref}

\newcolumntype{C}[1]{>{\centering\let\newline\\\arraybackslash\hspace{0pt}}m{#1}}

\def\put{Poznan University of Technology}

\usepackage{tablefootnote}

\NewEnviron{myeq}{%
\begin{equation}\begin{split}
  \BODY
\end{split}\nonumber\end{equation}}
\mathchardef\mhyphen="2D % Define a "math hyphen"

\renewcommand{\iff}{\Leftrightarrow}

\providecommand{\true}{}
\renewcommand{\true}{\mathsf{true}}

\newcommand\sx{\kern-1ex}

\newcommand{\key}{\mathsf{key}}

\newcommand{\maxx}{\mathsf{max}\xspace}

\newcommand{\pput}{\mathrm{put}}
\newcommand{\gget}{\mathrm{get}}
\newcommand{\rremove}{\mathrm{remove}}

\newcommand{\get}{\mathrm{get}}

\newcommand{\jiffy}{Jiffy\xspace}

\newcommand{\snaptree}{SnapTree\xspace}
\newcommand{\kary}{k-ary tree\xspace}
\newcommand{\catree}{CA-imm\xspace}
\newcommand{\caavl}{CA-AVL\xspace}
\newcommand{\casl}{CA-SL\xspace}
\newcommand{\lfca}{LFCA tree\xspace}
\newcommand{\jsl}{Java CSLM\xspace}
\newcommand{\kiwi}{KiWi\xspace}

\newcommand{\mops}{Mops/s\xspace}

\newcommand{\rtdscp}{\texttt{RDTSCP}\xspace}

\newcommand{\TSC}{\mathrm{TSC}}

\newcommand{\oldVersion}{\mathsf{oldVer}}
\newcommand{\optVersion}{\mathsf{optVer}}

\newcommand{\snapVersion}{\mathrm{snapVersion}}

\newcommand{\version}{\mathsf{version}}

\newcommand{\applyBatch}{\mathrm{batchUpdate}}
\newcommand{\batchUpdate}{\mathrm{batchUpdate}}

\newcommand{\keys}{\mathrm{keys}}

\newcommand{\nnext}{\mathrm{next}}

\newcommand{\rread}{\mathrm{read}}

\newcommand{\values}{\mathrm{values}}

\newcommand{\nanotime}{\texttt{System.nanoTime()}\xspace}

\newcommand{\newRevision}{\mathsf{newRev}}
\newcommand{\bnode}{\mathsf{node}}
\newcommand{\nnode}{\mathsf{nextNode}}

\newcommand{\findNodeForKey}{\mathrm{findNodeForKey}}
\newcommand{\TempSplitNode}{\mathrm{TempSplitNode}}
\newcommand{\helpMergeTerminator}{\mathrm{helpMergeTerminator}}
\newcommand{\helpSplit}{\mathrm{helpSplit}}
\newcommand{\helpTempSplitNode}{\mathrm{helpTempSplitNode}}
\newcommand{\findAndHelpMergeRevision}{\mathrm{findAndHelpMergeRevision}}
\newcommand{\findMergeRevision}{\mathrm{findMergeRevision}}

\newcommand{\bbreak}{\textbf{break}}
\newcommand{\continue}{\textbf{continue}}
\newcommand{\is}{\textbf{is}}

\newcommand{\autoscaler}{\mathsf{autoscaler}}
\newcommand{\finVersion}{\mathsf{finVer}}
\newcommand{\hashes}{\mathsf{hashes}}
\newcommand{\hhead}{\mathsf{head}}
\newcommand{\indices}{\mathsf{indices}}
\newcommand{\keyOfRightNode}{\mathsf{rightKey}}
\renewcommand{\keys}{\mathsf{keys}}
\newcommand{\length}{\mathsf{length}}
\newcommand{\mt}{\mathsf{mTerm}}
\newcommand{\lsr}{\mathsf{lRev}}
\newcommand{\rsr}{\mathsf{rRev}}
\newcommand{\rrev}{\mathsf{headRev}}
\newcommand{\rightNext}{\mathsf{rightNext}}
\newcommand{\revision}{\mathsf{revision}}
\newcommand{\sibling}{\mathsf{sibling}}
\newcommand{\terminated}{\mathsf{terminated}}
\newcommand{\updateType}{\mathsf{updateType}}
\newcommand{\vvalue}{\mathsf{value}}
\renewcommand{\values}{\mathsf{values}}

\newcommand{\regularUpdate}{\mathsf{REGULAR\_UPDATE}}
\newcommand{\nodeSplit}{\mathsf{NODE\_SPLIT}}
\newcommand{\nodeMerge}{\mathsf{NODE\_MERGE}}
\newcommand{\newestVersion}{\mathsf{NEWEST\_VERSION}}
\newcommand{\NOP}{\mathsf{NOP}}

\newcommand{\SplitRevision}{\mathsf{SplitRevision}}
\newcommand{\MergeRevision}{\mathsf{MergeRevision}}
\newcommand{\MergeTerminator}{\mathsf{MergeTerminator}}

\newcommand{\applyPutAndSplit}{\mathrm{putAndSplit}}
\newcommand{\cloneAndApplyPut}{\mathrm{cloneAndPut}}
\newcommand{\cloneAndApplyRemove}{\mathrm{cloneAndRemove}}
\newcommand{\CAS}{\mathrm{CAS}}

\newcommand{\helpPut}{\mathrm{helpPendingUpdate}}
\newcommand{\getNewestRevision}{\mathrm{getNewestRevision}}
\newcommand{\getRevision}{\mathrm{getRevision}}
\newcommand{\performGC}{\mathrm{performGC}}
\newcommand{\publishVersion}{\mathrm{waitUntil}}
\newcommand{\query}{\mathrm{query}}
\newcommand{\trySetVersion}{\mathrm{trySetVersion}}

\providetoggle{appendix}
\settoggle{appendix}{true}

\begin{document}
\title{\jiffy: A Lock-free Skip List with Batch Updates and
Snapshots}

%%
%% The "author" command and its associated commands are used to define the authors and their affiliations.
\author{Tadeusz Kobus}
\affiliation{%
%   \institution{Institute of Computing Science, Poznan University of Technology}
    \institution{\put}
%   \streetaddress{P.O. Box 1212}
  \city{Pozna\'n}
  \country{Poland}
}
\email{Tadeusz.Kobus@cs.put.edu.pl}
\author{Maciej~Kokoci\'nski}
\affiliation{%
%   \institution{Institute of Computing Science, Poznan University of Technology}
    \institution{\put}
%   \streetaddress{P.O. Box 1212}
  \city{Pozna\'n}
  \country{Poland}
}
\email{Maciej.Kokocinski@cs.put.edu.pl}
\author{Pawe{\l}~T.~Wojciechowski}
\affiliation{%
%   \institution{Institute of Computing Science, Poznan University of Technology}
    \institution{\put}
  \city{Pozna\'n}
  \country{Poland}
}
\email{Pawel.T.Wojciechowski@cs.put.edu.pl}

% \author{Lars Th{\o}rv{\"a}ld}
% \orcid{0000-0002-1825-0097}
% \affiliation{%
%   \institution{The Th{\o}rv{\"a}ld Group}
%   \streetaddress{1 Th{\o}rv{\"a}ld Circle}
%   \city{Hekla}
%   \country{Iceland}
% }
% \email{larst@affiliation.org}
% 
% \author{Valerie B\'eranger}
% \orcid{0000-0001-5109-3700}
% \affiliation{%
%   \institution{Inria Paris-Rocquencourt}
%   \city{Rocquencourt}
%   \country{France}
% }
% \email{vb@rocquencourt.com}
% 
% \author{J\"org von \"Arbach}
% \affiliation{%
%   \institution{University of T\"ubingen}
%   \city{T\"ubingen}
%   \country{Germany}
% }
% \email{jaerbach@uni-tuebingen.edu}
% \email{myprivate@email.com}
% \email{second@affiliation.mail}
% 
% \author{Wang Xiu Ying}
% \author{Zhe Zuo}
% \affiliation{%
%   \institution{East China Normal University}
%   \city{Shanghai}
%   \country{China}
% }
% \email{firstname.lastname@ecnu.edu.cn}
% 
% \author{Donald Fauntleroy Duck}
% \affiliation{%
%   \institution{Scientific Writing Academy}
%   \city{Duckburg}
%   \country{Calisota}
% }
% \affiliation{%
%   \institution{Donald's Second Affiliation}
%   \city{City}
%   \country{country}
% }
% \email{donald@swa.edu}

%%
%% The abstract is a short summary of the work to be presented in the
%% article.
\begin{abstract}
In this paper we introduce \jiffy, the first lock-free, linearizable ordered 
key-value index that offers both (1) batch updates, which are put and remove 
operations that are executed atomically, and (2) consistent snapshots used by, 
e.g., range scan operations. \jiffy is built as a multiversioned lock-free skip 
list and relies on CPU's Time Stamp Counter register to generate version 
numbers at minimal cost. For faster skip list traversals and better utilization 
of the CPU caches, key-value entries are grouped into immutable objects 
called \emph{revisions}. Moreover, by changing the size of revisions and thus
modifying the synchronization granularity, our index can adapt to varying 
contentions levels (smaller revisions are more suited for write-heavy workloads 
whereas large revisions benefit read-dominated workloads, especially when they 
feature many range scan operations). Structure modifications to the index, 
which result in changing the size of revisions, happen through (lock-free) 
skip list node split and merge operations that are carefully coordinated with 
the update operations. Despite rich semantics, \jiffy offers highly 
scalable performance, which is comparable or exceeds the performance of the 
state-of-the-art lock-free ordered indices that feature linearizable range 
scan operations. Compared to its (lock-based) rivals that also support batch 
updates, \jiffy can execute large batch updates up to 7.4$\times$ more 
efficiently.

\end{abstract}

\maketitle

\iftoggle{appendix}{}{
%%% do not modify the following VLDB block %%
%%% VLDB block start %%%
\begingroup\small\noindent\raggedright\textbf{PVLDB Reference Format:}\\
\vldbauthors. \vldbtitle. PVLDB, \vldbvolume(\vldbissue): \vldbpages, \vldbyear.\\
\href{https://doi.org/\vldbdoi}{doi:\vldbdoi}
\endgroup
\begingroup
\renewcommand\thefootnote{}\footnote{\noindent
This work is licensed under the Creative Commons BY-NC-ND 4.0 International License. Visit \url{https://creativecommons.org/licenses/by-nc-nd/4.0/} to view a copy of this license. For any use beyond those covered by this license, obtain permission by emailing \href{mailto:info@vldb.org}{info@vldb.org}. Copyright is held by the owner/author(s). Publication rights licensed to the VLDB Endowment. \\
\raggedright Proceedings of the VLDB Endowment, Vol. \vldbvolume, No. \vldbissue\ %
ISSN 2150-8097. \\
\href{https://doi.org/\vldbdoi}{doi:\vldbdoi} \\
}\addtocounter{footnote}{-1}\endgroup
%%% VLDB block end %%%
}

   \algnewcommand\algorithmicoperation{\textbf{operation}}
\algdef{SE}[MESSAGE]{Operation}{EndOperation}                                   
   [2]{\algorithmicoperation\ \textproc{#1}\ifthenelse{\equal{#2}{}}{}{(#2)}}%
   {\algorithmicend\ \algorithmicoperation}%                

\algnewcommand\algorithmicempty{}
\algdef{SE}[MESSAGE]{Empty}{EndEmpty}                                   
   [2]{\algorithmicempty\ \textproc{#1}\ifthenelse{\equal{#2}{}}{}{(#2)}}%
   {\algorithmicend\ \algorithmicempty}%                

\algnewcommand\algorithmicoperator{\textbf{operator}}
\algdef{SE}[OPERATOR]{Operator}{EndOperator}                                   
   [2]{\algorithmicoperator\ \textproc{#1}\ifthenelse{\equal{#2}{}}{}{(#2)}}%
   {\algorithmicend\ \algorithmicoperator}%                                      

\algnewcommand\algorithmicmessage{\textbf{message}}
\algdef{SE}[MESSAGE]{Message}{EndMessage}                                   
   [2]{\algorithmicmessage\ \textproc{#1}\ifthenelse{\equal{#2}{}}{}{(#2)}}%
   {\algorithmicend\ \algorithmicmessage}%                             
              
\algnewcommand\algorithmicreceive{\textbf{receive}}
\algdef{SE}[MESSAGE]{Receive}{EndReceive}                                   
   [2]{\algorithmicreceive\ \textproc{#1}\ifthenelse{\equal{#2}{}}{}{(#2)}}%
   {\algorithmicend\ \algorithmicreceive}%   

\algnewcommand\algorithmicupon{\textbf{upon}}
\algdef{SE}[UPON]{Upon}{EndUpon}                                   
   [2]{\algorithmicupon\ \textproc{#1}\ifthenelse{\equal{#2}{}}{}{(#2)}}%
   {\algorithmicend\ \algorithmicupon}%
   
\algnewcommand\algorithmicperiodically{\textbf{periodically}}
\algdef{SE}[PERIODICALLY]{Periodically}{EndPeriodically}                                   
   [2]{\algorithmicperiodically\ \textproc{#1}\ifthenelse{\equal{#2}{}}{}{(#2)}}%
   {\algorithmicend\ \algorithmicperiodically}%

\algnewcommand\senddesc{\textbf{send}}
\algnewcommand\Send{\senddesc{} }

\algnewcommand\rbcastdesc{\textbf{rbcast}}
\algnewcommand\Rbcast{\rbcastdesc{} }

\algnewcommand\tobcastdesc{\textbf{tobcast}}
\algnewcommand\Tobcast{\tobcastdesc{} }

\algnewcommand\structdesc{\textbf{struct}}
\algnewcommand\Struct{\structdesc{} }

\algnewcommand\vardesc{\textbf{var}}
\algnewcommand\Var{\vardesc{} }

\algnewcommand\lockstartdesc{\textbf{lock \{}}
\algnewcommand\LockStart{\lockstartdesc{} }

\algnewcommand\lockenddesc{\textbf{\}}}
\algnewcommand\LockEnd{\lockenddesc{} }

\algnewcommand\algindentdesc{\hspace{2.8em}}
\algnewcommand\AlgIndent{\algindentdesc{} }

\algnewcommand\algindentsmalldesc{\hspace{1.3em}}
\algnewcommand\AlgIndentSmall{\algindentsmalldesc{} }

\algnewcommand\algindentindentdesc{\hspace{4.0em}}
\algnewcommand\AlgIndentIndent{\algindentindentdesc{} }

\algnewcommand{\IIf}[1]{\State\algorithmicif\ #1\ \algorithmicthen}
\algnewcommand{\EElse}[1]{\algorithmicelse}
\algnewcommand{\EndIIf}{}%\unskip\ \algorithmicend\ \algorithmicif

\algnotext{EndFor}
\algnotext{EndIf}
% \algtext*{EndIf}
\algnotext{EndUpon}
\algnotext{EndOperator}
\algnotext{EndMessage}
\algnotext{EndOperation}
\algnotext{EndReceive}
\algnotext{EndFunction} 
\algnotext{EndProcedure} 
\algnotext{EndWhile}
\algnotext{EndEmpty}
\algnotext{EndPeriodically}

\newcommand{\algrule}[1][.2pt]{\par\vskip.5\baselineskip\hrule height
#1\par\vskip.5\baselineskip}

\newcommand{\LineComment}[1]{\hfill\textit{// #1}}

\section{Introduction} \label{sec:introduction}

Concurrent programming is inherently difficult. Hence, to develop applications 
and complex systems, such as database engines, which are optimized for modern 
multicore hardware, programmers often rely on \emph{concurrent data 
structures}. These structures expose a well defined interface and can be safely 
used in a multithreaded environment without additional synchronization (see, 
e.g., \cite{JavaConcurrent}). Under the hood, concurrent data structures 
feature sophisticated, often non-blocking synchronization algorithms optimized 
for performance. With the proliferation of multicore hardware in recent years, 
many new concurrent data structures, such as concurrent lists \cite{V95, H01}, 
sets \cite{EFRB10, SRP10, HJ12, BP12, S13, NM14}, (ordered) key-value indices 
(or maps, dictionaries)
% (or maps or dictionaries) 
\cite{MS95, ST03, ST04, F04, FR04, BCC+10, BH11, BBF+12, SGS12, PBG+12, ASS13, 
BBB+17, SW15a, SW18, WSJ18}, etc., have been proposed, each time improving the 
performance over the existing solutions and introducing new features, such as 
the support for consistent range scan operations or snapshots that provide a 
read-only, static and consistent view over the state of the entire dataset.

In this paper, we introduce \emph{\jiffy}, the first linearizable \cite{HW90}, 
lock-free ordered index (sorted key-value map) that besides offering consistent 
snapshots used, e.g., by range scans, provides support for 
\emph{batch updates}, which are put and remove operations that are executed 
atomically. We propose several innovations to make our algorithm highly 
scalable, despite the rich semantics it offers.

% Our novel algorithm uniquely 
% combines several concepts and techniques to deliver highly scalable 
% performance despite the rich semantics it offers.

% The novel design of our index is based on a multiversioned \cite{BG83} skip 
% list \cite{P90}. 
% Since a skip list is a probabilistic data structure, using it 
% as the cornerstone for \jiffy means that our index is less sensitive to skewed 
% workloads contrary to various unbalanced tree-based data structures, such as 
% \cite{EFRB10, BH11, BBF+12, SW15a, SW18, WSJ18}. 

% Unlike many existing multiversioned concurrent indices, which rely on a 
% single atomic counter to generate version numbers, e.g., \cite{LBDF+11, 
% LMLM+16, BBB+17}, \jiffy obtains version numbers by reading CPU's Time Stamp 
% Counter (TSC) register \cite{Intel08, DLS13}, a high-resolution clock, 
% that has been available for many years on popular hardware architectures, 
% such as x86\_64. Reading the TSC register is an extremely fast operation as 
% it does not involve a system call. In turn, \jiffy does not feature a single 
% point of contention and offers scalable performance on modern 40+ core CPUs. 

The novel design of our index is based on a multiversioned \cite{BG83} skip 
list \cite{P90}. However, unlike many existing multiversioned concurrent 
indices, which rely on a single atomic counter to generate version numbers, 
e.g., \cite{LBDF+11, LMLM+16, BBB+17}, \jiffy's concurrency control mechanism 
is specially designed to use version numbers obtained by reading 
CPU's Time Stamp Counter (TSC) register \cite{Intel08, DLS13}, a 
high-resolution clock available on the x86\_64 platform. Reading the TSC 
register is an extremely fast operation as it does not involve a system call. 
In turn, \jiffy does not feature a single point of contention and offers 
scalable performance on modern 40+ core CPUs. 

Key-value entries are grouped in \jiffy into immutable objects, called 
\emph{revisions}, which are tagged with a version number. The use of revisions 
instead of maintaining each key-value pair as a separate object has several 
benefits. Firstly, the use of revisions allows the index to be smaller and thus 
quicker to traverse. Secondly, accesses to individual key-value entries can be 
performed more efficiently through the use of a lightweight hash index inside 
each revision, whereas range scans can benefit from keys and values being 
stored in sorted arrays within the revision. Crucially, however, by growing or 
shrinking the skip list and thus modifying the sizes of revisions, we can 
optimize the synchronization granularity in \jiffy, which allows it to adapt to 
changing workloads.
% , similarly to
Smaller revisions are more suited for write-heavy workloads whereas large 
revisions benefit read-dominated workloads, especially when they feature many 
range scan operations. Automatic adaptation to the workload is accomplished
on per-revision basis through a simple, yet versatile policy based on 
monitoring the time concurrent threads spend executing update (i.e., put, 
remove and batch update) and read (i.e., lookup or range scan) operations, not 
by counting the number of operations performed or monitoring the contention on 
shared references, as in other existing approaches, e.g., \cite{SW15a, SW18, 
WSJ18}. 

% The layout of a revision and the hash index facilitate both ultra-fast key 
% lookup and well as efficient processing of a revision during a range scan.
% The use of revisions instead of maintaining each key-value pair as a separate 
% object has several benefits. Firstly, the use of revisions allows the index 
% to be smaller and thus quicker to traverse. 

The core contribution of our paper is, however, the novel lock-free algorithm 
that enables updates, reads, as well as index structure modifications, which 
facilitate varying the sizes of revisions. Structure modifications are 
streamlined with updates and happen through the skip list node \emph{split} and 
\emph{merge} operations based on the atomic compare-and-swap (CAS) operations. 
% that are carefully coordinated with update operations. 
Our algorithm is based on a few simple rules all threads in \jiffy must abide:
\begin{itemize}
\item always help to complete a structure modification when encountering one,
\item a node split happens \emph{towards} higher keys (a new node inherits the 
upper half of the key range of the node that undergoes a split operation), 
\item merges happen \emph{towards} lower keys (the preceding node inherits the 
key range of the node that undergoes a merge operation),
\item batch updates proceed from the largest keys included in a batch  
towards the lower keys.
\end{itemize}

We implemented \jiffy in Java and extensively tested it on various workloads 
against the state-of-the-art lock-free ordered indices that feature linearizable 
range scans \cite{BCC+10, BH11, BBF+12, SW18} and the (lock-based) 
ordered indices that also support batch updates \cite{WSJ18}. Our tests show
the highly scalable performance of \jiffy, which is comparable or exceeds the 
performance of the other systems. Crucially, due to its lock-free architecture, 
\jiffy can execute large batch updates much more efficiently compared to its 
(lock-based) rivals, with speedup in throughput ranging from 1.1$\times$ to 
7.4$\times$, depending on a test scenario.

% \subsection{Contributions}
% 
% We summerize the contributions of this paper below:
% \begin{itemize}
% \item df
% \end{itemize}

% \subsection{Paper structure}
% 
% In \Cref{sec:related_work} we discuss research related in our work. Then, 
% in \Cref{sec:design} we describe in detail our novel key-value index and argue 
% about its correctness. We present the results of experimental evaluation of 
% \jiffy in \Cref{sec:evaluation} and conclude in \Cref{sec:conclusions}.
% 

% \input{1a-background.tex}
% \input{2-related_work.tex}
\section{Related work} \label{sec:related_work}

A template for obtaining non-blocking algorithms for concurrent data
structures based on CAS was originally proposed by Herlihy \cite{H90} 
\cite{H91}. In practice, however, implementations based on this approach suffer
from low parallelism and high overhead due to excessive copying and reliance 
on a single global pointer accessed through CAS by all threads. Much better 
performing ordered index implementations can be achieved through purposefully 
designed (non-blocking) algorithms, which we discuss next. In particular, we 
focus on non-blocking skip lists and other high performance ordered indices that 
support snapshots and batch updates.

Skip lists were first introduced by Pugh \cite{P90}. Valois \cite{V95} was the 
first to sketch a lock-free algorithm for a skip list, although the first 
complete algorithm was proposed by Sundell and Tsigas \cite{ST04}, as an 
extension of their prior work on concurrent priority queues \cite{ST03}. Their 
implementation relied on the CAS and FAA (fetch-and-add)-based lock-free memory 
management scheme originally proposed by Valois \cite{V95} and later revised by 
Michael and Scott \cite{MS95}. 

Frasier \cite{F04} gave an alternative implementation of a lock-free skip list, 
which relies on Harris' CAS-based approach for implementing lock-free linked 
lists \cite{H01}. Fomitchev and Ruppert's implementation of a lock-free skip 
list \cite{FR04} combines the techniques of Valois and Harris.
% Harris' approach involves marking pointers to facilitate node 
% deletions, as opposed to Valois idea that involved using auxilary nodes.
The ubiquitous 
\texttt{ConcurrentSkipListMap} 
%by Doug Lea 
\cite{JavaConcurrent}, which is part of the standard Java 
\texttt{java.util.concurrent} library, draws from Freiser's, Fomitchev's and 
Sundell's work. All algorithms discussed above are linearizable 
\cite{HW90} except for range scans. Moreover, unlike \jiffy, they do not 
support batch updates or snapshots.

LeapList \cite{ASS13} and KiWi \cite{BBB+17} are skip list-based indices that 
provide linearizable range scans (but no fully linearizable snapshots, 
as \jiffy). LeapList relies on fine-grained locks and Software Transactional 
Memory (STM) for concurrency control whereas KiWi features a multiversioned 
architecture and CAS-based operations to provide lock-freedom (range 
scans are wait-free). However, not every update operation in KiWi creates 
a new version: without concurrent range scans, an update operation simply 
overwrites the old value in the index. Version numbers are managed through an 
atomic counter, which is bound to become a bottleneck (in \jiffy we rely on the 
TSC register for this purpose, see below). Each of the base nodes in both 
LeapList and KiWi holds $k$ key-value entries for cache-friendliness, but $k$ 
is fixed (unlike in \jiffy). 

Nitro \cite{LMLM+16} is a skip list-based index used in Couchbase. Nitro uses 
multiversioning to provide snapshots, but the creation of a new snapshot is not 
a thread-safe operation (it cannot be executed concurrently with put or remove 
operations).

Now we discuss tree-based ordered index data structures. SnapTree by Bronson 
\etal \cite{BCC+10} is a lock-based relaxed balance AVL tree. SnapTree uses a 
linearizable clone operation for atomic snapshots and range scans, which can 
severely slow down concurrent update operations (in \jiffy, creating a snapshot, 
which is also used for a range scan, is an O(1) operation that does not impact 
concurrent operations in any way). Brown \etal proposes k-ary search trees 
\cite{BH11, BBF+12}, which are a generalization of lock-free binary search 
trees by Ellen \etal \cite{EFRB10}. Range scans undergo a validation phase for 
ensuring linearizability and are restarted when a concurrent update is detected 
(in \jiffy, a range scan may help to complete some concurrent update operations, 
but is never restarted). CTrie \cite{PBG+12} is a lock-free concurrent hash trie 
based on CAS. Atomic snapshots are provided through a lazy copy-on-write 
operation, which slows down concurrent update operations. In CTrie no partial 
snapshots can be obtained. 
% Bw-tree \cite{LLS13} is a lock-free index based on a B-tree, that is 
% optimized for flash storage.
Minuet \cite{SGS12} is a distributed, in-memory B-tree index with linearizable 
snapshots. To create snapshots, Minuet also relies on a relatively expensive 
copy-on-write method, but allows snapshots to be shared across multiple range 
scans.

Sagonas \etal proposed a number of \emph{contention-adapting (CA)} tree-based 
data structures with linearizable range scans. The data structures
feature a lock-based \cite{SW15a, SW18} or a lock-free 
binary search tree \cite{WSJ18} as the main part of the index, where each leaf 
node is a variable-size \emph{container}, i.e., an AVL tree, a skip list or an 
immutable data structure that holds multiple key-value entries (which is 
similar to a \emph{revision} in \jiffy). The size of the container is adjusted 
to the observed contention level (we discuss the differences with our 
index autoscaling policy in \Cref{sec:design:autoscaling}). Linearizable range 
scans are achieved either through locking, optimistic scan and validation or 
replacing the leaf data structures using CAS with special objects that can be 
used by concurrent threads to help with completing the range scan (and to block 
update operations in the meantime). From all of the data structures we 
discussed so far, only the lock-based variants of the CA trees support batch 
update operations.

Besides works of Sagonas \etal on CA trees, we are aware of several works on 
data structures that dynamically adapt to changing contention levels, e.g., 
\cite{AKKM+12, CGR13}. Unlike CA trees, 
% however, 
none of the proposed algorithms support linearizable range scans or batch 
updates.

Finally, several researchers have investigated general techniques for adding 
% (non-blocking) 
linearizable range scans (but not batch updates) to existing concurrent data 
structures, e.g. \cite{MS98, PT13, MSFM15, C17, AB18}. 

% Petrank \etal \cite{PT13} propose a general technique for adding wait-free
% range scan operations to linearizable sets, so also skip lists. This approach
% has been further developed by Chatterjee \cite{C17} to enable multiple 
% concurrent snapshots on disjoint key ranges. The scalability of Chatterjee's 
% approach is limited due to a global sequential hot spot that can taccessed by  

The concurrency control mechanism implemented in \jiffy 
% \jiffy's concurrency control mechanism 
shares some 
similarities with the multiversioned transactional engine in \cite{LBDF+11}, 
which also relies on structures similar to our batch descriptors and 
CAS operations to ensure that all updates become visible to 
concurrent operations atomically. However, unlike 
% the mentioned 
this implementation, \jiffy is lock-free and no update operation, including 
batch updates, ever aborts. Crucially, instead of using a shared atomic counter 
to generate version numbers, 
% which supports only 
% hash indices, \kdb keeps data in a sorted order. Moreover, 
\jiffy relies on the CPU's Time Stamp Counter (TSC) register 
% (the \rtdscp instruction) 
\cite{Intel08}, which greatly helps to reduce contention between 
concurrent threads on modern 40+ core CPUs. TSC has been used for a similar 
purpose also in the context of transactional memory \cite{DLS13} 
\cite{GDV18}, a concurrent stack implementation \cite{DHK15}, and a 
serializable (but not linearizable) database engine \cite{LKA17}.

\section{Design of \jiffy} \label{sec:design}

In this section, we discuss the architecture of \jiffy, the crucial details 
regarding its implementation, and also argue about its correctness.

\begin{figure}[t]
\bgroup
\input{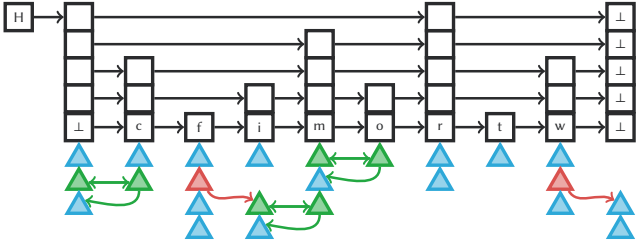}
% \begin{tabular}{cc}
% % \vspace{-2cm}
% \begin{minipage}{0.4\textwidth}
% \vspace{-4cm}
% % \multicolumn{1}{b{0.4\textwidth}}{
% \scalebox{0.7}{
% \input{figures/skiplist.tex}
% }
% \end{minipage}
% & 
% \scalebox{0.75}{
% \input{figures/shard.tex}
% }
% \end{tabular}

% \includegraphics[scale=0.2]{figures/skiplist.png}
\scalebox{0.8}{
\bgroup
\begin{tikzpicture}[font=\sf\footnotesize] 
\newcommand{\ventryDist}{1}
\newcommand{\indexShift}{0.03}
\newcommand{\indexDist}{0.45}
\newcommand{\revisionDist}{0.4}
\newcommand{\revisionShift}{-0.12}

\node (startNode) at (0,0) {};

\topEntry{e0}{$ (startNode) + (0 * \ventryDist, 0)$}{$\bot$}
\topEntry{e1}{$ (startNode) + (1 * \ventryDist, 0)$}{c}
\topEntry{e2}{$ (startNode) + (2 * \ventryDist, 0)$}{f}
\topEntry{e3}{$ (startNode) + (3 * \ventryDist, 0)$}{i}
\topEntry{e4}{$ (startNode) + (4 * \ventryDist, 0)$}{m}
\topEntry{e5}{$ (startNode) + (5 * \ventryDist, 0)$}{o}
\topEntry{e6}{$ (startNode) + (6 * \ventryDist, 0)$}{r}
\topEntry{e7}{$ (startNode) + (7 * \ventryDist, 0)$}{t}
\topEntry{e8}{$ (startNode) + (8 * \ventryDist, 0)$}{w}
\topEntry{e9}{$ (startNode) + (9 * \ventryDist, 0)$}{$\bot$}

\topIdx{i02}{$ (startNode) + (0 * \ventryDist, \indexShift + 1 * \indexDist)$}{}
\topIdx{i03}{$ (startNode) + (0 * \ventryDist, \indexShift + 2 * \indexDist)$}{}
\topIdx{i04}{$ (startNode) + (0 * \ventryDist, \indexShift + 3 * \indexDist)$}{}
\topIdx{i05}{$ (startNode) + (0 * \ventryDist, \indexShift + 4 * \indexDist)$}{}
 
\topIdx{i12}{$ (startNode) + (1 * \ventryDist, \indexShift + 1 * \indexDist)$}{}
\topIdx{i13}{$ (startNode) + (1 * \ventryDist, \indexShift + 2 * \indexDist)$}{}
 
\topIdx{i32}{$ (startNode) + (3 * \ventryDist, \indexShift + 1 * \indexDist)$}{}

\topIdx{i42}{$ (startNode) + (4 * \ventryDist, \indexShift + 1 * \indexDist)$}{}
\topIdx{i43}{$ (startNode) + (4 * \ventryDist, \indexShift + 2 * \indexDist)$}{}
\topIdx{i44}{$ (startNode) + (4 * \ventryDist, \indexShift + 3 * \indexDist)$}{}
 
\topIdx{i52}{$ (startNode) + (5 * \ventryDist, \indexShift + 1 * \indexDist)$}{}

\topIdx{i62}{$ (startNode) + (6 * \ventryDist, \indexShift + 1 * \indexDist)$}{}
\topIdx{i63}{$ (startNode) + (6 * \ventryDist, \indexShift + 2 * \indexDist)$}{}
\topIdx{i64}{$ (startNode) + (6 * \ventryDist, \indexShift + 3 * \indexDist)$}{}
\topIdx{i65}{$ (startNode) + (6 * \ventryDist, \indexShift + 4 * \indexDist)$}{}
                                                                         
\topIdx{i82}{$ (startNode) + (8 * \ventryDist, \indexShift + 1 * \indexDist)$}{}                                                                         
\topIdx{i83}{$ (startNode) + (8 * \ventryDist, \indexShift + 2 * \indexDist)$}{}                                                                         

\topIdx{i92}{$ (startNode) + (9 * \ventryDist, \indexShift + 1 * \indexDist)$}{$\bot$}                                                                         
\topIdx{i93}{$ (startNode) + (9 * \ventryDist, \indexShift + 2 * \indexDist)$}{$\bot$}                                                                         
\topIdx{i94}{$ (startNode) + (9 * \ventryDist, \indexShift + 3 * \indexDist)$}{$\bot$}                                                                         
\topIdx{i95}{$ (startNode) + (9 * \ventryDist, \indexShift + 4 * \indexDist)$}{$\bot$}                                                                         

\topIdx{h}{$ (startNode) + (-1 * \ventryDist, \indexShift + 4 * \indexDist)$}{H}

\draw[top arrow] (h) -- (i05);

\draw[top arrow] (e0) -- (e1);
\draw[top arrow] (e1) -- (e2);
\draw[top arrow] (e2) -- (e3);
\draw[top arrow] (e3) -- (e4);
\draw[top arrow] (e4) -- (e5);
\draw[top arrow] (e5) -- (e6);
\draw[top arrow] (e6) -- (e7);
\draw[top arrow] (e7) -- (e8);
\draw[top arrow] (e8) -- (e9);

\draw[top arrow] (i02) -- (i12);
\draw[top arrow] (i12) -- (i32);
\draw[top arrow] (i32) -- (i42);
\draw[top arrow] (i42) -- (i52);
\draw[top arrow] (i52) -- (i62);
\draw[top arrow] (i62) -- (i82);
\draw[top arrow] (i82) -- (i92);

\draw[top arrow] (i03) -- (i13);
\draw[top arrow] (i13) -- (i43);
\draw[top arrow] (i43) -- (i63);
\draw[top arrow] (i63) -- (i83);
\draw[top arrow] (i83) -- (i93);

\draw[top arrow] (i04) -- (i44);
\draw[top arrow] (i44) -- (i64);
\draw[top arrow] (i64) -- (i94);

\draw[top arrow] (i05) -- (i65);
\draw[top arrow] (i65) -- (i95);

\revisionN{r01}{$ (startNode) + (0 * \ventryDist, \revisionShift - 1 * \revisionDist) $}{}
\splitRevisionN{r02}{$ (startNode) + (0 * \ventryDist, \revisionShift - 2 * \revisionDist) $}{}
\revisionN{r03}{$ (startNode) + (0 * \ventryDist, \revisionShift - 3 * \revisionDist) $}{}

\revisionN{r11}{$ (startNode) + (1 * \ventryDist, \revisionShift - 1 * \revisionDist) $}{}
\splitRevisionN{r12}{$ (startNode) + (1 * \ventryDist, \revisionShift - 2 * \revisionDist) $}{}

\revisionN{r21}{$ (startNode) + (2 * \ventryDist, \revisionShift - 1 * \revisionDist) $}{}
\mergeRevisionN{r22}{$ (startNode) + (2 * \ventryDist, \revisionShift - 2 * \revisionDist) $}{}
\revisionN{r23}{$ (startNode) + (2 * \ventryDist, \revisionShift - 3 * \revisionDist) $}{}
\revisionN{r24}{$ (startNode) + (2 * \ventryDist, \revisionShift - 4 * \revisionDist) $}{}

\revisionN{r31}{$ (startNode) + (3 * \ventryDist, \revisionShift - 1 * \revisionDist) $}{}
\splitRevisionN{r32}{$ (startNode) + (3 * \ventryDist, \revisionShift - 3 * \revisionDist) $}{}
\revisionN{r33}{$ (startNode) + (3 * \ventryDist, \revisionShift - 4 * \revisionDist) $}{}

\splitRevisionN{r41}{$ (startNode) + (4 * \ventryDist, \revisionShift - 1 * \revisionDist) $}{}
\revisionN{r42}{$ (startNode) + (4 * \ventryDist, \revisionShift - 2 * \revisionDist) $}{}
\splitRevisionN{r43}{$ (startNode) + (4 * \ventryDist, \revisionShift - 3 * \revisionDist) $}{}

\splitRevisionN{r51}{$ (startNode) + (5 * \ventryDist, \revisionShift - 1 * \revisionDist) $}{}

\revisionN{r61}{$ (startNode) + (6 * \ventryDist, \revisionShift - 1 * \revisionDist) $}{}
\revisionN{r62}{$ (startNode) + (6 * \ventryDist, \revisionShift - 2 * \revisionDist) $}{}

\revisionN{r71}{$ (startNode) + (7 * \ventryDist, \revisionShift - 1 * \revisionDist) $}{}

\revisionN{r81}{$ (startNode) + (8 * \ventryDist, \revisionShift - 1 * \revisionDist) $}{}
\mergeRevisionN{r82}{$ (startNode) + (8 * \ventryDist, \revisionShift - 2 * \revisionDist) $}{}
\revisionN{r83}{$ (startNode) + (8 * \ventryDist, \revisionShift - 3 * \revisionDist) $}{}

\revisionN{r91}{$ (startNode) + (9 * \ventryDist, \revisionShift - 3 * \revisionDist) $}{}
\revisionN{r92}{$ (startNode) + (9 * \ventryDist, \revisionShift - 4 * \revisionDist) $}{}

% \draw[rev arrow short] (e0) -- (r01);
% \draw[rev arrow short] (r01) -- (r02);
% \draw[rev arrow short] (r02) -- (r03);
% 
% \draw[rev arrow short] (e1) -- (r11);
% \draw[rev arrow short] (r11) -- (r12);
% 
% \draw[rev arrow short] (e2) -- (r21);
% \draw[rev arrow short] (r21) -- (r22);
% \draw[rev arrow short] (r22) -- (r23);
% \draw[rev arrow short] (r23) -- (r24);
% 
% \draw[rev arrow short] (e3) -- (r31);
% \draw[rev arrow short] (r32) -- (r33);
% 
% \draw[rev arrow short] (e4) -- (r41);
% \draw[rev arrow short] (r41) -- (r42);
% \draw[rev arrow short] (r42) -- (r43);
% 
% \draw[rev arrow short] (e5) -- (r51);
% 
% \draw[rev arrow short] (e6) -- (r61);
% \draw[rev arrow short] (r61) -- (r62);
% 
% \draw[rev arrow short] (e7) -- (r71);
% 
% \draw[rev arrow short] (e8) -- (r81);
% \draw[rev arrow short] (r81) -- (r82);
% \draw[rev arrow short] (r82) -- (r83);
% 
% \draw[rev arrow short] (r91) -- (r92);

\draw[split arrow] (r02) -- (r12);
\draw[split arrow] (r32) -- (r43);
\draw[split arrow] (r41) -- (r51);

\draw[split arrow] (r12) -- (r02);
\draw[split arrow] (r43) -- (r32);
\draw[split arrow] (r51) -- (r41);

\draw[split arrow] (r12.south) to[out=270, in=350, looseness=1.0] (r03.north east);
\draw[split arrow] (r43.south) to[out=270, in=350, looseness=1.0] (r33.north east);
\draw[split arrow] (r51.south) to[out=270, in=350, looseness=1.0] (r42.north east);

\draw[merge arrow] (r22.south east) to[out=300, in=150, looseness=1.0] (r32.north west);
\draw[merge arrow] (r82.south east) to[out=300, in=150, looseness=1.0] (r91.north west);

\end{tikzpicture}
\egroup
}
\egroup
\caption{The multiversioned architecture of \jiffy. Each node of the 
lowest-level list of the skip list manages a range of keys, e.g., $(-\infty, 
c)$, $[c, f)$, $[f, i)$, etc. Key-value entries are kept in immutable revisions 
(triangles), each in a concrete version, with newest at the top. The skip list 
grows and shrinks by splitting or merging nodes and through split and merge 
revisions (colored green and red, respectively).}
\label{fig:skiplist}
\end{figure}

\subsection{The architecture overview} \label{sec:design:overview}

\jiffy is a multiversioned \cite{BG83} skip list \cite{P90}, where each 
\emph{node} (an object on the lowest-level linked list of the skip list) 
manages a continuous range of keys (see \Cref{fig:skiplist}). More 
precisely, each node stores (1) a \emph{node key}, i.e., a key that represents 
the lower end of the managed key range (the exclusive upper end is defined by 
the node key of the successor node), and (2) a reference to the head of a 
\emph{revision list}. The revision list consists of \emph{revisions}, 
immutable objects that store key-value entries that fit the node's range (we 
discuss the layout of data in a revision in \Cref{sec:design:revision}). Each 
revision is tagged with a version number, which thus serves as the version 
number for each key-value entry stored in the revision. Unlike in a classic 
skip list, the first node, called the \emph{base} node, is not just a sentinel 
but also manages a range of entries (its key is $\bot$, and thus its key range 
is $(-\infty, c)$ in our example). Update operations, such as put, remove or 
batch update use the compare-and-swap (CAS)\footnote{$\CAS(\mathit{val}, 
\mathit{oldVal}, \mathit{newVal})$ atomically replaces $\mathit{val}$ with 
$\mathit{newVal}$ only if $\mathit{val} = \mathit{oldVal}$. The operation 
returns a boolean value that indicates if the 
operation was successful.} operation to add a new revision as the head of the 
revision list (we simply say that a new revision is added to the node) and cut 
the list short whenever the internal garbage collector indicates that certain 
revisions will not be needed any more.

In \jiffy, \emph{structure modifications}, i.e., changes to the index, are 
more involved compared to a typical lock-free skip list, such as 
\cite{JavaConcurrent}, where nodes are added or removed upon inserting new keys 
or removing the existing ones. In our approach, the index grows by 
splitting a node into two and shrinks by merging two nodes into one (see 
details in \Cref{sec:design:structure_modifications}). The index 
starts with a single \emph{base} node (with key $\bot$). During a split of a 
node with key $n$ (referred to as node $n$), a new node $n'$, where $n' > n$ is 
added directly after node $n$ (or node $\bot$ if the base node undergoes a 
split). Node $n'$ inherits the upper half of the key range originally assigned 
to node $n$ (the key of node $n$ does not change). On the other hand, during a 
merge operation of node $n$, it is merged with the node directly preceding 
node $n$ in the index (so with a node with a strictly lower key; the base node 
cannot undergo a merge operation and is never removed). As in a classic skip 
list, the \emph{index nodes} (i.e., the nodes on all but the lowest-level 
linked lists, which facilitate fast traversals of the data structure), are 
inserted to the higher-level linked lists probabilistically (in our 
implementation, the probability of inserting index nodes up to a certain level 
is the same as in \cite{JavaConcurrent}). Operations on higher-level linked 
lists are also performed using CAS.

\begin{figure}[t]
\bgroup

\newcommand{\separation}{1.45cm}
\input{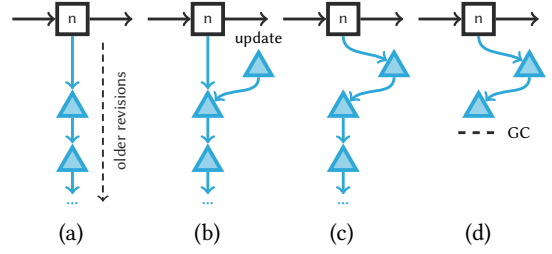}
\hspace{1.2cm}\begin{tabular}{p{\separation}p{\separation}p{\separation}p{\separation}}
% \begin{minipage}{0.4\textwidth}
% \vspace{-4cm}
\multicolumn{4}{c}{
\hspace{-1.1cm}
\scalebox{0.9}{
\bgroup
\begin{tikzpicture}[font=\sf\footnotesize] 
\newcommand{\ventryDist}{1}
\newcommand{\indexShift}{0.03}
\newcommand{\indexDist}{0.45}
\newcommand{\revisionDist}{1.2}
\newcommand{\shortRevisionDist}{0.8}
\newcommand{\revisionShift}{-0.12}

\newcommand{\scopeShift}{2.0}

\begin{scope}[shift={(0, 0)}]
    \node (startNode) at (0,0) {};

    \node (left) at ($ (startNode) + (-1 * \ventryDist, 0)$) {};
    \topEntry{e0}{$ (startNode) + (0 * \ventryDist, 0)$}{n}
    \node (right) at ($ (startNode) + (1 * \ventryDist, 0)$) {};

    \draw[top arrow] (left) -- (e0);
    \draw[top arrow] (e0) -- (right);

    \revisionN{r01}{$ (startNode) + (0 * \ventryDist, \revisionShift - 1 * \revisionDist) $}{}
    \revisionN{r02}{$ (startNode) + (0 * \ventryDist, \revisionShift - 1 * \revisionDist - 1 * \shortRevisionDist) $}{}
    \node[text=myBlue] (r03) at ($ (startNode) + (0 * \ventryDist, \revisionShift - 1 * \revisionDist - 1.75 * \shortRevisionDist) $) {\textbf{...}};
    
    \draw[rev arrow] (e0) -- (r01);
    \draw[rev arrow] (r01) -- (r02);
    \draw[rev arrow] (r02) -- (r03);
    
    \draw[thick, draw=myGrey, ->, densely dashed] ($ (startNode) + (0.45, -0.35)$) -- ($ (startNode) + (0.45, -2.7)$);
    \node[text=myGrey, rotate=90] at ($ (startNode) + (0.7, -1.5)$) {older revisions};
\end{scope}

\begin{scope}[shift={(1 * \scopeShift, 0)}]
    \node (startNode) at (0,0) {};

    \node (left) at ($ (startNode) + (-1 * \ventryDist, 0)$) {};
    \topEntry{e0}{$ (startNode) + (0 * \ventryDist, 0)$}{n}
    \node (right) at ($ (startNode) + (1 * \ventryDist, 0)$) {};

    \draw[top arrow] (left) -- (e0);
    \draw[top arrow] (e0) -- (right);

    \revisionN{r01}{$ (startNode) + (0 * \ventryDist, \revisionShift - 1 * \revisionDist) $}{}
    \revisionN{r02}{$ (startNode) + (0 * \ventryDist, \revisionShift - 1 * \revisionDist - 1 * \shortRevisionDist) $}{}
    \node[text=myBlue] (r03) at ($ (startNode) + (0 * \ventryDist, \revisionShift - 1 * \revisionDist - 1.75 * \shortRevisionDist) $) {\textbf{...}};
    
    \draw[rev arrow] (e0) -- (r01);
    \draw[rev arrow] (r01) -- (r02);
    \draw[rev arrow] (r02) -- (r03);
    
    \revisionN{rn}{$ (startNode) + (0.75 * \ventryDist, \revisionShift - 0.5 * \revisionDist) $}{}
    \node (gc) at ($ (startNode) + (0.75 * \ventryDist, \revisionShift - 0.15 * \revisionDist) $) {update};
    
    \draw[rev arrow] (rn.south) to[out=270, in=50, looseness=1.0] (r01.north east);
\end{scope}

\begin{scope}[shift={(2 * \scopeShift, 0)}]
    \node (startNode) at (0,0) {};

    \node (left) at ($ (startNode) + (-1 * \ventryDist, 0)$) {};
    \topEntry{e0}{$ (startNode) + (0 * \ventryDist, 0)$}{n}
    \node (right) at ($ (startNode) + (1 * \ventryDist, 0)$) {};

    \draw[top arrow] (left) -- (e0);
    \draw[top arrow] (e0) -- (right);

    \revisionN{r01}{$ (startNode) + (0 * \ventryDist, \revisionShift - 1 * \revisionDist) $}{}
    \revisionN{r02}{$ (startNode) + (0 * \ventryDist, \revisionShift - 1 * \revisionDist - 1 * \shortRevisionDist) $}{}
    \node[text=myBlue] (r03) at ($ (startNode) + (0 * \ventryDist, \revisionShift - 1 * \revisionDist - 1.75 * \shortRevisionDist) $) {\textbf{...}};
    
%     \draw[rev arrow] (e0) -- (r01);
    \draw[rev arrow] (r01) -- (r02);
    \draw[rev arrow] (r02) -- (r03);
    
    \revisionN{rn}{$ (startNode) + (0.75 * \ventryDist, \revisionShift - 0.5 * \revisionDist) $}{}
    
    \draw[rev arrow] (rn.south) to[out=270, in=50, looseness=1.0] (r01.north east);
    \draw[rev arrow] (e0.south) to[out=270, in=150, looseness=1.0] (rn.north west);
\end{scope}

\begin{scope}[shift={(3 * \scopeShift, 0)}]
    \node (startNode) at (0,0) {};

    \node (left) at ($ (startNode) + (-1 * \ventryDist, 0)$) {};
    \topEntry{e0}{$ (startNode) + (0 * \ventryDist, 0)$}{n}
    \node (right) at ($ (startNode) + (1 * \ventryDist, 0)$) {};

    \draw[top arrow] (left) -- (e0);
    \draw[top arrow] (e0) -- (right);

    \revisionN{r01}{$ (startNode) + (0 * \ventryDist, \revisionShift - 1 * \revisionDist) $}{}
%     \revisionN{r02}{$ (startNode) + (0 * \ventryDist, \revisionShift - 1 * \revisionDist - 1 * \shortRevisionDist) $}{}
%     \node[text=myBlue] (r03) at ($ (startNode) + (0 * \ventryDist, \revisionShift - 1 * \revisionDist - 1.75 * \shortRevisionDist) $) {\textbf{...}};
    
%     \draw[rev arrow] (e0) -- (r01);
%     \draw[rev arrow] (r01) -- (r02);
%     \draw[rev arrow] (r02) -- (r03);
    
    \revisionN{rn}{$ (startNode) + (0.75 * \ventryDist, \revisionShift - 0.5 * \revisionDist) $}{}
    
    \draw[rev arrow] (rn.south) to[out=270, in=50, looseness=1.0] (r01.north east);
    \draw[rev arrow] (e0.south) to[out=270, in=150, looseness=1.0] (rn.north west);
    
    \draw[dashed, draw=myGrey,-, very thick] ($ (startNode) + (0 * \ventryDist - 0.3, \revisionShift - 1.3 * \revisionDist) $) -- ($ (startNode) + (0 * \ventryDist + 0.3, \revisionShift - 1.3 * \revisionDist) $);
    \node (gc) at ($ (startNode) + (0 * \ventryDist + 0.6, \revisionShift - 1.3 * \revisionDist) $) {GC};
\end{scope}

\end{tikzpicture}
\egroup
}
} \\
(a) & (b) & (c) & (d)\\
\end{tabular}
\egroup
\vspace{-0.3cm}
\caption{Regular update operation: (a) initial state, (b) create a new revision, 
(c) add the new revision to the node (CAS), (d) garbage collect obsolete 
revisions.}
\label{fig:update}
\end{figure}

A node split or a merge can occur only upon some update operation, i.e., 
put, remove or batch update (which we discuss in detail in 
Sections~\ref{sec:design:updates}-\ref{sec:design:batchUpdates}). When an update 
operation of some key $k$ is performed and the appropriate node is found (i.e., 
node $n$, where $k \ge n$ and there does not exist a node $n'$ where $k \ge 
n'$), an \emph{autoscaling policy} decides how the update is to be performed 
(we discuss the details of our autoscaling policy in 
\Cref{sec:design:autoscaling}). 
In majority of cases, a \emph{regular} update is performed (see 
\Cref{fig:update}). Regular update involves copying the head of the revision 
list at node $n$, applying the update on the copied revision, 
adding it to the revision list and, if necessary, garbage collecting obsolete 
revisions, i.e., revisions that will never be read again, 
including in any snapshot. Otherwise, a node split or a merge is performed. In 
case of a node split, the update operation is reflected in one of the two new 
\emph{split} revisions (\emph{left split revision} inserted as the head of the 
revision list on node $n$ and \emph{right split revision} as the 
head of the revision list on the new node). On the other hand, in case 
of a merge, the new \emph{merge revision} (on the node directly preceding 
node $n$ in the index) includes the update to $k$, as well the entries for all 
other keys previously stored within the two nodes. 
% SHORT
% Note that 
Node splits and merges % of nodes 
mean that now revision lists are not just simple linked lists: through split 
and merge revisions, revision lists branch and join.

\jiffy is a lock-free data structure, which means that it guarantees 
system-wide progress. To this end, threads occasionally \emph{help} one another 
in completing other (update) operations (in case, e.g., some thread is 
preempted for a long time). Doing so may involve a number of steps, especially 
in case of batch updates or updates that result in node splits or merges. To 
ensure orderly execution of all update operations, we define the following 
rules:
\begin{enumerate}
\item any operation (so also a lookup or a range scan) that encounters
a node split or a merge, helps to complete the operation that invoked the split 
or merge,
% \item before a merge of nodes $n$ and $n'$, $n < n'$, can happen, there 
% cannot be pending (not yet completed) operations on node $n$,
\item an operation can add a new revision $r$ to the revision list 
at some node $n$ only if there is no pending operation at node $n$ (the thread 
helps to complete the pending operations before adding $r$),
\item the execution of a batch update, which comprises of a set of 
put and remove operations, starts by updating the highest key in the 
batch and always continues towards lower keys.
\end{enumerate}
Rule (1) means that our index returns to a stable state (i.e., without ongoing 
structure changes) as soon as possible, so that subsequent operations 
(including lookups and range scans) can be performed efficiently. Rules (2) and 
(3) enforce a consistent order of performing updates (also across 
batch updates), thus allowing \jiffy to guarantee linearizability.
Rules (2) and (3) also give precedence to operations that happen on nodes with 
lower keys, thus preventing live-locks (e.g., two threads operating on 
neighboring nodes, with one constantly attempting to perform a split, the
other a merge).\footnote{Recall that a merge operation on some node $n$ 
involves adding a merge revision to the existing node directly preceding 
node $n$ in the index, which is a much more complex operation than adding a new 
node in a split operation.}

The lock-free nature of \jiffy inevitably means that under some highly 
unfavorable workloads, helping other threads will have a convoying effect which 
results in all threads attempting to complete the same updates/splits/merges 
thus wasting resources. This, however, is unavoidable if we are to guarantee
system-wide progress.

\subsection{Version numbers} \label{sec:design:versions}

We already briefly stated that \jiffy is a multiversioned data structure.
Now we discuss how version numbers are generated and used.

To provide linearizable behavior \cite{HW90} (intuitively, all operations 
appear as if they were executed sequentially on a single CPU), threads in a 
multiversion system typically synchronize on a shared (atomic) 
counter, which is used to generate version numbers (see, e.g., \cite{BBB+17}). 
This, however, introduces a point of contention that quickly becomes a 
bottleneck.\footnote{Reading the atomic counter is also necessary to create 
snapshots of the dataset. The first version of \jiffy that relied an atomic 
counter to generate version numbers did not scale past 4-8 threads.} In \jiffy 
we avoid such a bottleneck by relying on a high-resolution clock supported by 
CPU. More precisely, version numbers are obtained by reading the Time Stamp 
Counter (TSC), a 64-bit register (available on the x86\_64 architecture since 
2008), which functions as a CPU-cycle-level resolution wall-clock for the 
entire multi-CPU machine (see the \texttt{constant\_tsc} and 
\texttt{nonstop\_tsc} flags in Linux's \texttt{/proc/cpu\_info}) \cite{Intel08,
IntelTscSync, DLS13}. TSC is reset to 0 upon machine restart and 
then advances with constant rate.\footnote{TSC registers across CPU sockets 
must be synchronized using a synchronous \texttt{RESET} signal, which is 
commonly the case on modern hardware \cite{IntelTscSync, KernelTscSupport}.} 
Reading the TSC register (e.g., using the \rtdscp instruction) is an extremely 
fast operation as it does not involve a system call (in our tests, \rtdscp takes 
about 10~ns to complete). 

Since \jiffy is implemented in Java, we do not access the TSC register 
directly. Instead, we use the \nanotime method \cite{JavaNanotime}, which 
on the popular Java Virtual Machines (JVMs) for the x86\_64 platforms, e.g., 
\cite{OpenJDK, OracleJDK}, internally relies on TSC. By specification, 
\nanotime is a thread-safe operation that for all invocations of this method in 
an instance of JVM returns a monotonically increasing 8B 
integer.\footnote{Assume for now that \nanotime always returns a positive 
value.}
In our pseudocodes, we will use the $\TSC.\rread()$ function to retrieve
values from the TSC register.

\begin{comment}
\rtdscp ensures both \emph{local} and \emph{global monotonicity} (see also 
\cite{DLS13}). The former means that given two \rtdscp executions $e_1$, $e_2$ 
on the same thread, which return $v_1$ and $v_2$, respectively, the following 
holds: $e_1 \rightarrow e_2$ ($e_1$ completed before $e_2$ was issued) $\iff 
v_1 < v_2$. The latter means that, given $e_1$ and $e_2$ happen on threads $t_1$ 
and $t_2$, respectively, after performing $e_1$, $t_1$ writes $x$ to a shared 
variable, and $t_2$ reads $x$ from the variable before performing $e_2$, the 
following holds: $e_1 \rightarrow e_2 \iff v_1 < v_2$. Without the accesses to 
the shared variable only $e_1 \rightarrow e_2 \Rightarrow v_1 < v_2$ holds. It 
is guaranteed that \rtdscp completes only after all preceding load 
instructions are completed. However, CPU may reorder \rtdscp with the preceding 
store instructions as well as any instructions that follow \rtdscp in the code, 
unless there is a data dependency between the operations. In our algorithm such 
% SHORT
% a data 
dependency is always present. E.g., upon snapshot creation, a thread 
reads the TSC register and stores the read value as the snapshot version $s$. 
The thread then uses $s$ to select only the relevant (to the snapshot) entries 
of the index.
\end{comment}

We use the values generated by TSC in the following way. Each update operation 
(put, remove or batch update) and each revision created by such an operation is 
associated with two version numbers: in the beginning a temporary one, which we 
call \emph{an optimistic version number} and, eventually, the \emph{final 
version number}, which never changes again. An optimistic version number is 
negative, which signals a concurrent thread that encounters a revision with 
such a version number about the pending update operation (which the thread 
might now have to complete). Moreover, there is a special relationship between 
the optimistic and the final version numbers, which allows us to better handle 
lookups and range scans that are performed on snapshots.

More precisely, an update operation commences with an optimistic version number 
$v = -(t+1)$, where $t$ is obtained by reading the TSC register. The 
name \emph{optimistic version number} comes from the fact that $|v|$ 
corresponds to the lowest possible final version number with which the 
update operation can complete. 
% (and thus which can be assigned to the created revision or revisions). 
Hence we define an invariant $v' \ge |v|$, where $v'$ 
is the final version number assigned to the revision. 
For correctness of our algorithm, 
% it is necessary that 
revisions in each revision must have unique version numbers. Since the values 
read by a thread from TSC are not guaranteed to be strictly monotonically 
increasing, we add $1$ to $t$, and before we assign the final version number to 
the revision, we ensure that the current value of the TSC register is greater or 
equal $v'$. 

Lookup and range scan operations (see details in 
\Cref{sec:design:lookups}) use the version numbers stored in revisions to 
retrieve the correct revision and, from it, the value for the searched key. 
The read operations can be performed also on a snapshot acquired earlier by the 
thread. Snapshot creation consists of recording the current value of TSC as the 
\emph{snapshot version} and storing it in a special (lock-free) linked-list 
shared between the threads. A snapshot with snapshot version $s$
corresponds to the state of the dataset at time $s$.

Assume that we have already found the appropriate node and evaluate the 
revisions in its revision list. The most recent value for some key $k$ can be 
found in the most recently completed revision, i.e., the revision with the 
greatest positive version number. On the other hand, for lookups and range 
scans performed on a snapshot (with snapshot version $s$), 
% SHORT
% we need to perform more complex logic. 
when evaluating a revision $r$ with version number $v$, we do the following:
\begin{itemize}
\item if $|v| > s$, skip reading $r$,
\item if $v > 0 \wedge v \le s$, and the revision list contains no 
revision with version number $v'$, s.t. $v < v' \le s$, then 
retrieve $r$, or
\item if $v < 0 \wedge -v \le s$, help to complete the update operation that 
created $r$, resolve the final version number for $r$, and act accordingly.
\end{itemize}

\subsection{Implementation details}

% Now we dive into more details concerting the implementation of \jiffy. We 
% start with structure modifications to our index, then we discuss how update 
% and read-only operations are implemented and 

\subsubsection{Structure modifications} \label{sec:design:structure_modifications}

% \begin{figure*}
% % \includegraphics[scale=0.23]{figures/split.png}
% \bgroup
% \newcommand{\separation}{2.85cm}
% \input{figures/macros.tex}
% \begin{tabular}{p{\separation}p{\separation}p{\separation}p{\separation}p{\separation}}
% \multicolumn{5}{c}{
% \hspace{-0.4cm}
% \scalebox{0.8}{
% \input{figures/split.tex}
% }
% } \\
% \centering{(a)} & \centering{(b)} & \centering{(c)} & \centering{(d)} & \centering{(e)}\\
% \end{tabular}
% % \includegraphics[scale=0.2]{figures/skiplist.png}
% \egroup
% \vspace{-0.7cm}
% \caption{Node split operation of node $k$: (a) initial state, (b) create split 
% revisions, add the left split revision to node $k$ (CAS), and create a temporary 
% split node $o$, (c) add the temporary split node $o$ to the index (CAS), (d) 
% create node $o$ with the right split revision, (e) add node $o$ to the index 
% (CAS) and garbage collect the temporary split node.}
% \label{fig:split}
% \end{figure*}

\newcommand{\lsrshort}{\mathit{lsr}}
\newcommand{\rsrshort}{\mathit{rsr}}
\newcommand{\mtshort}{\mathit{mt}}
\newcommand{\mrshort}{\mathit{mr}}

\begin{figure*}
\bgroup
\newcommand{\separation}{2.65cm}
\input{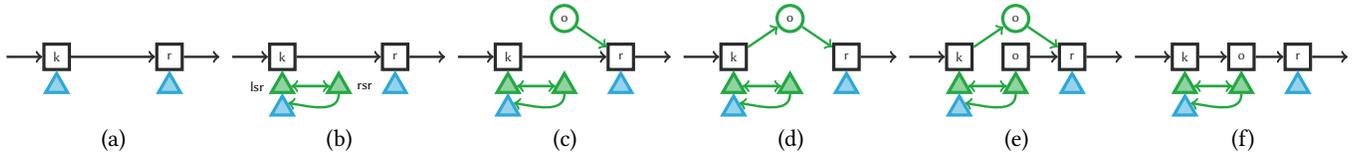}
\begin{tabular}{p{\separation}p{\separation}p{\separation}p{\separation}p{\separation}p{\separation}}
\multicolumn{6}{c}{
\hspace{-0.4cm}
\scalebox{0.75}{
\bgroup
\begin{tikzpicture}[font=\sf\footnotesize] 
\newcommand{\ventryDist}{1}
\newcommand{\indexShift}{0.03}
\newcommand{\indexDist}{0.45}
\newcommand{\revisionDist}{0.4}
\newcommand{\revisionShift}{-0.12}

\newcommand{\scopeShift}{4.0}

\begin{scope}[shift={(0, 0)}]
    \node (startNode) at (0,0) {};

    \node (left) at ($ (startNode) + (-1 * \ventryDist, 0)$) {};
    \topEntry{e0}{$ (startNode) + (0 * \ventryDist, 0)$}{k}
    \topEntry{e2}{$ (startNode) + (2 * \ventryDist, 0)$}{r}
    \node (right) at ($ (startNode) + (3 * \ventryDist, 0)$) {};

    \draw[top arrow] (left) -- (e0);
    \draw[top arrow] (e0) -- (e2);
    \draw[top arrow] (e2) -- (right);

    \revisionN{r01}{$ (startNode) + (0 * \ventryDist, \revisionShift - 1 * \revisionDist) $}{}
    \revisionN{r21}{$ (startNode) + (2 * \ventryDist, \revisionShift - 1 * \revisionDist) $}{}
\end{scope}

\begin{scope}[shift={(1 * \scopeShift, 0)}]
    \node (startNode) at (0,0) {};

    \node (left) at ($ (startNode) + (-1 * \ventryDist, 0)$) {};
    \topEntry{e0}{$ (startNode) + (0 * \ventryDist, 0)$}{k}
    \topEntry{e2}{$ (startNode) + (2 * \ventryDist, 0)$}{r}
    \node (right) at ($ (startNode) + (3 * \ventryDist, 0)$) {};

    \draw[top arrow] (left) -- (e0);
    \draw[top arrow] (e0) -- (e2);
    \draw[top arrow] (e2) -- (right);

    \splitRevisionN{r01}{$ (startNode) + (0 * \ventryDist, \revisionShift - 1 * \revisionDist) $}{}
    \splitRevisionN{r11}{$ (startNode) + (1 * \ventryDist, \revisionShift - 1 * \revisionDist) $}{}
    \revisionN{r02}{$ (startNode) + (0 * \ventryDist, \revisionShift - 2 * \revisionDist) $}{}
    \revisionN{r21}{$ (startNode) + (2 * \ventryDist, \revisionShift - 1 * \revisionDist) $}{}
    
    \draw[split arrow] (r01) -- (r11);
    \draw[split arrow] (r11) -- (r01);
    \draw[split arrow] (r11.south) to[out=270, in=350, looseness=1.0] (r02.north east);
            
    \node (x) [left=0.05cm of r01] {lsr};
    \node (x) [right=0.05cm of r11] {rsr};
\end{scope}

\begin{scope}[shift={(2 * \scopeShift, 0)}]
    \node (startNode) at (0,0) {};

    \node (left) at ($ (startNode) + (-1 * \ventryDist, 0)$) {};
    \topEntry{e0}{$ (startNode) + (0 * \ventryDist, 0)$}{k}
    \topEntry{e2}{$ (startNode) + (2 * \ventryDist, 0)$}{r}
    \node (right) at ($ (startNode) + (3 * \ventryDist, 0)$) {};

    \draw[top arrow] (left) -- (e0);
    \draw[top arrow] (e0) -- (e2);
    \draw[top arrow] (e2) -- (right);

    \splitRevisionN{r01}{$ (startNode) + (0 * \ventryDist, \revisionShift - 1 * \revisionDist) $}{}
    \splitRevisionN{r11}{$ (startNode) + (1 * \ventryDist, \revisionShift - 1 * \revisionDist) $}{}
    \revisionN{r02}{$ (startNode) + (0 * \ventryDist, \revisionShift - 2 * \revisionDist) $}{}
    \revisionN{r21}{$ (startNode) + (2 * \ventryDist, \revisionShift - 1 * \revisionDist) $}{}
    
    \draw[split arrow] (r01) -- (r11);
    \draw[split arrow] (r11) -- (r01);
    \draw[split arrow] (r11.south) to[out=270, in=350, looseness=1.0] (r02.north east);
    
    \splitNodeN{s}{$ (startNode) + (1 * \ventryDist, \revisionShift + 2 * \revisionDist) $}{o}
    
    \draw[split arrow] (s) -- (e2);
\end{scope}

\begin{scope}[shift={(3 * \scopeShift, 0)}]
    \node (startNode) at (0,0) {};

    \node (left) at ($ (startNode) + (-1 * \ventryDist, 0)$) {};
    \topEntry{e0}{$ (startNode) + (0 * \ventryDist, 0)$}{k}
    \topEntry{e2}{$ (startNode) + (2 * \ventryDist, 0)$}{r}
    \node (right) at ($ (startNode) + (3 * \ventryDist, 0)$) {};

    \draw[top arrow] (left) -- (e0);
%     \draw[top arrow] (e0) -- (e2);
    \draw[top arrow] (e2) -- (right);

    \splitRevisionN{r01}{$ (startNode) + (0 * \ventryDist, \revisionShift - 1 * \revisionDist) $}{}
    \splitRevisionN{r11}{$ (startNode) + (1 * \ventryDist, \revisionShift - 1 * \revisionDist) $}{}
    \revisionN{r02}{$ (startNode) + (0 * \ventryDist, \revisionShift - 2 * \revisionDist) $}{}
    \revisionN{r21}{$ (startNode) + (2 * \ventryDist, \revisionShift - 1 * \revisionDist) $}{}
    
    \draw[split arrow] (r01) -- (r11);
    \draw[split arrow] (r11) -- (r01);
    \draw[split arrow] (r11.south) to[out=270, in=350, looseness=1.0] (r02.north east);
    
    \splitNodeN{s}{$ (startNode) + (1 * \ventryDist, \revisionShift + 2 * \revisionDist) $}{o}
    
    \draw[split arrow] (e0) -- (s);
    \draw[split arrow] (s) -- (e2);
\end{scope}

\begin{scope}[shift={(4 * \scopeShift, 0)}]
    \node (startNode) at (0,0) {};

    \node (left) at ($ (startNode) + (-1 * \ventryDist, 0)$) {};
    \topEntry{e0}{$ (startNode) + (0 * \ventryDist, 0)$}{k}
    \topEntry{e1}{$ (startNode) + (1 * \ventryDist, 0)$}{o}
    \topEntry{e2}{$ (startNode) + (2 * \ventryDist, 0)$}{r}
    \node (right) at ($ (startNode) + (3 * \ventryDist, 0)$) {};

    \draw[top arrow] (left) -- (e0);
%     \draw[top arrow] (e0) -- (e2);

    \draw[top arrow] (e1) -- (e2);
    \draw[top arrow] (e2) -- (right);

    \splitRevisionN{r01}{$ (startNode) + (0 * \ventryDist, \revisionShift - 1 * \revisionDist) $}{}
    \splitRevisionN{r11}{$ (startNode) + (1 * \ventryDist, \revisionShift - 1 * \revisionDist) $}{}
    \revisionN{r02}{$ (startNode) + (0 * \ventryDist, \revisionShift - 2 * \revisionDist) $}{}
    \revisionN{r21}{$ (startNode) + (2 * \ventryDist, \revisionShift - 1 * \revisionDist) $}{}
    
    \draw[split arrow] (r01) -- (r11);
    \draw[split arrow] (r11) -- (r01);
    \draw[split arrow] (r11.south) to[out=270, in=350, looseness=1.0] (r02.north east);
    
    \splitNodeN{s}{$ (startNode) + (1 * \ventryDist, \revisionShift + 2 * \revisionDist) $}{o}
    
    \draw[split arrow] (e0) -- (s);
    \draw[split arrow] (s) -- (e2);
\end{scope}

\begin{scope}[shift={(5 * \scopeShift, 0)}]
    \node (startNode) at (0,0) {};

    \node (left) at ($ (startNode) + (-1 * \ventryDist, 0)$) {};
    \topEntry{e0}{$ (startNode) + (0 * \ventryDist, 0)$}{k}
    \topEntry{e1}{$ (startNode) + (1 * \ventryDist, 0)$}{o}
    \topEntry{e2}{$ (startNode) + (2 * \ventryDist, 0)$}{r}
    \node (right) at ($ (startNode) + (3 * \ventryDist, 0)$) {};

    \draw[top arrow] (left) -- (e0);
    \draw[top arrow] (e0) -- (e1);
    \draw[top arrow] (e1) -- (e2);
    \draw[top arrow] (e2) -- (right);

    \splitRevisionN{r01}{$ (startNode) + (0 * \ventryDist, \revisionShift - 1 * \revisionDist) $}{}
    \splitRevisionN{r11}{$ (startNode) + (1 * \ventryDist, \revisionShift - 1 * \revisionDist) $}{}
    \revisionN{r02}{$ (startNode) + (0 * \ventryDist, \revisionShift - 2 * \revisionDist) $}{}
    \revisionN{r21}{$ (startNode) + (2 * \ventryDist, \revisionShift - 1 * \revisionDist) $}{}
    
    \draw[split arrow] (r01) -- (r11);
    \draw[split arrow] (r11) -- (r01);
    \draw[split arrow] (r11.south) to[out=270, in=350, looseness=1.0] (r02.north east);
    
%     \splitNodeN{s}{$ (startNode) + (1 * \ventryDist, \revisionShift + 2 * \revisionDist) $}{o}
    
%     \draw[split arrow] (e0) -- (s);
%     \draw[split arrow] (s) -- (e2);
\end{scope}

\end{tikzpicture}
\egroup
}
} \\
\centering{(a)} & \centering{(b)} & \centering{(c)} & \centering{(d)} & \centering{(e)} & \centering{(f)}\\
\end{tabular}
\egroup
\vspace{-0.7cm}
\caption{Node split operation of node $k$: (a) initial state, (b) create split 
revisions ($\lsrshort$, $\rsrshort$), add the left split revision ($\lsrshort$) 
to node $k$ (CAS), (c) create a temporary split node $o$, (d) add the temporary 
split node $o$ to the index (CAS), (e) create node $o$ with the right split 
revision, (f) add node $o$ to the index (CAS) and garbage collect the temporary 
split node.}
\label{fig:split}
\end{figure*}

\begin{figure*}
\bgroup
\newcommand{\separation}{2.65cm}
\input{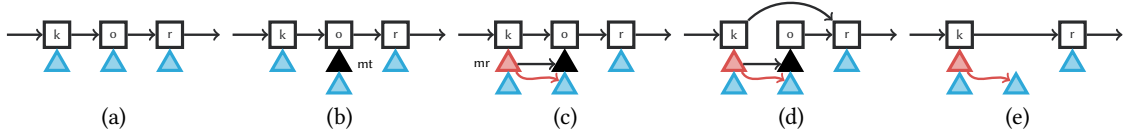}
\begin{tabular}{p{\separation}p{\separation}p{\separation}p{\separation}p{\separation}}
% \begin{minipage}{0.4\textwidth}
\multicolumn{5}{c}{
\hspace{-0.4cm}
\scalebox{0.75}{
\bgroup
\begin{tikzpicture}[font=\sf\footnotesize] 
\newcommand{\ventryDist}{1}
\newcommand{\indexShift}{0.03}
\newcommand{\indexDist}{0.45}
\newcommand{\revisionDist}{0.4}
\newcommand{\revisionShift}{-0.12}

\newcommand{\scopeShift}{4.0}

\begin{scope}[shift={(0, 0)}]
    \node (startNode) at (0,0) {};

    \node (left) at ($ (startNode) + (-1 * \ventryDist, 0)$) {};
    \topEntry{e0}{$ (startNode) + (0 * \ventryDist, 0)$}{k}
    \topEntry{e1}{$ (startNode) + (1 * \ventryDist, 0)$}{o}
    \topEntry{e2}{$ (startNode) + (2 * \ventryDist, 0)$}{r}
    \node (right) at ($ (startNode) + (3 * \ventryDist, 0)$) {};

    \draw[top arrow] (left) -- (e0);
    \draw[top arrow] (e0) -- (e1);
    \draw[top arrow] (e1) -- (e2);
    \draw[top arrow] (e2) -- (right);

    \revisionN{r01}{$ (startNode) + (0 * \ventryDist, \revisionShift - 1 * \revisionDist) $}{}
    \revisionN{r11}{$ (startNode) + (1 * \ventryDist, \revisionShift - 1 * \revisionDist) $}{}
    \revisionN{r21}{$ (startNode) + (2 * \ventryDist, \revisionShift - 1 * \revisionDist) $}{}
\end{scope}

\begin{scope}[shift={(1 * \scopeShift, 0)}]
    \node (startNode) at (0,0) {};

    \node (left) at ($ (startNode) + (-1 * \ventryDist, 0)$) {};
    \topEntry{e0}{$ (startNode) + (0 * \ventryDist, 0)$}{k}
    \topEntry{e1}{$ (startNode) + (1 * \ventryDist, 0)$}{o}
    \topEntry{e2}{$ (startNode) + (2 * \ventryDist, 0)$}{r}
    \node (right) at ($ (startNode) + (3 * \ventryDist, 0)$) {};

    \draw[top arrow] (left) -- (e0);
    \draw[top arrow] (e0) -- (e1);
    \draw[top arrow] (e1) -- (e2);
    \draw[top arrow] (e2) -- (right);

    \revisionN{r01}{$ (startNode) + (0 * \ventryDist, \revisionShift - 1 * \revisionDist) $}{}
    \mergeTerminatorN{r11}{$ (startNode) + (1 * \ventryDist, \revisionShift - 1 * \revisionDist) $}{}
    \revisionN{r12}{$ (startNode) + (1 * \ventryDist, \revisionShift - 2 * \revisionDist) $}{}
    \revisionN{r21}{$ (startNode) + (2 * \ventryDist, \revisionShift - 1 * \revisionDist) $}{}
    
    \node (x) [right=0.05cm of r11] {mt};
\end{scope}

\begin{scope}[shift={(2 * \scopeShift, 0)}]
    \node (startNode) at (0,0) {};

    \node (left) at ($ (startNode) + (-1 * \ventryDist, 0)$) {};
    \topEntry{e0}{$ (startNode) + (0 * \ventryDist, 0)$}{k}
    \topEntry{e1}{$ (startNode) + (1 * \ventryDist, 0)$}{o}
    \topEntry{e2}{$ (startNode) + (2 * \ventryDist, 0)$}{r}
    \node (right) at ($ (startNode) + (3 * \ventryDist, 0)$) {};

    \draw[top arrow] (left) -- (e0);
    \draw[top arrow] (e0) -- (e1);
    \draw[top arrow] (e1) -- (e2);
    \draw[top arrow] (e2) -- (right);

    \mergeRevisionN{r01}{$ (startNode) + (0 * \ventryDist, \revisionShift - 1 * \revisionDist) $}{}
    \revisionN{r02}{$ (startNode) + (0 * \ventryDist, \revisionShift - 2 * \revisionDist) $}{}
    \mergeTerminatorN{r11}{$ (startNode) + (1 * \ventryDist, \revisionShift - 1 * \revisionDist) $}{}
    \revisionN{r12}{$ (startNode) + (1 * \ventryDist, \revisionShift - 2 * \revisionDist) $}{}
    \revisionN{r21}{$ (startNode) + (2 * \ventryDist, \revisionShift - 1 * \revisionDist) $}{}
    
    \draw[merge arrow] (r01.south east) to[out=300, in=150, looseness=1.0] (r12.north west);
    \draw[top arrow] (r01.east) -- (r11.west);
    
    \node (x) [left=0.05cm of r01] {mr};
\end{scope}

\begin{scope}[shift={(3 * \scopeShift, 0)}]
    \node (startNode) at (0,0) {};

    \node (left) at ($ (startNode) + (-1 * \ventryDist, 0)$) {};
    \topEntry{e0}{$ (startNode) + (0 * \ventryDist, 0)$}{k}
    \topEntry{e1}{$ (startNode) + (1 * \ventryDist, 0)$}{o}
    \topEntry{e2}{$ (startNode) + (2 * \ventryDist, 0)$}{r}
    \node (right) at ($ (startNode) + (3 * \ventryDist, 0)$) {};

    \draw[top arrow] (left) -- (e0);
    \draw[top arrow] (e1) -- (e2);
    \draw[top arrow] (e2) -- (right);

    \mergeRevisionN{r01}{$ (startNode) + (0 * \ventryDist, \revisionShift - 1 * \revisionDist) $}{}
    \revisionN{r02}{$ (startNode) + (0 * \ventryDist, \revisionShift - 2 * \revisionDist) $}{}
    \mergeTerminatorN{r11}{$ (startNode) + (1 * \ventryDist, \revisionShift - 1 * \revisionDist) $}{}
    \revisionN{r12}{$ (startNode) + (1 * \ventryDist, \revisionShift - 2 * \revisionDist) $}{}
    \revisionN{r21}{$ (startNode) + (2 * \ventryDist, \revisionShift - 1 * \revisionDist) $}{}
    
    \draw[merge arrow] (r01.south east) to[out=300, in=150, looseness=1.0] (r12.north west);
    
    \draw[top arrow] (e0) to[out=45, in=135, looseness=1.0] (e2);
    \draw[top arrow] (r01.east) -- (r11.west);
\end{scope}

\begin{scope}[shift={(4 * \scopeShift, 0)}]
    \node (startNode) at (0,0) {};

    \node (left) at ($ (startNode) + (-1 * \ventryDist, 0)$) {};
    \topEntry{e0}{$ (startNode) + (0 * \ventryDist, 0)$}{k}
    \topEntry{e2}{$ (startNode) + (2 * \ventryDist, 0)$}{r}
    \node (right) at ($ (startNode) + (3 * \ventryDist, 0)$) {};

    \draw[top arrow] (left) -- (e0);
    \draw[top arrow] (e0) -- (e2);
    \draw[top arrow] (e2) -- (right);

    \mergeRevisionN{r01}{$ (startNode) + (0 * \ventryDist, \revisionShift - 1 * \revisionDist) $}{}
    \revisionN{r02}{$ (startNode) + (0 * \ventryDist, \revisionShift - 2 * \revisionDist) $}{}
%     \mergeTerminatorN{r11}{$ (startNode) + (1 * \ventryDist, \revisionShift - 1 * \revisionDist) $}{}
    \revisionN{r12}{$ (startNode) + (1 * \ventryDist, \revisionShift - 2 * \revisionDist) $}{}
    \revisionN{r21}{$ (startNode) + (2 * \ventryDist, \revisionShift - 1 * \revisionDist) $}{}
    
    \draw[merge arrow] (r01.south east) to[out=300, in=150, looseness=1.0] (r12.north west);
\end{scope}

\end{tikzpicture}
\egroup
}
} \\
\centering{(a)} & \centering{(b)} & \centering{(c)} & \centering{(d)} & \centering{(e)}\\
\end{tabular}
\egroup
\vspace{-0.7cm}
\caption{Node merge operation of node $o$: (a) initial state, (b) add a merge 
terminator ($\mtshort$) to node $o$ (CAS), (c) add a merge revision 
($\mrshort$) to node $k$ (CAS), (d) unlink node $o$ from the index (CAS), (e) 
garbage collect node $o$ and the merge terminator.}
\label{fig:merge}
\end{figure*}

We start the descriptions of structure modifications in \jiffy with a 
node split operation. For simplicity we abstract away from the fact that 
in \jiffy all structure modifications are streamlined with the update 
operations. Consider the example in \Cref{fig:split}, in which we show how node 
$k$, that manages a range of keys $[k, r)$, is split so a new node $o$ (whose 
range is $[o, r)$) is to be inserted between node $k$ and node $r$. To this end 
we first create two special revisions, called \emph{left} ($\lsrshort$) and 
\emph{right split revisions} ($\rsrshort$). Each split revision contains half 
of the entries from the revision that was the head of the revision list at node 
$k$ in the beginning. We use a CAS operation to add the left split revision 
($\lsrshort$) to the revision list at node $k$ (\Cref{fig:split}b). Next we 
create a \emph{temporary split node}, whose next pointer is set to $r$ 
(\Cref{fig:split}c). We use CAS to swing the next pointer from node $k$ to the 
temporary split node and thus add it into the index (\Cref{fig:split}d). Note 
that the node has key $o$, so, e.g., concurrent lookups searching for keys in 
range $[o,r)$ will be able to find it and help to complete the split operation 
(information necessary to complete the split operation is accessible through 
split revisions and the temporary split node).
Next, we create node $o$ with the right split revision as the sole 
revision on the node's revision list. The next pointer of node $o$ is set to 
node $r$ (\Cref{fig:split}e). Finally, we use CAS to swing the next pointer of 
node $k$ from the temporary split node to node $o$, garbage collect the 
temporary split node and write the final version number to split revisions 
(\Cref{fig:split}f).

Why could not we simply insert node $o$ in-between nodes $k$ and $r$ using a 
single CAS operation, as in a simple lock-free linked list \cite{H01}? It is 
because the entire split operation involves adding a revision to node $k$ 
\emph{and} creating node $o$. Without a temporary split node $o$ an ABA problem 
is possible. Imagine two threads, A and B. Thread A acquires the reference to 
node $r$, adds a left split revision to the revision list at node $k$, and is 
preempted. Then, thread B that tries to add a revision at node $k$, observes  
a pending split operation and adds node $o$ with the right split revision. 
Suppose that subsequently node $o$ is merged back to node $k$, so the next 
pointer at node $k$ again points to node $r$. When thread A continues its 
execution, it incorrectly adds node $o$ in-between nodes $k$ and $r$, which may 
corrupt lookup and range scan operations. In our scheme the ABA problem on the 
temporary split node is still possible, but we can recover from it without 
corrupting concurrent operations. If thread A observes that some other thread 
already set the final version number in the left split revision, it means that 
node $o$ must have already been created (and merged into node $k$, assigning 
the final version number is the last operation of a node split). In such case 
the temporary split node can be safely removed.

Now let us consider the node merge operation. In the example in 
\Cref{fig:merge}b a merge terminator ($\mtshort$) is added to the revision 
list at node $o$, thus initiating the merge operation. No other revision can 
now be added to the revision list at node $o$, hence also a split operation 
cannot be invoked on node $o$. In the next step, we invoke the 
$\helpMergeTerminator$ function to find the node directly preceding node $o$ 
and, if necessary, complete all pending operations at this node (in some cases 
we need to perform the search for the preceding node again). Once we find node 
$k$, we create a merge revision ($\mrshort$) that encompasses the entries from 
the merge terminator's successor revision in the revision list as well as the 
head of the revision list at node $k$ (\Cref{fig:merge}c). Note that the 
merge revision joins the revision lists at node $k$ and node $o$ (excluding the 
merge terminator), and so has two successors: left (default, same as in 
an ordinary revision) and right. Next we use CAS to swing the next pointer at 
node $k$ to node $r$, thus unlinking node $o$ from the index 
(\Cref{fig:merge}d). Finally, we mark node $o$ as \emph{terminated}, which means 
that now it can be garbage collected together with the merge terminator 
(\Cref{fig:merge}e).

In our implementation, structure changes to the index are driven by update 
operations. E.g., a put operation may cause a node split. In such case, one of 
the split revisions reflects also the put operation that caused the node split 
in the first place. This way no revisions are created unnecessarily.

\begin{comment}
In our implementation, structure changes to the index are driven by update 
operations. E.g., split operation not only divides one revision into 
two separate ones that belong to neighboring nodes, but the 
revisions also reflect the update operation that caused the node split in the 
first place. This way structure changes are streamlined with update operations 
and no revisions are created unnecessarily.
\end{comment}

As we mentioned earlier, completing node splits and merges is performed 
also by lookup or range scan operations that happen to encounter a not yet 
completed structure modification operation. The rather complex logic of dealing 
with various stages of node splits and merges is hidden in the $\helpSplit$, 
$\helpTempSplitNode$, $\helpMergeTerminator$ and $\findAndHelpMergeRevision$ 
functions, which we use in the operations we discuss next.

\subsubsection{The put and remove operations} \label{sec:design:updates}

\begin{algorithm*}[t] 
\caption{The $\pput$ and $\rremove$ operations in \jiffy}
\label{alg:put_remove}  
% \algsetup{linenosize=\tiny}
%   \footnotesize
\small
\vspace{-0.55cm}
\begin{multicols*}{2}
\begin{algorithmic}[1]
\Procedure{$\pput$}{$\key$, $\vvalue$} 
    \State{\Var $\newRevision = \bot$}
    \While{$\true$}
        \State{\Var $\bnode = \findNodeForKey(\key)$} \LineComment{unlinks terminated nodes} \label{alg:put:find} \label{alg:put:whileFirstLine}
        \State{\Var $\nnode = \bnode.\nnext$} \label{alg:put:getNext}
        \If{$\bnode\ \is\ \TempSplitNode$} \LineComment{middle of node split}
            \State{$\helpTempSplitNode(\bnode)$} \label{alg:put:helpSplitNode} \LineComment{corresponds to \Cref{fig:split}e-f}
            \State{\continue}
        \EndIf
        \State{\Var $\rrev = \bnode.\hhead$} \LineComment{first revision from the revision list}
            \label{alg:put:getHead}
        \If{$\bnode.\terminated$} \LineComment{ready to unlink} \label{alg:put:terminated}
            \State{\continue}
        \EndIf
        \If{$\rrev.\version < 0$} \LineComment{pending update operation} \label{alg:put:helpStart}
            \State{$\helpPut(\rrev)$} \LineComment{complete the operation} 
            \State{\continue}\label{alg:put:helpEnd} 
        \EndIf
        \If{$\bnode.\nnext \neq \nnode$} \LineComment{a split or merge happened} \label{alg:put:nextChanged}
            \State{\continue}
        \EndIf \label{alg:put:beforeOptVersion}
        \State{\Var $\optVersion = -1 * (\TSC.\rread() + 1)$} \label{alg:put:optimistic}       
        \State{\Var $\updateType = \autoscaler.\query(\rrev, \key)$}
        \If{$\updateType = \regularUpdate$} \label{alg:put:regularUpdateStart}
            \State{$\newRevision = \rrev.\cloneAndApplyPut(\key, \vvalue, \optVersion)$}
            \If{$\CAS(\bnode.\hhead, \rrev, \newRevision)$} \LineComment{successful} \label{alg:put:casRegular}
                \State{\bbreak}
            \EndIf \label{alg:put:regularUpdateEnd}
        \Else \LineComment{$\updateType = \nodeSplit$}
            \State{\Var $(\lsr, \rsr) = \rrev.\applyPutAndSplit(\key, \vvalue, \optVersion)$}\label{alg:put:applyPutAndSplit}
            \If{$\CAS(\bnode.\hhead, \rrev, \lsr)$} \LineComment{successful} \label{alg:put:casSplit}
                \State{$\helpSplit(\lsr)$} \LineComment{corresponds to \Cref{fig:split}c-f}
                \State{$\newRevision = \lsr$}
                \State{\bbreak}
            \EndIf
        \EndIf
    \EndWhile
%     \State{\Var $\finVersion = \trySetVersion(\newRevision, \TSC.\rread() + 1)$} \label{alg:put:trySetFinalVersion}
     \State{\Var $\finVersion = \maxx(\TSC.\rread(), -\optVersion)$} \label{alg:put:finVersion} \LineComment{to ensure the invariant}
    \State{$\publishVersion(\finVersion)$} \label{alg:put:publishVersion} 
    \State{$\finVersion = \trySetVersion(\newRevision, \finVersion)$} \label{alg:put:trySetFinalVersion}
    \If{$\newRevision\ \is\ \SplitRevision$} \LineComment{set $\finVersion$ on both revisions}
        \State{$\newRevision.\sibling.\version = \finVersion$} 
    \EndIf \label{alg:put:setSiblingVersion}
    \State{$\performGC(\newRevision)$} \LineComment{cut revision list short if necessary} \label{alg:put:gc}

\EndProcedure
\Procedure{$\rremove$}{$\key$}
    \State{\Var $\newRevision = \bot$}
    \While{$\true$}
        \State{...} \LineComment{same as lines \ref{alg:put:whileFirstLine}--\ref{alg:put:beforeOptVersion} in $\pput$}
        \If{$\rrev.\gget(\key) = \bot$} \label{alg:remove:returnEarly} \LineComment{nothing to do}
            \State{\Return}
        \EndIf
        \State{\Var $\optVersion = -1 * (\TSC.\rread() + 1)$} 
        \State{\Var $\updateType = \autoscaler.\query(\rrev, \key)$}
        \If{$\updateType = \regularUpdate$}
            \State{$\newRevision = \rrev.\cloneAndApplyRemove(\key, \optVersion)$}
            \If{$\CAS(\bnode.\hhead, \rrev, \newRevision)$} \LineComment{successful}
                \State{\bbreak}
            \EndIf
        \Else \LineComment{$\updateType = \nodeMerge$} \label{alg:remove:nodeMergeStart}
            \State{\Var $\mt = \MergeTerminator(\rrev, \key, \optVersion)$}
            \If{$\CAS(\bnode.\hhead, \rrev, \mt)$} \LineComment{successful}
                \State{$\helpMergeTerminator(\mt)$} \LineComment{corresp. to \Cref{fig:merge}c-e}
                \State{$\newRevision = \mt$}
                \State{\bbreak}
            \EndIf
        \EndIf \label{alg:remove:nodeMergeEnd}
    \EndWhile    
    \If{$\newRevision\ \is\ \MergeTerminator$} 
        \State{$\newRevision = \findAndHelpMergeRevision(\newRevision)$} \label{alg:remove:findAndHelpMergeRevision} \LineComment{\Cref{fig:merge}d-e}
    \EndIf
    \State{\Var $\finVersion = \maxx(\TSC.\rread(), -\optVersion)$} \label{alg:remove:finVersion} \LineComment{to ensure the invariant}
    \State{$\publishVersion(\finVersion)$} \label{alg:remove:publishVersion}
    \State{$\finVersion = \trySetVersion(\newRevision, \finVersion)$} \label{alg:remove:trySetFinalVersion}
    \State{$\performGC(\newRevision)$} \LineComment{cut revision list short if necessary}
\EndProcedure
\Function{$\trySetVersion$}{$\revision, \version$} \LineComment{set the final version, but}
  \State{\Var $\oldVersion = \revision.\version$} \LineComment{only if not already set}
  \If{$\oldVersion > 0$}
    \State{\Return $\oldVersion$}
  \EndIf
  \If{$\CAS(\revision.\version, \oldVersion, \version)$} \LineComment{successful,} \label{alg:trySetVersion:cas} 
    \State{\Return $\version$} \LineComment{linearization point}
  \EndIf
  \State{\Return $\revision.\version$}
\EndFunction
\Procedure{$\publishVersion$}{$\version$} \LineComment{wait until TSC advances enough}
  \While{$\TSC.\rread() < \version$} \LineComment{in practice, false right away} \label{alg:publishVersion:activeWait}
    \State{$\NOP$}
  \EndWhile
\EndProcedure
\end{algorithmic}
\end{multicols*}
\vspace{-0.35cm}
\end{algorithm*}

Consider the pseudocodes of the put and remove operations in 
\Cref{alg:put_remove}. The pseudocodes require small changes to 
accommodate the batch update operations. We will discuss these changes in the 
next section, which is devoted to batch updates.

A thread that performs $\pput(\key,\vvalue)$ first finds the appropriate node 
(line~\ref{alg:put:find}) and acquires a reference to the neighboring 
node (the succeeding node in the index, line~\ref{alg:put:getNext}). This 
reference will be required later to ensure that we adequately handle all 
concurrent node splits and merges. Now we perform a series of checks. In case 
any condition is satisfied, we always start over by searching $\key$ again. 
First we check if we found ourselves in a temporary split node. If so, we help 
with completing the split operation (and start over, 
line~\ref{alg:put:helpSplitNode}). Next, we retrieve the head of the revision 
list ($\rrev$, line~\ref{alg:put:getHead}) and check if the node has been 
terminated (through a node merge operation, line~\ref{alg:put:terminated}). 
Then, we check the version number of $\rrev$, and if necessary, help to 
complete the update operation that added $\rrev$ to the node and start over 
(lines \ref{alg:put:helpStart}-\ref{alg:put:helpEnd}, $\helpPut$ uses the same 
logic as put, remove or batch update to complete a pending update
operation). Finally, we check if the neighboring node did not change in the 
meantime (line~\ref{alg:put:nextChanged}). 

By reaching line~\ref{alg:put:optimistic} we know that we are in the correct 
node and thus we can safely try to add a new revision. To this end, we acquire 
the optimistic version number $\optVersion$ from TSC and query the autoscaler 
to determine the type of update we need to perform: a regular update or a node 
split. In the former case (lines 
\ref{alg:put:regularUpdateStart}-\ref{alg:put:regularUpdateEnd}), we clone 
$\rrev$ and modify the value for $\key$ through the $\cloneAndApplyPut$ 
function on $\rrev$. We then try to add such created $\newRevision$ to the 
revision list (using CAS). If we fail, we start over. On the other hand, 
if we were successful, we obtain the final version number $\finVersion$ from TSC 
(line~\ref{alg:put:finVersion}), wait until the current value of the TSC 
register is greater or equal $|\optVersion|$ (to ensure our invariant, see 
\Cref{sec:design:versions}), and set the final version number $\finVersion$ on 
$\newRevision$ (and its sibling, if necessary, lines 
\ref{alg:put:trySetFinalVersion}-\ref{alg:put:setSiblingVersion}). Because of 
TSC's high resolution, in our tests we have never encountered a situation in 
which the active wait in line~\ref{alg:publishVersion:activeWait} was necessary.
Finally we go through the revision list to see if some revisions can be removed 
(line~\ref{alg:put:gc}, we discuss how the garbage collection mechanism works 
when we discuss snapshots in \Cref{sec:design:lookups}). 

In case of a node split, we create a pair $(\lsr, \rsr)$ of new split revisions 
through $\applyPutAndSplit$ function on $\rrev$ 
(line~\ref{alg:put:applyPutAndSplit}). 
The left ($\lsr$) and right ($\rsr$) split revisions reference each other 
through the $\sibling$ field. We attempt to add $\lsr$ to the 
revision list. Again, if we fail, we start over. On the other hand, if we are 
successful, we complete the update operation by calling the $\helpSplit$ 
function and, eventually, also assigning $\finVersion$ to $\rsr$.

There are a few details worth pointing out:
\begin{itemize} 
\item The $\findNodeForKey$ function, also used by the remove, batch update, 
lookup and range scan operations, during traversing the index unlinks all 
terminated nodes (nodes, whose merge operation completed), as well as 
the appropriate index nodes (nodes in the higher levels of the skip list that 
point to the terminated node).
\item If CAS fails in either line~\ref{alg:put:casRegular} or 
\ref{alg:put:casSplit}, no other thread is aware of our attempt to perform 
$\pput$. Thus, we can safely start over. 
\item In certain situation we do not have to traverse the entire index 
again to find the appropriate node, but for brevity we skip this optimization in 
our pseudocode. 
\item If $\pput$ successfully added $\lsr$ to the revision list, then $\pput$ 
will try to complete the node split operation (other threads might help in this 
operation as well). By the time $\pput$ reaches 
line~\ref{alg:put:finVersion}, we can be certain that the node split 
has finished. 
\item CAS in line~\ref{alg:trySetVersion:cas} can fail only if some other 
thread already assigned the final version number to $\newRevision$. Assigning 
the final version number to $\newRevision$ is the linearization point for the 
entire $\pput$ operation.
\end{itemize}

The $\rremove(\key)$ operation proceeds similarly to $\pput$, but with two 
differences. Firstly, $\rremove$ returns early if $\rrev$ does not contain
a value for $\key$ (line~\ref{alg:remove:returnEarly}). Secondly, $\rremove$ 
might result in a node merge instead of a node split, as $\rremove$ decreases 
the size of the revision at the head of the revision list (lines 
\ref{alg:remove:nodeMergeStart}-\ref{alg:remove:nodeMergeEnd}). In such a case, 
$\rremove$ creates a merge terminator $\mt$ and tries to add it to
the revision list. If CAS was successful, we complete the merge operation, 
as discussed in \Cref{sec:design:structure_modifications}. If some other thread 
helped to complete the node merge, we find the merge revision that it 
created and unlink the terminated node, if necessary. 
% SHORT
% For simplicity, in
In our 
pseudocode we always perform the full search for merge revision and ensure that 
the node merge is completed (line~\ref{alg:remove:findAndHelpMergeRevision}).

So far for simplicity we assumed that $\TSC.\rread()$ returns only positive 
values. \nanotime, which we use in our implementation to retrieve 
values from TSC, can return negative values. To adhere to our earlier 
assumption, for every \nanotime operation invoked in \jiffy, we subtract from
the returned value the value of \nanotime obtained upon creation of our index.

\subsubsection{The batch update operation} \label{sec:design:batchUpdates}

A batch update comprises of a number of $\pput$ and $\rremove$ operations that 
are to be performed atomically. The $\batchUpdate$ function in \jiffy relies
on the same logic as $\pput$ and $\rremove$, except for a few differences:
\begin{enumerate}
\item All put and remove operations to be executed by $\batchUpdate$ are stored 
within a \emph{batch descriptor}, which also manages a $\version$ field which 
initially contains the optimistic version number and, eventually, the final 
version number. Thus, reading the version number in a revision created by a 
$\batchUpdate$ happens indirectly through the batch descriptor (this is the 
difference we mentioned in the beginning of \Cref{sec:design:updates}).
\item Each revision created by $\batchUpdate$ reflects the changes to all 
keys managed by $\bnode$, which are included in the batch.
\item Execution of $\batchUpdate$ can result in both node splits and node 
merges, as determined by the autoscaler.
\item In order to (help) complete a $\batchUpdate$, a thread must add all
necessary revisions to appropriate nodes (in descending order of keys) and 
only then try to assign the final version number to the $\version$ field in the 
batch descriptor.
\item Suppose that $\batchUpdate$ happened to find a $\rrev$ (in an appropriate 
node) in which the value for key $k$ is not present. If batch includes the 
$\rremove(k)$ operation, we need to clone $\rrev$ and add it to the node, 
unlike in case of a simple $\rremove(k)$ operation, where we could return early 
without modifying the revision list. 
\end{enumerate}
The order of updates performed by $\batchUpdate$ 
% (from higher to lower keys) 
naturally follows from our design assumption for the node merge operation
to proceed towards lower keys. Assume that $\batchUpdate$ proceeds in the 
opposite order (from lower to higher keys). Then it is possible that 
$\batchUpdate$ adds a revision to some node $n_i$, and then proceeds to node 
$n_j$ that directly follows $n_i$ in the index and decides to perform a node 
merge operation on $n_j$. Consequently, a new (merge) revision would have to be 
created on $n_i$, which is suboptimal.

To explain why adding a revision in situation described in (5) is necessary 
consider otherwise. Suppose a concurrent $\batchUpdate$ add a new revision 
with an update of key $k$ at the same node and finishes with a lower
final version number than the $\batchUpdate$ from (5). Such a situation would 
represent the lost update (lost remove) anomaly: a lookup on a 
snapshot that includes both batch updates would incorrectly return a value for 
$k$ instead of $\bot$.

\subsubsection{Lookup operations and range scans} \label{sec:design:lookups}

\begin{algorithm*}[t] 
\caption{The $\get$ operations in \jiffy}
\label{alg:get}  
% \algsetup{linenosize=\tiny}
%   \footnotesize
\small
\vspace{-0.55cm}
\begin{multicols*}{2}
\begin{algorithmic}[1]
% \Function{$\effectiveVersion$}{$\revision$} \label{alg:get:effectiveVersion}
%     \If{$\revision$ created by $\batchUpdate$} % \LineComment{unfinished $\VolatileBatchEntry$}
%         \State{\Return $\revision.\descriptor.\version$}
%     \EndIf
%     \State{\Return $\revision.\version$}
% \EndFunction \label{alg:get:effectiveVersionEnd}

\Function{$\get$}{$\key$} \label{alg:getNewest} \LineComment{get the most recent value for $\key$} \label{alg:getNewest}
    \State{\Return $\get(\key, \newestVersion)$}
\EndFunction \label{alg:getNewestEnd}

\Function{$\get$}{$\key$, $\snapVersion$} \label{alg:getSnapshot} \LineComment{get the value for $\key$ in a snapshot}
    \While{$\true$} \label{alg:get:startSearch}
        \State{\Var $\bnode = \findNodeForKey(\key)$} \LineComment{unlinks terminated nodes} \label{alg:get:find} \label{alg:get:whileFirstLine}
        \State{\Var $\nnode = \bnode.\nnext$} \label{alg:get:getNext}
        \If{$\bnode\ \is\ \TempSplitNode$} \LineComment{middle of node split}
            \State{$\helpTempSplitNode(\bnode)$} \label{alg:get:helpSplitNode} \LineComment{corresponds to \Cref{fig:split}e-f}
            \State{\continue}
        \EndIf
        \State{\Var $\rrev = \bnode.\hhead$} \LineComment{first revision from the revision list}
            \label{alg:put:getHead}
        \If{$\rrev\ \is\ \MergeTerminator$} \label{alg:get:terminated}
            \State{$\helpMergeTerminator(\rrev)$} \LineComment{corr. to \Cref{fig:merge}c-e}
            \State{\continue}
        \EndIf
        \If{$\bnode.\nnext \neq \nnode$} \LineComment{a split or merge happened} \label{alg:get:nextChanged}
            \State{\continue}
        \EndIf
        \State{\bbreak}
  \EndWhile \label{alg:get:endSearch}
  \State{\Var $\revision = \bot$}
  \If{$\snapVersion = \newestVersion$}
      \State{$\revision = \getNewestRevision(\rrev, \key)$} \label{alg:get:getNewestRevision}
  \Else
      \State{$\revision = \getRevision(\rrev, \key, \snapVersion)$} \label{alg:get:getRevision}
  \EndIf
  \If{$\revision = \bot$} \label{alg:get:extractValue}
      \State{\Return $\bot$}
  \EndIf
  \State{\Return $\revision.\get(\key)$} \label{alg:get:extractValueEnd}
\EndFunction \label{alg:getSnapshotEnd}

\Function{$\getNewestRevision$}{$\rrev, \key$} \label{alg:getNewestRevision}
  \State{\Var $\revision = \rrev$}
  \While{$\revision \neq \bot$} \LineComment{iterate over revision list}
    \If{$\revision.\version > 0$} \LineComment{first from a completed update} \label{alg:getNewestRevision:positiveVersion}
      \State{\bbreak}
    \EndIf
    \If{$\revision\ \is\ \MergeRevision \wedge \key \ge \revision.\keyOfRightNode$} \label{alg:getNewestRevision:mergeRevision}
      \State{$\revision = \revision.\rightNext$} \LineComment{choose the right successor}
    \Else
      \State{$\revision = \revision.\nnext$} \LineComment{choose the only (or left) successor}
    \EndIf
  \EndWhile
  \State{\Return $\revision$}  
\EndFunction \label{alg:getNewestRevisionEnd}
  
\Function{$\getRevision$}{$\rrev, \key, \snapVersion$} \label{alg:getRevision}
  \State{\Var $\revision = \rrev$}
  \While{$\revision \neq \bot$} \LineComment{iterate over revision list}
    \State{\Var $\version = \revision.\version$} 
    \If{$\revision.\version > 0 \wedge \version \le \snapVersion$} 
      \State{\bbreak}
    \EndIf
    \If{$\version < 0 \wedge -\version \le \snapVersion$}
        \State{$\helpPut(\revision)$}  \LineComment{complete the operation} 
        \State{$\version = \revision.\version$}
        \If{$\revision\ \is\ \MergeTerminator$}
            \State{$\revision = \findMergeRevision(\revision)$} \label{alg:getRevision:findMergeRevision}
        \EndIf
        % \If{$\revision = \bot \vee \version \le \snapVersion$}
        \If{$\version \le \snapVersion$} 
            \State{\bbreak}
        \EndIf
    \EndIf
    \If{$\revision\ \is\ \MergeRevision \wedge \key \ge \revision.\keyOfRightNode$}
      \State{$\revision = \revision.\rightNext$} \LineComment{choose the right successor}
    \Else
      \State{$\revision = \revision.\nnext$} \LineComment{choose the only (or left) successor}
    \EndIf
  \EndWhile
  \State{\Return $\revision$}  
\EndFunction  \label{alg:getRevisionEnd}
\end{algorithmic}
\end{multicols*}
\vspace{-0.35cm}
\end{algorithm*}

Before we discuss how the lookups and range scans are implemented, 
let us focus on the way snapshots are maintained. \jiffy is implemented in 
Java, which means that we do not have to manage the memory manually. However, 
we still need to track snapshots acquired by threads to let the 
JVM's garbage collector reclaim revisions that are no longer useful (will 
never be read again). To this end, a thread that acquires a snapshot 
\emph{registers} in the index by adding a special object to the \emph{snapshot 
list}, which is a lock-free linked list. Each object on the snapshot list 
contains a publicly available snapshot version $\snapVersion$ acquired from 
TSC upon thread registration. A snapshot with $\snapVersion$ 
corresponds to the state of the dataset at time $\snapVersion$. \jiffy's inner 
garbage collector periodically scans the list to obtain the lowest 
$\snapVersion$, so it knows which entries can be safely disposed of (removing 
unnecessary revisions happens upon every update operation, see, e.g., 
line~\ref{alg:put:gc} in $\pput$ in \Cref{alg:put_remove}). A thread can easily 
refresh the snapshot by querying again the TSC register and writing the new 
value in the thread's entry on the list (this operation does not even require a 
CAS operation, as 8B values are written atomically on the x86\_64 
architecture). Note that this operation has to be performed immediately after 
registering because \jiffy's inner garbage collector could have already freed 
some entries that would be visible to the reader thread. A reader thread should 
regularly refresh its snapshot to allow the garbage collector to progress, and 
\emph{unregister} (remove its object from the list) when it will not use 
snapshots any more. Note that if a thread wants to use several snapshots at the 
same time, it suffices that the value $\snapVersion$ stored in the thread's 
entry in snapshot list represents the smallest snapshot version of all thread's 
snapshots.

A lookup operation (see \Cref{alg:get}) comes in two variants: 
$\get(\key)$, which is used to retrieve the newest entry for some key $\key$ 
(lines \ref{alg:getNewest}-\ref{alg:getNewestEnd}), and $\get(\key, 
\snapVersion)$, used when operating on a snapshot $\snapVersion$ (lines 
\ref{alg:getSnapshot}-\ref{alg:getSnapshotEnd}). In fact, the former function 
calls the latter function with the special value $\newestVersion$.

The $\get(\key, \snapVersion)$ function starts similarly to $\pput$ and 
$\rremove$ by finding the appropriate node for $\key$ (lines 
\ref{alg:get:startSearch}-\ref{alg:get:endSearch}). However, unlike those 
functions, it helps only in completing pending structure modifications (which 
are rare), not regular updates. Then, depending on the value of $\snapVersion$, 
either the $\getNewestRevision$ or $\getRevision$ function is invoked (lines 
\ref{alg:get:getNewestRevision} and \ref{alg:get:getRevision}, respectively). 
Finally, the value for $\key$ is retrieved from the revision, unless the 
revision is $\bot$. In such case, $\get$ returns $\bot$ as well (lines 
\ref{alg:get:extractValue}-\ref{alg:get:extractValueEnd}).

The $\getNewestRevision$ function (lines 
\ref{alg:getNewestRevision}-\ref{alg:getNewestRevisionEnd}) simply iterates 
over the revision list and returns the first revision with a positive version 
number (line~\ref{alg:getNewestRevision:positiveVersion}).\footnote{Note that 
for simplicity we abstract away from the fact that version numbers of revisions 
created by $\batchUpdate$ operations have to be accessed indirectly, through 
the batch descriptor.} Since revisions in the revision list are kept in 
descending order of the absolute values of their version numbers, the function 
will return a revision from the most recently completed update operation at 
this node. Note that when we reach a merge revision which the function does not 
return (line~\ref{alg:getNewestRevision:mergeRevision}), we need to decide 
whether to proceed to the left or to the right successor of the merge revision. 
To this end, we compare $\key$ with the $\keyOfRightNode$ field of the merge 
revision, which stores the key of the node that underwent a merge operation 
that resulted in the merge revision.

The $\getRevision$ function (lines 
\ref{alg:getRevision}-\ref{alg:getRevisionEnd}) performs more complex logic, 
which corresponds to the rules we already discussed in 
\Cref{sec:design:versions}. Note that when $\getRevision$ encounters a merge 
terminator and helps to complete the merge operation, it needs to find the 
corresponding merge revision (on a node that precedes in the index the node 
with the merge terminator, line~\ref{alg:getRevision:findMergeRevision}). 
% It may happen that \jiffy's internal garbage collector already disposed of 
% this revision, if it is not required in snapshot $\snapVersion$. In such 
% case, $\getRevision$ returns $\bot$.

Range scans (which always operate on some snapshot) rely on the same 
logic as $\getRevision$, except for one difference. Recall that in 
$\getRevision(\key, \snapVersion)$ in some cases we use $\key$ to decide
whether to proceed to the left or the right successor of a merge revision.
A range scan intends to retrieve all key-value entries from the 
appropriate revision. Hence, if a range scan encounters a merge revision when 
evaluating a revision list at some node, it retrieves a \emph{bulk 
revision} that is constructed by recursively traversing all successors of all 
the encountered merge revisions. In practice, bulk revisions are created 
extremely rarely. In our tests (see \Cref{sec:evaluation}), revision lists 
contain at most 3-4 revisions at a time, and usually only 2. Moreover, node 
merges are rare, so there are few merge revisions that would necessitate in 
creating bulk revisions.

\subsubsection{Revision layout} \label{sec:design:revision}

So far we treated a revision as an immutable object that holds a range of 
key-value entries in a concrete version. Now we discuss, how revisions are 
implemented.

A revision holds two arrays: $\keys$ and $\values$. Data in both arrays is 
sorted according to the keys. This way we can perform lookup operations in a 
cache-friendly manner. Transforming one revision into a new one, as required by 
the update operations, involves copying the arrays and updating/removing the 
appropriate keys/values. Since all keys and values are kept in a contiguous 
range of memory, such copy operations are fast.

Our tests have shown, that threads spend a significant amount
of time performing binary search in revisions. Thus, we added a lightweight 
hash index for a fast key lookup. More precisely, in each revision, we 
maintain two additional arrays. The first array, $\indices$, contains 2B 
values and is twice the length of the $\keys$ array. Upon 
creation of a revision, the $\indices$ array is populated so that for each $k = 
\keys[i]$, $i$ is written to either $\indices[2 * t]$ or $\indices[2 * t + 1]$, 
where $t = h(k)\ \mathit{mod}\ \length(\keys)$, for some hash function $h$. A 
lookup operation for some key $k$ calculates $h(k)$ and $t$, and then checks if 
$k$ is stored in $\keys[\indices[2 * t]]$. If not, it looks for $k$ again in 
$\keys[\indices[2 * t + 1]]$. If either of these checks were successful, we can 
return the value for $k$ by returning either $\values[\indices[2 * t]]$ or 
$\values[\indices[2 * t + 1]]$. On the other hand, if either $\indices[2 * t]$
or $\indices[2 * t + 1]$ was empty, $k$ is not present in the revision. 
Finally, if $k$ was not found under either index (because at least two other 
keys in the revision had the same hash value), a binary search is performed on 
$\keys$. To speed up populating the $\indices$ array, in the second array, 
$\hashes$, we store 2B hashes of keys calculated using the $h$ function. Upon 
creation of a new revision, the $\hashes$ array can be efficiently copied, 
similarly to the $\keys$ and $\values$ arrays. Since hashes are 2B, a revision 
can contain up to 65K key-value entries, which is more than enough (in the 
tests from \Cref{sec:evaluation} each revision stores 25-300 entries).

\subsubsection{Autoscaling policy} \label{sec:design:autoscaling}

% Before we outline our \emph{autoscaling policy} that enables \jiffy to 
% automatically adapt the size of its inner index to the workload, we share 
% a few observations. 

Determining the optimal size of a revision is problematic, because smaller 
revisions are better for updates, as less copying is needed, whereas 
larger revisions better suit reads, i.e., lookups and range scans, as the 
index is smaller and range scans can efficiently read large, sorted arrays 
of entries. Our experiments showed that the sizes of revisions should be 
between 25-300 entries, depending on the workload.\footnote{\jiffy is a generic 
Java data structure, which means that arrays in the revisions store references 
to key/value objects, and not the keys/values themselves. Hence, the size of 
a revision does not depend on the types of keys/values, as could be the case if 
\jiffy were implemented in, e.g., C++.} Moreover, we noticed that adding the 
lightweight hash indices to revisions not only improved the overall performance, 
but reduced the relative performance differences when we tested \jiffy with 
different predefined revision sizes. However, the size of revisions still 
impacts the relative performance of updates and reads. Hence, we need some way 
of monitoring the workload to automatically control the sizes of revisions. 

We cannot simply monitor the number of updates (or reads) in a unit of time, 
and adjust the sizes of revisions, because of a positive feedback loop: in a 
read dominated workload revisions are larger, which negatively impacts the 
execution of updates. Hence, fewer updates are executed. In turn, the ratio of 
reads to updates increases, which leads to the further increase of the 
revisions sizes. An analogous case can be made for write dominated workloads.

% Secondly,  Smaller revisions implicate high overhead resulting 
% from managing many small objects, whereas larger revisions cannot be copied 
% (and modified) efficiently. Each array in a revision that holds 100 key/value 
% entries is few hundred bytes in total, so can be efficiently copied by CPU 
% using 256b and 512b copy instructions, such as \texttt{AVX512} \cite{??}. 

\begin{comment}
Thirdly, a workload consisting of updates and reads (lookup/range scan 
operations) requires a compromise in terms of revision sizes, as updates 
perform better with smaller revisions whereas large revisions better suit 
reads. Moreover, we cannot simply monitor the number of updates (or reads) in a 
unit of time, and adjust the sizes of revisions, because of a positive feedback 
loop: a read dominated workload means larger revisions, which negatively 
impacts the execution of updates. Hence, fewer updates are executed. In turn, 
the ratio of reads to updates increases, which leads to the further increase of 
the revisions sizes. An analogous case can be made for write dominated 
workloads. 
\end{comment}

\newcommand{\preads}{\mathsf{pReads}}
\newcommand{\pupdates}{\mathsf{pUpdates}}

Our \emph{autoscaling policy} works as follows. Each revision maintains two 
exponential moving averages $\preads$ and $\pupdates$ that roughly correspond 
to the amount of time spent by threads performing reads and updates in 
the revisions in any node given. To this end, instead of using a constant, we 
weight both moving averages using the time that passed since the thread last 
performed any read or update, respectively. We use the ratio of these values 
and a simple linear function to calculate the suitable revision size from range 
[25, 300], with smaller revisions when the majority of operations are updates.

More precisely, when a thread adds a new revision $r$, its $\pupdates$ $= t + (1 
- t) * u$ and $\preads = (1 - t) * p$, where $u$ and $p$ are the values of 
$\pupdates$ and $\preads$, respectively, from $r$'s successor in the revision 
list and $0 < t \le 1$ is the time (in seconds) between last such operation 
performed by the thread and the creation of $r$ (both values are obtained from 
TSC). In a batch update, the weight $t$ is divided between all created 
revisions. Upon a read, a thread modifies the moving averages in the first 
revision on the revision list in a similar way ($\pupdates = (1 - t) * u$ and 
$\preads = t + (1 - t) * p$), but $t$ corresponds to the time that passed since 
the last read performed by the thread. Updating the moving averages by 
concurrent threads results in a race condition, which is harmless, as we are 
just gathering some statistics. To reduce the load on the reader threads as well 
as the chances of a race condition happening, reader threads update the moving 
averages only every 100 read operations (with $t$ corresponding to the time it 
took the thread to perform 100 reads). Range scans update the moving averages 
only once per revision despite reading many entries, because retrieving the 
revision from the index requires much more effort compared to reading entries 
from a revision.

\begin{comment}
More precisely, when a thread adds a new revision $r$, its $\pupdates$ is 
calculated as $t + (1 - t) * u$, where $u$ is the value of $\pupdates$ from the 
revision from the head of the revision list and $0 < t \le 1$ is the time (in 
seconds) between last such operation performed by the thread and the creation 
of $r$ (both values are obtained from TSC). In a batch update, the 
weight $t$ is divided between all created revisions. The value of $\preads$ (in 
the first revision on the revision list) is modified by threads that perform 
reads in a similar way to $\pupdates$ but $t$ corresponds to the time that 
passed since the last update of any $\preads$ by the thread. Updating the 
$\preads$ by concurrent threads results in a race condition, which is 
harmless, as we are just gathering some statistics. To reduce the load on the 
reader threads as well as the chances of a race condition happening, threads 
update $\preads$ only every 100 read operations (with $t$ corresponding to the 
time it took the thread to perform 100 reads). Range scans update $\preads$ 
only once per revision despite reading many entries, because retrieving the 
revision from the index requires much more effort compared to reading entries 
from a revision.
\end{comment}

Our autoscaling policy is completely different from the one from \cite{SW15a, 
SW18, WSJ18}, which relies on observing contention on shared references to 
containers (revisions). The CAS operations on these references are performed by 
updates and range scans. Interestingly, with a single thread, the mentioned 
approach leads to ever increasing revision sizes, which is problematic for 
updates.

\subsection{Correctness}

Now we argue that \jiffy ensures linearizability \cite{HW90}. For simplicity, 
we abstract from node splits and merges. It is easy to see that $\pput$, 
$\rremove$, and $\applyBatch$ operations (on the same keys) are serialized 
because (1) no revision can be added to the revision list if there are pending 
operations at this node (2) when encountering a pending operation at some node, 
a thread helps to complete the operation before proceeding with its own update. 
The final version numbers of revisions in the revision list of each node 
monotonically decrease when iterating from the head of the revision list (recall 
that an optimistic version number equals $-v$, where $v = \TSC.\rread() + 1$, 
so it represents a moment in time in the future, and the final version number 
$v'$, which is also acquired from TSC is such that $v' \ge v$). All 
$\applyBatch$ operations update keys in the descending order of keys, thus 
ensuring that no two batch updates with intersecting key sets update revision 
lists of two nodes in a different order. The entries created by every update 
operation can be read by other threads once the final version number is 
established. The final (positive) version number is written to the $\version$ 
field of the entry or to the batch descriptor, using an atomic operation (CAS). 
Entries created by the same $\applyBatch$ operation appear as added atomically 
because all entries share the same batch descriptor. The assignment of the 
final version number is the linearization point for updates.

Now we discuss the $\get$ operations. Observe the following:
\begin{enumerate} 
\item Entries (within revisions) for any key $k$ are arranged in the revision 
list of the node responsible for a key range that includes $k$, according to 
their (final) version numbers in descending order (as we argued above), and the 
$\gget$ operation always evaluates the entries in that order.
\item For each key $k$ at any given moment there can be only a single pending 
update operation that modifies $k$ (a revision without the final version number 
established), which precedes in the revision list all other revisions that 
might include $k$.
% \item Version numbers are generated by reading the TSC register whose values 
% are monotonically increasing \cite{Intel07}.  
% \item For any key $k$ there are no two entries with the same final version 
% number (see the wait in lines \ref{alg:put:wait} and \ref{alg:batchUpdate:wait} 
% in \Cref{alg:put_remove} and \Cref{alg:applyBatch}, respectively). 
\end{enumerate}

% The $\get(k)$ operation always evaluates entries in the order specified in 
% observation 1. 
% If $\get(k)$ is not performed on a snapshot ($\snapVersion = \bot$), 
According to the linearizable semantics, the $\get(k)$ operation
% of such an operation, it 
must return the newest value written for the key $k$ and it may or may not 
observe the effects of the concurrent operations (operations that have not 
completed before $\get(k)$ started). The inclusion or exclusion of a concurrent
update depends on whether its linearization point lies before or after the one
for the get operation. Hence, $\get(k)$ can safely skip reading an entry in a 
revision whose final version number is not yet determined, and thus return the 
value from the entry (for key $k$) from the first revision whose final version 
number is positive.

Now consider 
% a reader thread with $s$ as its snapshot version. 
the $\get(k,s)$ operation. This time, the linearization point of the snapshot 
creation or update determines which value should be returned by the get 
operation. The value $s$ is obtained from the TSC register upon registering or 
updating the snapshot. Entries written by update operations that finished 
prior to the acquisition of $s$ have the final version number $v \le s$ (recall 
the wait of the update operations until the TSC register indicates 
$t \ge v$). On the other hand, entries written by operations executed 
concurrently with the snapshot creation/update may (but not necessarily must) 
have final version numbers $v' > s$. We choose the linearization point for the 
snapshot creation/update so that it precedes all such concurrent operations.

% The $\get(k,s)$ operation has to return the newest value written for $K$ by an update 
% operation which finished prior the snapshot creation/update, or a value written 
% by an update operation concurrent with the snapshot creation/update. In the 
% former case the returned entry's final version number $v \le s$. In the latter case we 
% also limit considered entries to those whose final version number $v$ satisfies 
% the condition $v \le s$. 

The $\get(k,s)$ operation chooses the entry for key $k$ from a revision with 
the greatest final version number $v \le s$. Recall our observation (1). For a 
revision $r$ with version number $v$, such that $|v| > s$, we can skip reading 
$r$, because if $v < 0$, due to our invariant (see \Cref{sec:design:versions}), 
the final version number for this revision will be at least $|v|$, so also 
greater than $s$. If $v < 0$ and $-v \le s$, $\get(k,s)$ helps to complete the 
update operation. It means that $\get(k,s)$ will be able to determine the final 
version number for this revision. For the first revision with a final 
(positive) version number $v \le s$, $\get(k,s)$ extracts the value for key $k$ 
and returns it.

\section{Evaluation} \label{sec:evaluation}

% In this section, we present the results of the experimental evaluation of 
% \jiffy. Since providing rich semantics (in the form of batch updates and 
% snapshots) while ensuring good multithreaded performance is notoriously 
% difficult, our primary focus is on the scalability of \jiffy under varied 
% workloads.

\subsection{Test environment} \label{sec:evaluation:hardware}

We implemented \jiffy in Java and experimentally compared it with \snaptree 
\cite{BCC+10}, \kary \cite{BH11, BBF+12}, \catree (lock-based 
contention-adapting tree with immutable containers) \cite{SW15a}, \caavl and 
\casl (lock-based CA trees with mutable containers based on AVL trees and skip 
lists, respectively) \cite{SW18} and \lfca (lock-free CA tree with immutable 
containers) \cite{WSJ18} (see also \Cref{sec:related_work}). All of these 
ordered indices feature linearizable range scans. \caavl and \casl also 
support linearizable batch updates. For reference, we also include the 
ubiquitous \texttt{ConcurrentSkipListMap} (\jsl) \cite{JavaConcurrent}, which 
does not support either consistent range scans nor atomic batch updates. In 
some tests we also include \kiwi \cite{BBB+17} whose available codebase 
\cite{KiWi} supports only 4~B integer keys.\footnote{Hence comparing \kiwi's 
performance with the performance of other indices in our tests is difficult 
(all other indices are generic, so they work with keys and values of different 
types and store them as Java objects, not values of primitive types).}

We conducted our tests on a server equipped with two Intel Xeon Gold 6252N 
CPUs, 192 GB of DRAM and running OpenSUSE Tumbleweed (version 20200815) with 
kernel 5.8. Each CPU has 24 cores (48 hyperthreads), is clocked at 2.3 GHz and 
features 36 MB of L3 cache. We ran our tests on OpenJDK 14.0.2.

\subsection{Microbenchmark and test scenarios} \label{sec:evaluation:scenarios}

% Since providing rich semantics (in the form of batch updates and snapshots) 
% while ensuring good multithreaded performance is notoriously difficult, in 
% our first test, 
To show that our novel system can achieve good multithreaded performance 
despite providing rich semantics, we use a custom microbenchmark to assess how 
\jiffy (and its competitors) perform under multithreaded workloads with varied 
levels of contention. Each microbenchmark thread issues only one type of 
operations, i.e., either \emph{updates} (put/remove/batch update operations), 
\emph{lookups} (get operations) or \emph{range scans}, so that certain 
operations, such as long-running scans or batch updates do not stifle the 
execution of operations of other types. We vary the percentage of threads that
perform each kind of operations to uncover the characteristics of all tested
indices.
% \footnote{Batch updates are enabled only 
% for \jiffy and \leveldb, as the other tested implementations do not support 
% such 
% operations. Scan operations in \jiffy as well as \leveldb and \fptree are 
% atomic.}

In total we consider four test scenarios: an \emph{update-only} 
scenario, an \emph{update-lookup} scenario (25\% of threads do 
updates, 75\% of threads do lookups) and two \emph{mixed} scenarios (25\%  
threads do updates, 50\% threads do lookups, 25\% threads do range scans, but 
range scans are either \emph{short} or \emph{long}, i.e., cover 100 or 10000 
subsequent key-value entries, starting from a randomly chosen key). 

To assess the performance of batch updates in \jiffy, we test it in five 
variants. In the default variant, \jiffy performs all updates as single put or 
remove operations. Other variants correspond to results obtained when \jiffy 
executes all update operations in 10-operation batch updates or large, 
100-operation batches. To demonstrate the performance of batch updates in the 
extreme cases, they are either \emph{sequential} (update consecutive 
key-value entries) or \emph{random} (update randomly chosen key-value entries). 
In a similar way we test CA-AVL and CA-SL, which also support batch updates.

The dataset has the average size of 10M entries (20M unique keys). \jiffy 
is multiversioned, so it typically maintains more entries at any given moment. 
The sizes of key/value sizes are set to 16/100~B and 4/4~B  (typical for such 
tests, see, e.g., \cite{CSTR+10, AXFJ+12}). We examine the systems 
when keys are randomly chosen with a uniform and a Zipfian distribution 
(distribution skew is 0.99, which is the same as in the YCSB benchmark in the 
default settings \cite{CSTR+10}). 
% ; so about 10\% keys are referenced in about 85\% of cases,).

% (\Cref{fig:throughput}a-d)
% (\Cref{fig:throughput}e-h)

The results are reported in (millions of) \emph{basic} operations per second, 
i.e., put, remove or get operations on a single key (a scan over 
10 key-value entries counts as 10 get operations). 
% Due to space constraints,
% we include plots only for a subset of all tests we conducted (see Appendix for
% additional results).

% \input{4a-throughput_put_16_100.tex}
% \input{4a-throughput_put_b10_16_100.tex}

\providecommand{\plotscale}{0.71}
\providecommand{\STAB}[1]{\vspace{0.5cm}#1}

\bgroup
\def\arraystretch{1.2}
\def\colwidth{1.6in}
\def\toptrim{0.32cm}
\def\lefttrim{0.35cm}
\begin{figure*}[t]
% \begin{table}
%   \caption{Throughput.}
    \scalebox{0.95}{
  \begin{tabular}{C{0.2in}C{1.7in}C{\colwidth}C{\colwidth}C{\colwidth}}
%   \multicolumn{2}{c|}{Uniform distribution} & \multicolumn{2}{c}{Zipfian distribution (skewness: 0.99)}\\
%   \multicolumn{2}{c|}{Uniform distribution} & \multicolumn{2}{c}{Zipfian distribution}\\
%   \hline
    & (a) 100\% threads: put/remove & (b) 25\% threads: put/remove & \multicolumn{2}{c}{25\% threads: put/remove, 50\% threads: get, 25\% threads: scan} \\
    & & 75\% threads: get & (c) Short scans (100 ops) & (d) Long scans (10000 ops)\\ 
    
%     (e) 100\% threads: put/remove & (f)
%     75\% threads: get & 
    
%     100\% threads: get\\
    \STAB{\rotatebox[origin=c]{90}{Simple put/remove}}
    &
    \includegraphics[scale=\plotscale, trim={0 0 0 \toptrim}, clip]{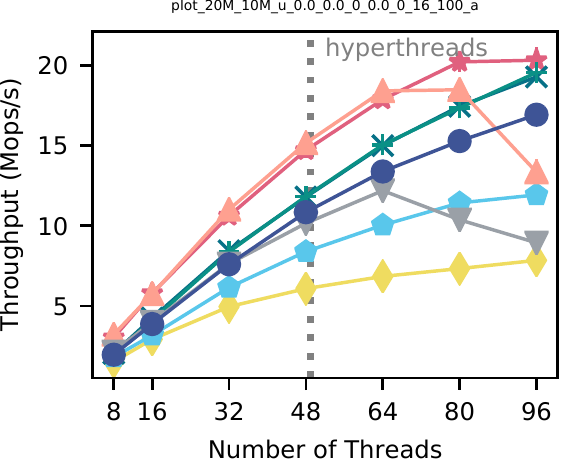}
    & 
    \includegraphics[scale=\plotscale, trim={{\lefttrim} 0 0 \toptrim}, clip]{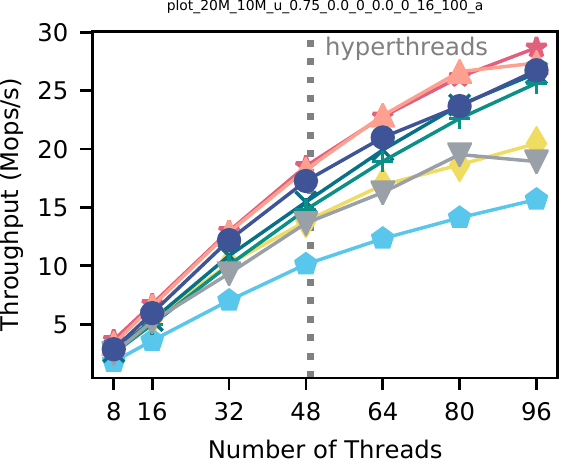} 
    & 
    \includegraphics[scale=\plotscale, trim={{\lefttrim} 0 0 \toptrim}, clip]{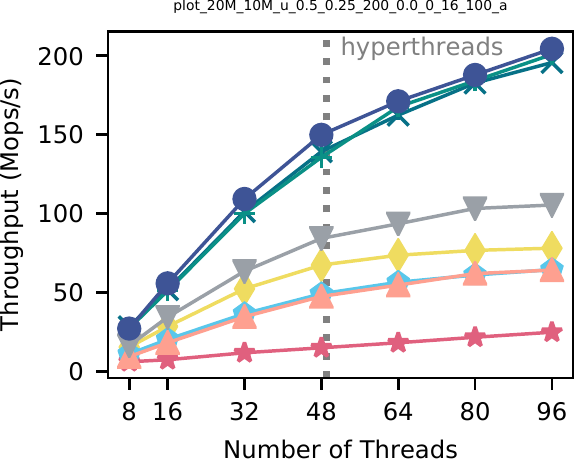}
    & 
    \includegraphics[scale=\plotscale, trim={{\lefttrim} 0 0 \toptrim}, clip]{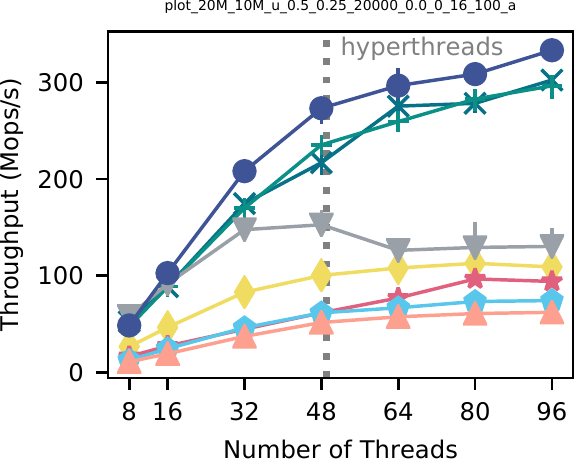} \\
    
    & \multicolumn{4}{l}{\hspace{2cm}\includegraphics[scale=\plotscale, trim={0 0 2cm 0}, clip]{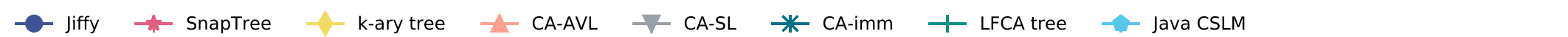}}\\
    
    \STAB{\rotatebox[origin=c]{90}{10-op. batch updates}}
    &
    \includegraphics[scale=\plotscale, trim={0 0 0 \toptrim}, clip]{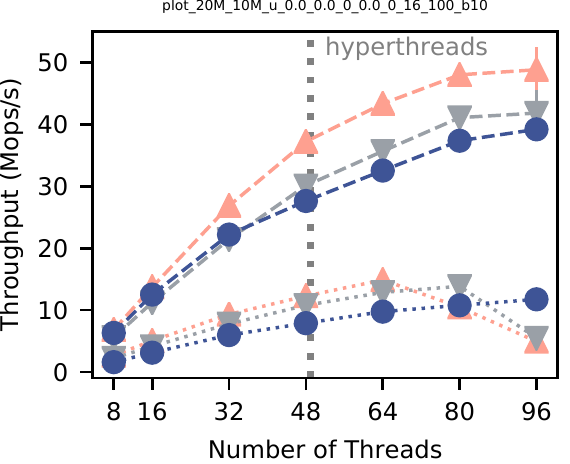}
    & 
    \includegraphics[scale=\plotscale, trim={{\lefttrim} 0 0 \toptrim}, clip]{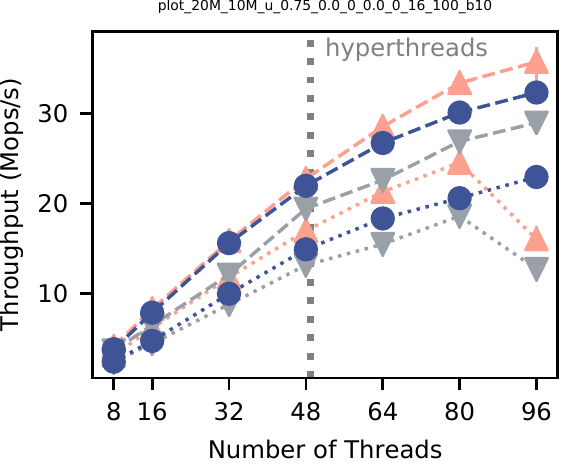} 
    & 
    \includegraphics[scale=\plotscale, trim={{\lefttrim} 0 0 \toptrim}, clip]{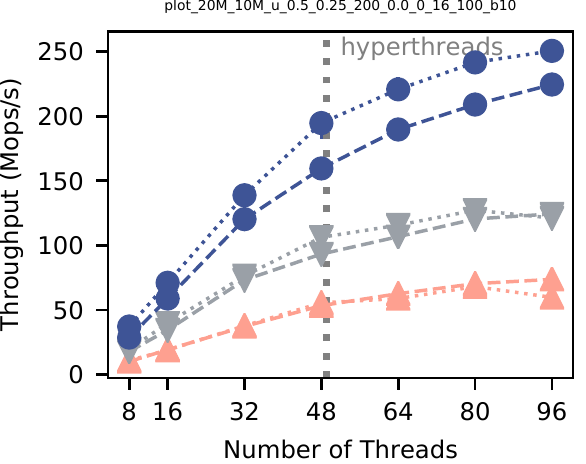}
    & 
    \includegraphics[scale=\plotscale, trim={{\lefttrim} 0 0 \toptrim}, clip]{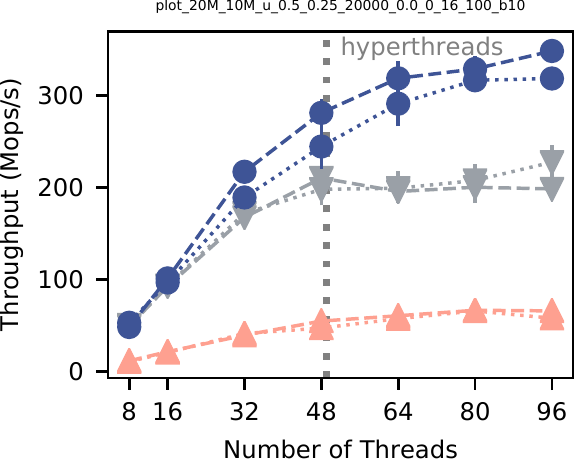} \\
    
    \STAB{\rotatebox[origin=c]{90}{100-op. batch updates}}
    &
    \includegraphics[scale=\plotscale, trim={0 0 0 \toptrim}, clip]{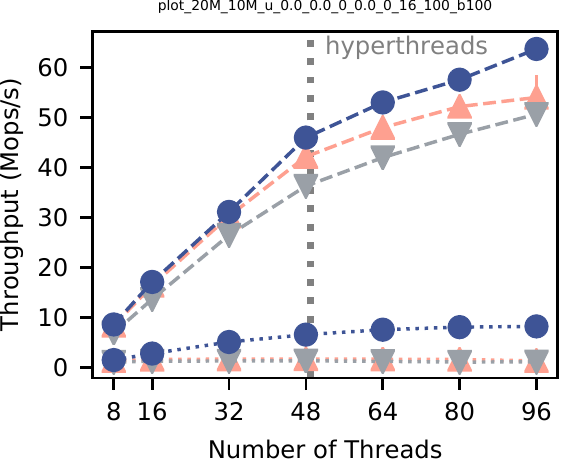}
    & 
    \includegraphics[scale=\plotscale, trim={{\lefttrim} 0 0 \toptrim}, clip]{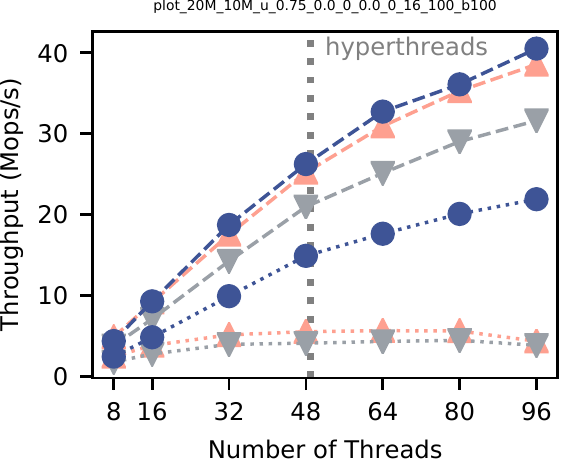} 
    & 
    \includegraphics[scale=\plotscale, trim={{\lefttrim} 0 0 \toptrim}, clip]{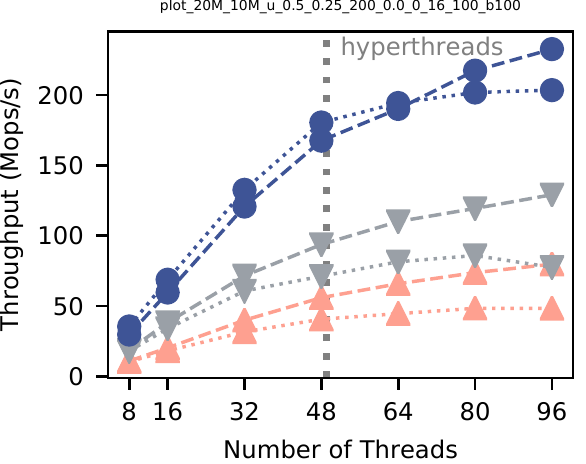}
    & 
    \includegraphics[scale=\plotscale, trim={{\lefttrim} 0 0 \toptrim}, clip]{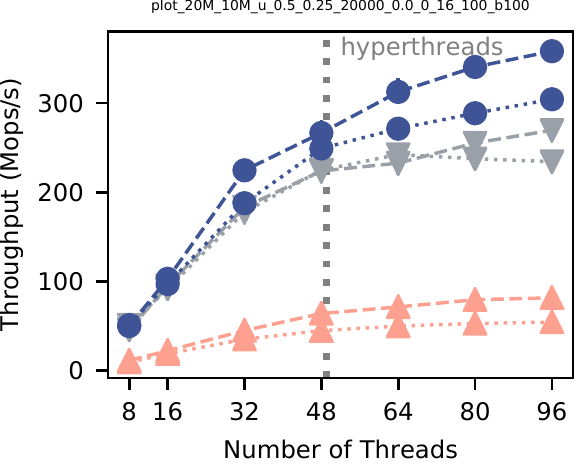} \\
    
    &\multicolumn{4}{l}{\hspace{3cm}\includegraphics[scale=\plotscale, trim={0 0 2cm 0}, clip]{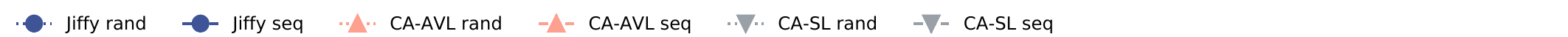}}\\

  \end{tabular}
  }
% \end{table}
\caption{Throughput scalability results (16~B keys, 100~B values, keys chosen with uniform distribution). 
% \kdb 
% appears in 5 variants (see \Cref{sec:evaluation:scalability:scenarios} for 
% details). 
% % In the default variant (red line, fully filled circle 
% % marker), \kdb performs all put/remove operations as separate updating operations 
% % (see \Cref{sec:design:single}). In the other varians, \kdb performs put/remove 
% % operations in batches (see \Cref{sec:design:batch}) counting either 10 keys 
% % (light red line, empty circle marker) or 100 keys (dark red line, half-filled 
% % circle marker). To demonstrate the performance of batches in the extreme cases,
% % batches update either 10/100 consecutive key-value entries (dashed line) or
% % randomly chosen 10/100 key-value entries (dotted line). 
% All raported values are in millions of basic operations per second (Mops/s). 
% E.g., an atomic batch update of 10 key-value entries counts as 10 put/remove 
% operations and a 1000-element scan counts as 1000 get operations.
}
\label{fig:throughput}
\end{figure*}
\egroup

\providecommand{\plotscale}{0.71}
\providecommand{\STAB}[1]{\vspace{0.5cm}#1}

\bgroup
\def\arraystretch{1.2}
\def\colwidth{1.6in}
\def\toptrim{0.32cm}
\def\lefttrim{0.35cm}
\begin{figure*}[t]
% \begin{table}
%   \caption{Throughput.}
    \scalebox{0.95}{
  \begin{tabular}{C{0.2in}C{1.7in}C{\colwidth}C{\colwidth}C{\colwidth}}
%   \multicolumn{2}{c|}{Uniform distribution} & \multicolumn{2}{c}{Zipfian distribution (skewness: 0.99)}\\
%   \multicolumn{2}{c|}{Uniform distribution} & \multicolumn{2}{c}{Zipfian distribution}\\
%   \hline
    & (a) 100\% threads: put/remove & (b) 25\% threads: put/remove & \multicolumn{2}{c}{25\% threads: put/remove, 50\% threads: get, 25\% threads: scan} \\
    & & 75\% threads: get & (c) Short scans (100 ops) & (d) Long scans (10000 ops)\\ 
    
%     (e) 100\% threads: put/remove & (f)
%     75\% threads: get & 
    
%     100\% threads: get\\
    \STAB{\rotatebox[origin=c]{90}{Simple put/remove}}
    &
    \includegraphics[scale=\plotscale, trim={0 0 0 \toptrim}, clip]{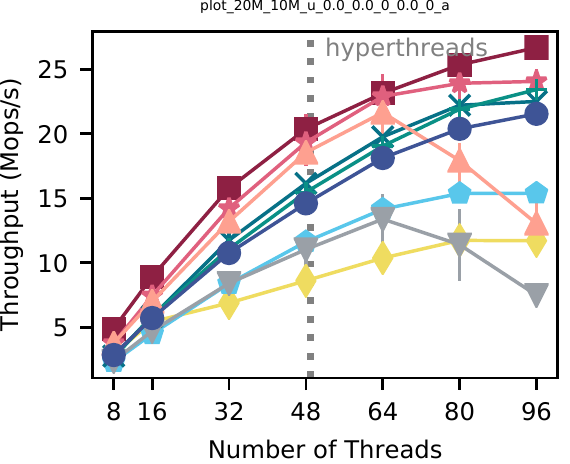}
    & 
    \includegraphics[scale=\plotscale, trim={{\lefttrim} 0 0 \toptrim}, clip]{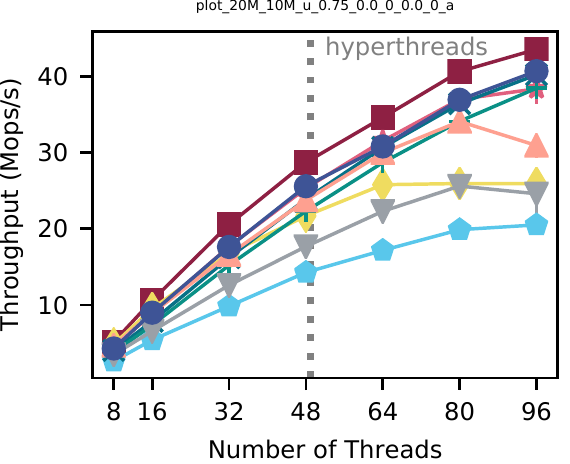} 
    & 
    \includegraphics[scale=\plotscale, trim={{\lefttrim} 0 0 \toptrim}, clip]{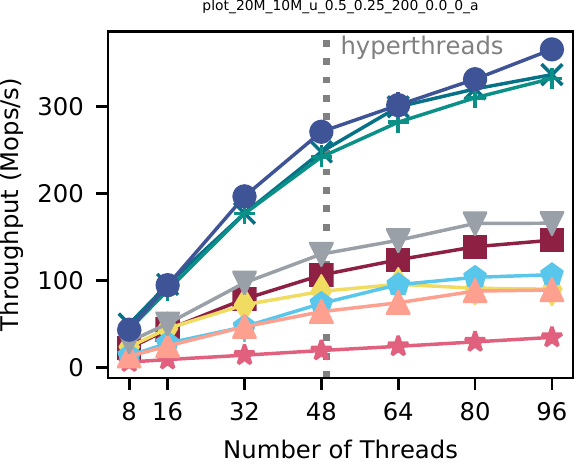}
    & 
    \includegraphics[scale=\plotscale, trim={{\lefttrim} 0 0 \toptrim}, clip]{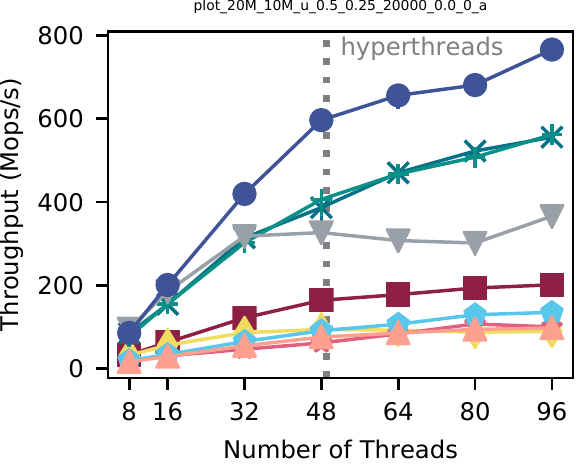} \\
    
    & \multicolumn{4}{l}{\hspace{2cm}\includegraphics[scale=\plotscale, trim={0 0 1cm 0}, clip]{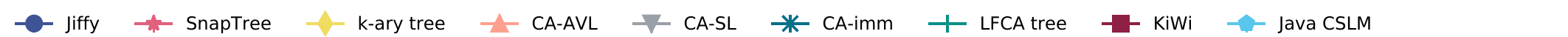}}\\
    
    \STAB{\rotatebox[origin=c]{90}{10-op. batch updates}}
    &
    \includegraphics[scale=\plotscale, trim={0 0 0 \toptrim}, clip]{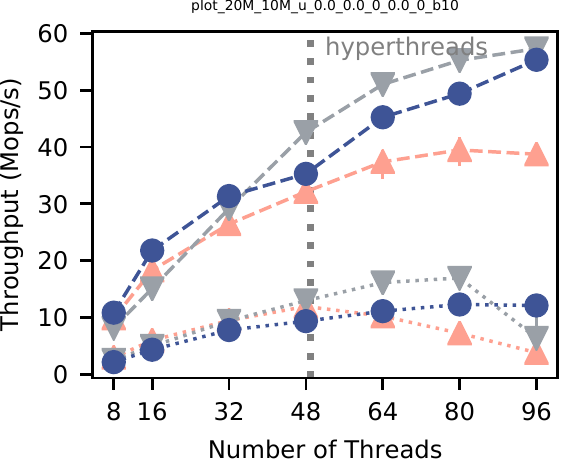}
    & 
    \includegraphics[scale=\plotscale, trim={{\lefttrim} 0 0 \toptrim}, clip]{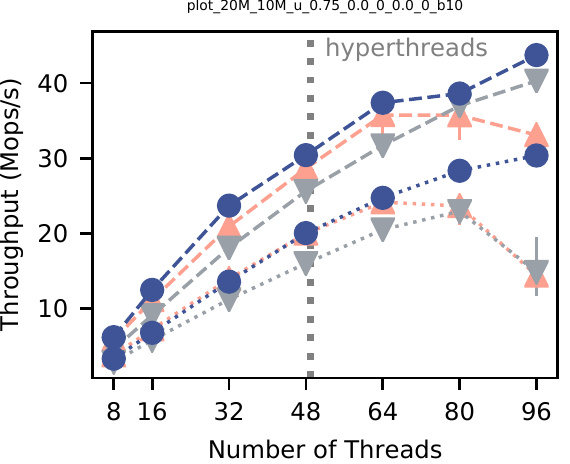} 
    & 
    \includegraphics[scale=\plotscale, trim={{\lefttrim} 0 0 \toptrim}, clip]{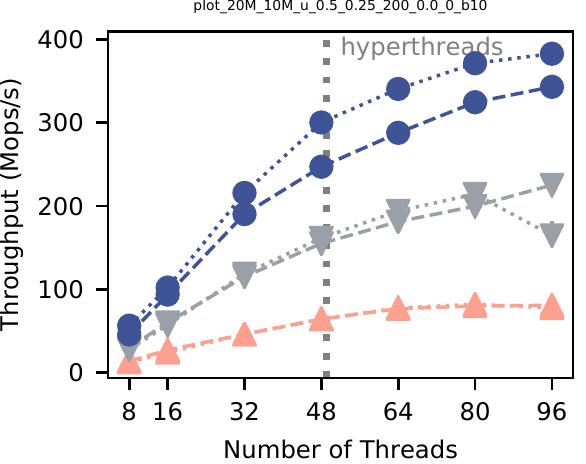}
    & 
    \includegraphics[scale=\plotscale, trim={{\lefttrim} 0 0 \toptrim}, clip]{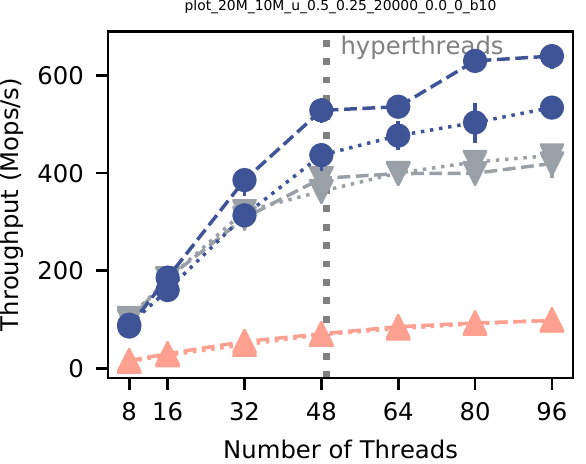} \\
    
    \STAB{\rotatebox[origin=c]{90}{100-op. batch updates}}
    &
    \includegraphics[scale=\plotscale, trim={0 0 0 \toptrim}, clip]{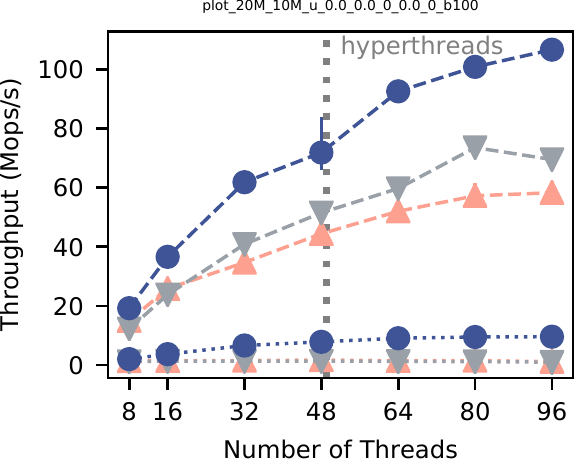}
    & 
    \includegraphics[scale=\plotscale, trim={{\lefttrim} 0 0 \toptrim}, clip]{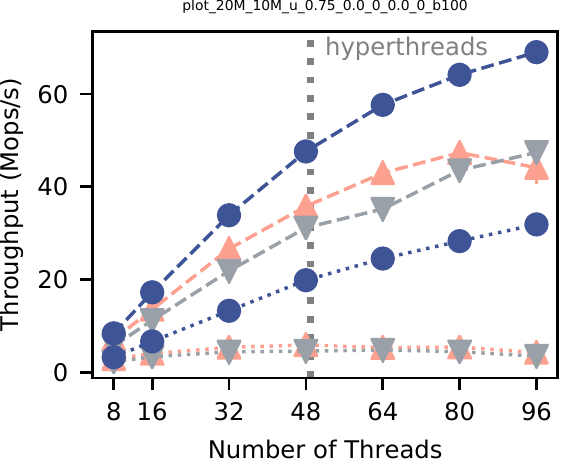} 
    & 
    \includegraphics[scale=\plotscale, trim={{\lefttrim} 0 0 \toptrim}, clip]{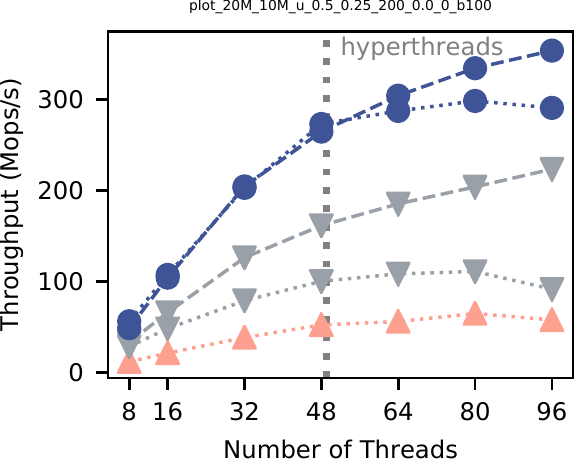}
    & 
    \includegraphics[scale=\plotscale, trim={{\lefttrim} 0 0 \toptrim}, clip]{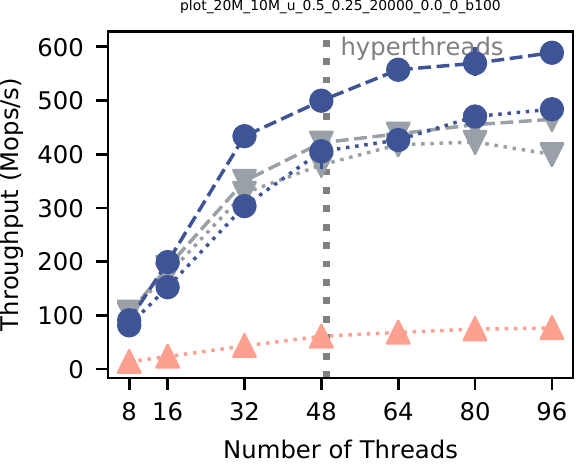} \\
    
    &\multicolumn{4}{l}{\hspace{3cm}\includegraphics[scale=\plotscale, trim={0 0 2cm 0}, clip]{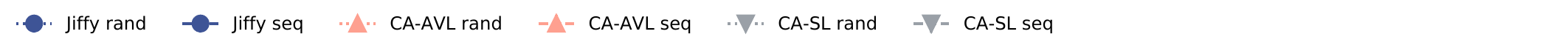}}\\
    
    \end{tabular}
  }
% \end{table}
\caption{Throughput scalability results (4~B keys, 4~B values, keys chosen with uniform distribution). 
% \kdb 
% appears in 5 variants (see \Cref{sec:evaluation:scalability:scenarios} for 
% details). 
% % In the default variant (red line, fully filled circle 
% % marker), \kdb performs all put/remove operations as separate updating operations 
% % (see \Cref{sec:design:single}). In the other varians, \kdb performs put/remove 
% % operations in batches (see \Cref{sec:design:batch}) counting either 10 keys 
% % (light red line, empty circle marker) or 100 keys (dark red line, half-filled 
% % circle marker). To demonstrate the performance of batches in the extreme cases,
% % batches update either 10/100 consecutive key-value entries (dashed line) or
% % randomly chosen 10/100 key-value entries (dotted line). 
% All raported values are in millions of basic operations per second (Mops/s). 
% E.g., an atomic batch update of 10 key-value entries counts as 10 put/remove 
% operations and a 1000-element scan counts as 1000 get operations.
}
\label{fig:throughput_4}
\end{figure*}
\egroup

\subsection{Results}

We start by discussing the results of tests in which key/value sizes were set 
to 16/100~B and keys were chosen with uniform distribution (see 
\Cref{fig:throughput}). In all tested scenarios, \jiffy exhibits scalable 
behaviour. Single put/remove operations in \jiffy are slightly more expensive 
than in some other systems, e.g., \snaptree, \catree, \lfca, \caavl (in the 
write-only scenario by about 30\% in the worst case at 64 threads and by 15\% 
for 96 threads, see the top row plot in \Cref{fig:throughput}a). The increased 
cost of updates comes from the multiversioned architecture of \jiffy. Each 
update that adds a revision to some node requires at least two CAS operations: 
one to add a revision to the revision list at the node and one to set the final 
version number to the revision. In other lock-free indices only one CAS is 
necessary: when update is performed in place (e.g., \jsl) or to replace an old 
key-value entry container with a new one (e.g., \lfca). Note that in \jiffy 
there is also an additional overhead resulting from managing lightweight hash 
indices inside revisions. As the hash indices boost the performance of lookups, 
the performance differences between \jiffy and the mentioned systems is smaller 
when lookups are introduced to the workload (see \Cref{fig:throughput}b). Our 
autoscaling policy set the revision sizes to around 35 entries in the 
write-only scenario vs 130 entries in the update-lookup scenarios. The revision 
size adjustment time was about 10 second (and about a second on a 1M entries 
dataset).

\jiffy executes range scans much more efficiently than its competitors (see 
\Cref{fig:throughput}c-d). \lfca and \catree are about 10\% slower than 
\jiffy's, whereas the only two other indices that, similarly to \jiffy, support
batch updates, i.e., \caavl and \casl, at best achieve only half of the total 
throughput of \jiffy. Interestingly, range scans are especially problematic in 
lock-based \snaptree, which performed best in the first two scenarios (see 
also \Cref{sec:related_work} for discussion on range scans in \snaptree).

Now let us consider the performance of batch updates in \jiffy. When batch 
updates are small (each includes 10 put/remove operations, see the plots in the 
middle row of \Cref{fig:throughput}), batch updates in \jiffy are slightly 
slower compared to \caavl's and \casl's, due to the same reasons, which we 
discussed earlier when explaining the performance of put/remove operations in 
\jiffy. Notice that with random batch updates the performance of lock-based 
\caavl and \casl starts to diminish towards the higher number of concurrent 
threads, whereas \jiffy continues to scale thanks to its lock-free 
architecture. The differences between the lock-based and the lock-free approach 
start to become apparent when all updates are executed as large batch updates 
(each includes 100 put/remove operations, see the plots in the bottom row of 
\Cref{fig:throughput}). When batch updates are sequential, the performance of 
\jiffy is about 15\% better than the performance of either \caavl or \casl (in 
the write-only scenario). However, with random batch updates, \jiffy's 
maximal throughput is 4.9$\times$ and 6.1$\times$ of the maximal throughput of 
\caavl and \casl, respectively.
% In the mixed scenarios we can clearly 
% observe that range scans are much more expensive in \caavl than \casl.

Notice the somewhat surprising way in which small batch updates impact 
the performance of \jiffy in the mixed scenario with small range scans 
(\Cref{fig:throughput}c, middle row). Using random batch updates results in 
a slightly better overall performance, compared to sequential batch updates, 
which are on average much cheaper to execute (each sequential batch update 
creates on average 1-2 revisions vs $n$ revisions for a random batch update
with $n$ put/remove operations). This phenomenon can be explained by examining 
the throughput of update operations \iftoggle{appendix}{(see 
\Cref{fig:a_throughput_u_16_100} in Appendix for the plot with update-only 
throughput)}{(plots for update-only throughput are available in the extended 
version of our paper \cite{KKW21-arxiv})}: with small sequential batch updates, 
\jiffy executes four times as many updates compared to the same test with random 
batch updates. In turn, in the former test, \jiffy has to manage many more 
revisions, which translates into slightly worse performance of lookups and 
scans. 

Let us now consider similar tests but conducted with 4~B key/value sizes (see 
\Cref{fig:throughput_4}). In the first two scenarios \kiwi (whose 
implementation is optimized for 4~B integer keys and does not support other 
key/value types) beats other indices, but not by much (in the write-only 
scenario the second best performing \snaptree is 10\% slower compared to \kiwi, 
whereas \jiffy is 20\% slower compared to \kiwi). Overall, the relative 
differences between the performance of the tested indices stay largely the 
same, except for two small differences. Firstly, with smaller key/value sizes, 
the performance of  lock-based \caavl and \casl starts to diminish earlier 
(with a smaller number of concurrent threads). Secondly, we can observe a much 
more substantial advantage of \jiffy in workloads that feature range scans. In 
the mixed scenario with long range scans, \jiffy beats the second-best 
performing indices \catree and \lfca by 30\%. Moreover, when updates are 
executed as large batch updates, our index achieves much better performance 
than the lock-based competitors. In the write-only scenarios, compared to tests 
with 16/100~B key/value sizes, \jiffy's advantage in speedup over competitors 
increases from 1.1$\times$ to 1.5$\times$ when batch updates are sequential and 
from 4.9$\times$/6.1$\times$ to 5.7$\times$/7.4$\times$ when batch updates are 
random (for \caavl and \casl, respectively).

We conducted similar tests but with keys chosen with Zipfian 
distribution \iftoggle{appendix}{(see \Cref{fig:a_throughput_z_16_100} and 
\Cref{fig:a_throughput_z_4_4} in Appendix)}{(due to space constraints, we refer 
the reader to \cite{KKW21-arxiv} for detailed results)}. The performance 
differences between the tested indices were largely the same as in the results 
presented above, although \kiwi no longer was the best performing index in the 
write-only scenario (in the update-lookup scenario \kiwi's performance was 
matched by \snaptree, \lfca, \catree and was a few percent better than \jiffy's). 
The biggest difference in performance could be observed when updates were 
executed as random batch updates. A skewed workload results in much higher 
contention levels which are further amplified when put/remove operations are 
performed as batch updates, each of which creates many new revisions 
(containers in \caavl and \casl). Such workloads were almost equally bad for 
\jiffy and its lock-based competitors. In the write-only scenario, the observed 
throughput for \jiffy, \caavl and \casl was about 1.5-2 \mops for small random 
batch updates and 0.3-0.5 \mops for large random batch updates.

\section{Conclusions} \label{sec:conclusions}

In this paper, we presented \jiffy, the first lock-free, linearizable ordered 
key-value index with batch updates and snapshots. Despite its rich 
functionality, \jiffy offers scalable performance across various workloads, 
often exceeding the performance of the state-of-the-art indices with less 
flexible semantics. Crucially, our novel lock-free, multiversioned algorithm 
that powers \jiffy allows it to execute batch updates more efficiently compared 
to its (lock-based) rivals, with speedup in throughput ranging from 1.1$\times$ 
to 7.4$\times$, depending on a test scenario.
% These results clearly validate our approach. 

\jiffy's codebase soon will be available on our github 
repository\iftoggle{appendix}{.}{ (we included the code as the supplemental 
material).}

% \begin{figure}
%   \centering
%   \includegraphics[width=\linewidth]{figures/duck}
%   \caption{An illustration of a Mallard Duck. Picture from Mabel Osgood Wright, \textit{Birdcraft}, published 1897.}
%   \label{fig:duck}
% \end{figure}
% 
% \begin{table*}[t]
%   \caption{A double column table.}
%   \label{tab:commands}
%   \begin{tabular}{ccl}
%     \toprule
%     A Wide Command Column & A Random Number & Comments\\
%     \midrule
%     \verb|\tabular| & 100& The content of a table \\
%     \verb|\table|  & 300 & For floating tables within a single column\\
%     \verb|\table*| & 400 & For wider floating tables that span two columns\\
%     \bottomrule
%   \end{tabular}
% \end{table*}

% \autoref{tab:freq}. 

% \begin{table}[hb]% h asks to places the floating element [h]ere.
%   \caption{Frequency of Special Characters}
%   \label{tab:freq}
%   \begin{tabular}{ccl}
%     \toprule
%     Non-English or Math & Frequency & Comments\\
%     \midrule
%     \O & 1 in 1000& For Swedish names\\
%     $\pi$ & 1 in 5 & Common in math\\
%     \$ & 4 in 5 & Used in business\\
%     $\Psi^2_1$ & 1 in 40\,000 & Unexplained usage\\
%   \bottomrule
% \end{tabular}
% \end{table}

\begin{acks}
This work was supported by the Foundation for Polish Science, within the
TEAM programme co-financed by the European Union under the European
Regional Development Fund (grant No. POIR.04. 04.00-00-5C5B/17-00).
% We would like to thank Lucas Lersch \etal for sharing their PMDK-based 
% implementations of \nvtree and \fptree and Intel Poland for providing us
% with hardware resources.
We thank Intel Poland for providing us with hardware resources.
%  This work was supported by the [...] Research Fund of [...] (Number [...]). Additional funding was provided by [...] and [...]. We also thank [...] for contributing [...].
\end{acks}

\clearpage

\balance

\bibliographystyle{ACM-Reference-Format}
\bibliography{bibliography}

\iftoggle{appendix}{
\clearpage
\section*{Appendix}

In \Cref{fig:a_throughput_z_16_100}, \Cref{fig:a_throughput_z_16_100}, 
\Cref{fig:a_throughput_u_4_4}, and \Cref{fig:a_throughput_z_4_4} we present the 
additional results of our scalability tests. Besides the total throughput, we
include the plots for throughput of update operations, so the data can be
easier to interpret.

\providecommand{\plotscale}{0.6}
\providecommand{\STAB}[1]{\vspace{0.5cm}#1}

\bgroup
\def\arraystretch{1.2}
\def\colwidth{1.65in}
\def\toptrim{0.15cm}
\def\lefttrim{0.35cm}
\begin{figure*}[t]
% \begin{table}
%   \caption{Throughput.}
    \scalebox{0.90}{
  \begin{tabular}{C{0.2in}C{0.2in}C{1.7in}C{\colwidth}C{\colwidth}C{\colwidth}}
%   \multicolumn{2}{c|}{Uniform distribution} & \multicolumn{2}{c}{Zipfian distribution (skewness: 0.99)}\\
%   \multicolumn{2}{c|}{Uniform distribution} & \multicolumn{2}{c}{Zipfian distribution}\\
%   \hline
    & & (a) 100\% threads: put/remove & (b) 25\% threads: put/remove & \multicolumn{2}{c}{25\% threads: put/remove, 50\% threads: get, 25\% threads: scan} \\
    & & & 75\% threads: get & (c) Short scans (100 ops) & (d) Long scans (10000 ops)\\ 
    
    \hline
%     (e) 100\% threads: put/remove & (f)
%     75\% threads: get & 
    
%     100\% threads: get\\
    \multirow{2}{*}{\STAB{\rotatebox[origin=c]{90}{Simple put/remove}}}
    & 
    \STAB{\rotatebox[origin=c]{90}{Total throughput}}
    &
    \includegraphics[scale=\plotscale, trim={0 0 0 \toptrim}, clip]{plots/plot_20M_10M_u_0_0_0_0_0_0_0_0_16_100_a.pdf}
    & 
    \includegraphics[scale=\plotscale, trim={{\lefttrim} 0 0 \toptrim}, clip]{plots/plot_20M_10M_u_0_75_0_0_0_0_0_0_16_100_a.pdf} 
    & 
    \includegraphics[scale=\plotscale, trim={{\lefttrim} 0 0 \toptrim}, clip]{plots/plot_20M_10M_u_0_5_0_25_200_0_0_0_16_100_a.pdf}
    & 
    \includegraphics[scale=\plotscale, trim={{\lefttrim} 0 0 \toptrim}, clip]{plots/plot_20M_10M_u_0_5_0_25_20000_0_0_0_16_100_a.pdf} \\
    
    &
    \STAB{\rotatebox[origin=c]{90}{Update throughput}}
    &
    \includegraphics[scale=\plotscale, trim={0 0 0 \toptrim}, clip]{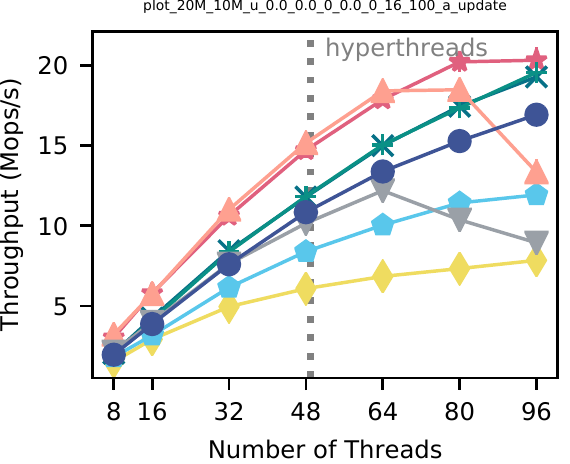}
    & 
    \includegraphics[scale=\plotscale, trim={{\lefttrim} 0 0 \toptrim}, clip]{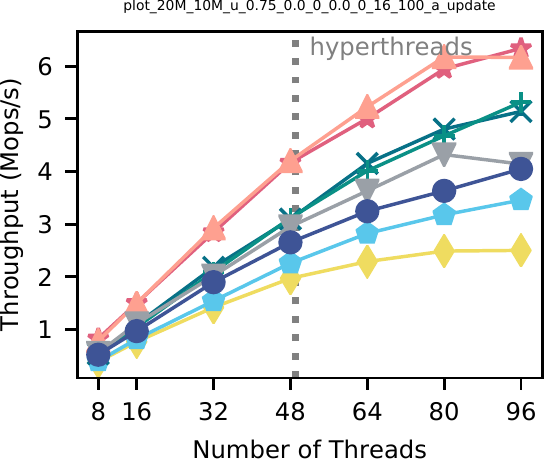} 
    & 
    \includegraphics[scale=\plotscale, trim={{\lefttrim} 0 0 \toptrim}, clip]{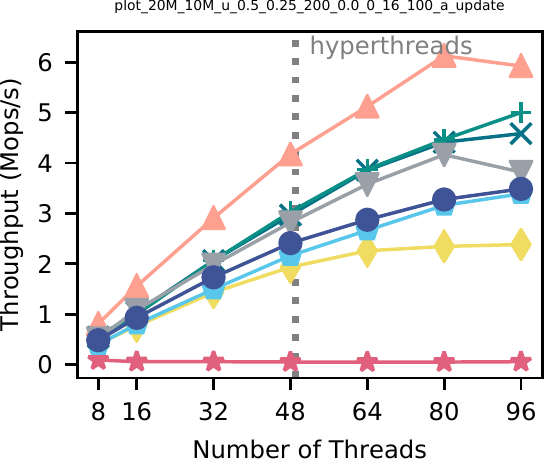}
    & 
    \includegraphics[scale=\plotscale, trim={{\lefttrim} 0 0 \toptrim}, clip]{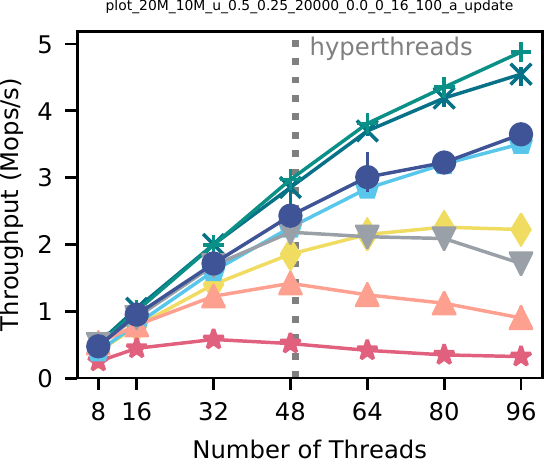} \\
    
   & & \multicolumn{4}{l}{\hspace{2cm}\includegraphics[scale=\plotscale, trim={0 0 2cm 0}, clip]{plots/plot_20M_10M_u_0_0_0_0_0_0_0_0_16_100_a_legend_single_row.pdf}}\\[2pt]
    
   \hline\\[-5pt]
    
    \multirow{2}{*}{\STAB{\rotatebox[origin=c]{90}{10-op. batch updates}}}
    &
    \STAB{\rotatebox[origin=c]{90}{Total throughput}}
    &
    \includegraphics[scale=\plotscale, trim={0 0 0 \toptrim}, clip]{plots/plot_20M_10M_u_0_0_0_0_0_0_0_0_16_100_b10.pdf}
    & 
    \includegraphics[scale=\plotscale, trim={{\lefttrim} 0 0 \toptrim}, clip]{plots/plot_20M_10M_u_0_75_0_0_0_0_0_0_16_100_b10.pdf} 
    & 
    \includegraphics[scale=\plotscale, trim={{\lefttrim} 0 0 \toptrim}, clip]{plots/plot_20M_10M_u_0_5_0_25_200_0_0_0_16_100_b10.pdf}
    & 
    \includegraphics[scale=\plotscale, trim={{\lefttrim} 0 0 \toptrim}, clip]{plots/plot_20M_10M_u_0_5_0_25_20000_0_0_0_16_100_b10.pdf} \\
    
    &
    \STAB{\rotatebox[origin=c]{90}{Update throughput}}
    &
    \includegraphics[scale=\plotscale, trim={0 0 0 \toptrim}, clip]{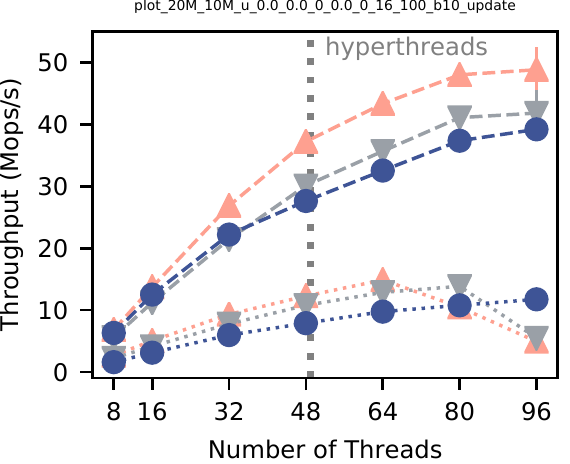}
    & 
    \includegraphics[scale=\plotscale, trim={{\lefttrim} 0 0 \toptrim}, clip]{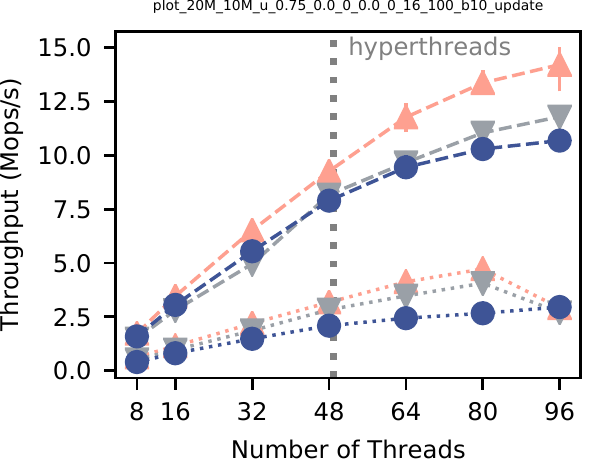} 
    & 
    \includegraphics[scale=\plotscale, trim={{\lefttrim} 0 0 \toptrim}, clip]{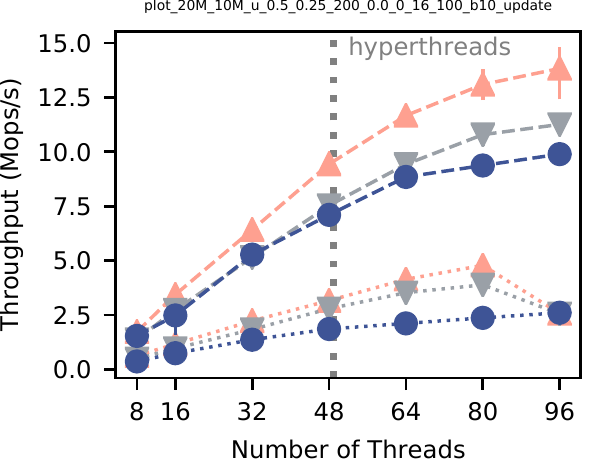}
    & 
    \includegraphics[scale=\plotscale, trim={{\lefttrim} 0 0 \toptrim}, clip]{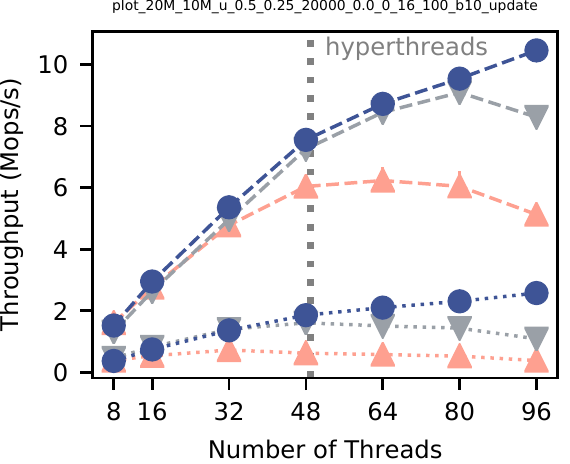} \\
    
%     & \multicolumn{4}{l}{\hspace{2cm}\includegraphics[scale=\plotscale, trim={0 0 2cm 0}, clip]{plots/{plot_20M_10M_u_0_0_0_0_0_0_0_0_16_100_b10_legend_single_row.pdf}}\\
    
    \hline\\[-5pt]
    
    \multirow{2}{*}{\STAB{\rotatebox[origin=c]{90}{100-op. batch updates}}}
    &
    \STAB{\rotatebox[origin=c]{90}{Total throughput}}
    &
    \includegraphics[scale=\plotscale, trim={0 0 0 \toptrim}, clip]{plots/plot_20M_10M_u_0_0_0_0_0_0_0_0_16_100_b100.pdf}
    & 
    \includegraphics[scale=\plotscale, trim={{\lefttrim} 0 0 \toptrim}, clip]{plots/plot_20M_10M_u_0_75_0_0_0_0_0_0_16_100_b100.pdf} 
    & 
    \includegraphics[scale=\plotscale, trim={{\lefttrim} 0 0 \toptrim}, clip]{plots/plot_20M_10M_u_0_5_0_25_200_0_0_0_16_100_b100.pdf}
    & 
    \includegraphics[scale=\plotscale, trim={{\lefttrim} 0 0 \toptrim}, clip]{plots/plot_20M_10M_u_0_5_0_25_20000_0_0_0_16_100_b100.pdf} \\
    
    &
    \STAB{\rotatebox[origin=c]{90}{Update throughput}}
    &
    \includegraphics[scale=\plotscale, trim={0 0 0 \toptrim}, clip]{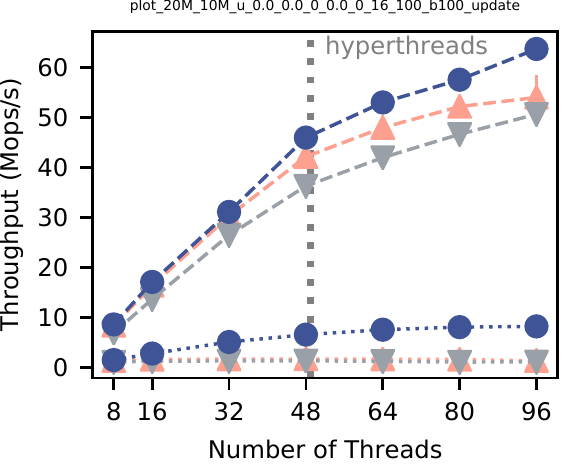}
    & 
    \includegraphics[scale=\plotscale, trim={{\lefttrim} 0 0 \toptrim}, clip]{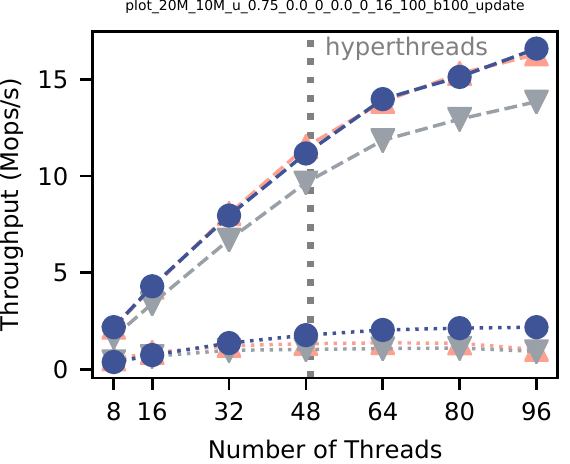} 
    & 
    \includegraphics[scale=\plotscale, trim={{\lefttrim} 0 0 \toptrim}, clip]{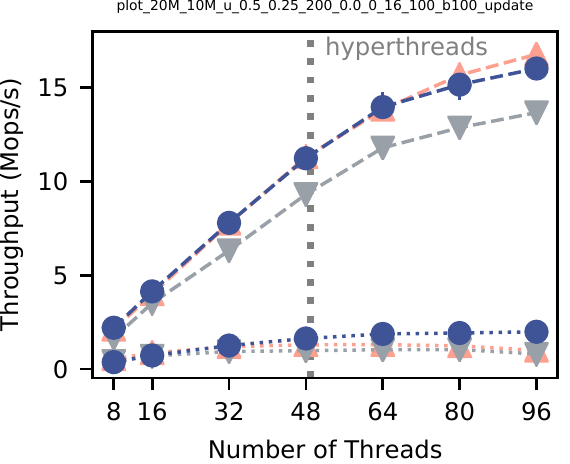}
    & 
    \includegraphics[scale=\plotscale, trim={{\lefttrim} 0 0 \toptrim}, clip]{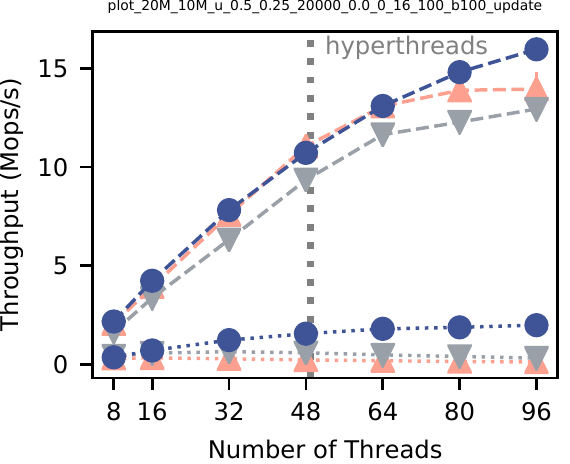} \\
    
    & & \multicolumn{4}{l}{\hspace{2cm}\includegraphics[scale=\plotscale, trim={0 0 2cm 0}, clip]{plots/plot_20M_10M_u_0_0_0_0_0_0_0_0_16_100_b10_legend_single_row.pdf}}\\
  \end{tabular}
  }
% \end{table}
\caption{Throughput scalability results (16~B keys, 100~B values, keys chosen with uniform distribution).} 
\label{fig:a_throughput_u_16_100}
\end{figure*}
\egroup

\providecommand{\plotscale}{0.6}
\providecommand{\STAB}[1]{\vspace{0.5cm}#1}

\bgroup
\def\arraystretch{1.2}
\def\colwidth{1.65in}
\def\toptrim{0.15cm}
\def\lefttrim{0.35cm}
\begin{figure*}[t]
% \begin{table}
%   \caption{Throughput.}
    \scalebox{0.90}{
  \begin{tabular}{C{0.2in}C{0.2in}C{1.7in}C{\colwidth}C{\colwidth}C{\colwidth}}
%   \multicolumn{2}{c|}{Uniform distribution} & \multicolumn{2}{c}{Zipfian distribution (skewness: 0.99)}\\
%   \multicolumn{2}{c|}{Uniform distribution} & \multicolumn{2}{c}{Zipfian distribution}\\
%   \hline
    & & (a) 100\% threads: put/remove & (b) 25\% threads: put/remove & \multicolumn{2}{c}{25\% threads: put/remove, 50\% threads: get, 25\% threads: scan} \\
    & & & 75\% threads: get & (c) Short scans (100 ops) & (d) Long scans (10000 ops)\\ 
    
    \hline
%     (e) 100\% threads: put/remove & (f)
%     75\% threads: get & 
    
%     100\% threads: get\\
    \multirow{2}{*}{\STAB{\rotatebox[origin=c]{90}{Simple put/remove}}}
    & 
    \STAB{\rotatebox[origin=c]{90}{Total throughput}}
    &
    \includegraphics[scale=\plotscale, trim={0 0 0 \toptrim}, clip]{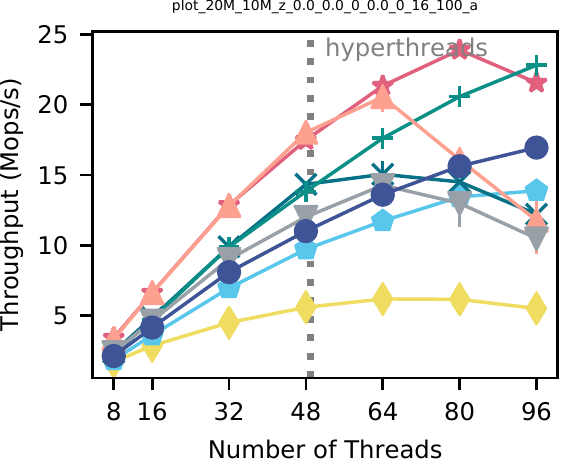}
    & 
    \includegraphics[scale=\plotscale, trim={{\lefttrim} 0 0 \toptrim}, clip]{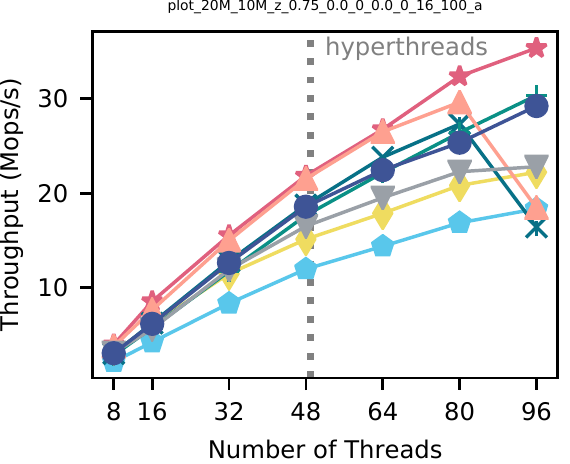} 
    & 
    \includegraphics[scale=\plotscale, trim={{\lefttrim} 0 0 \toptrim}, clip]{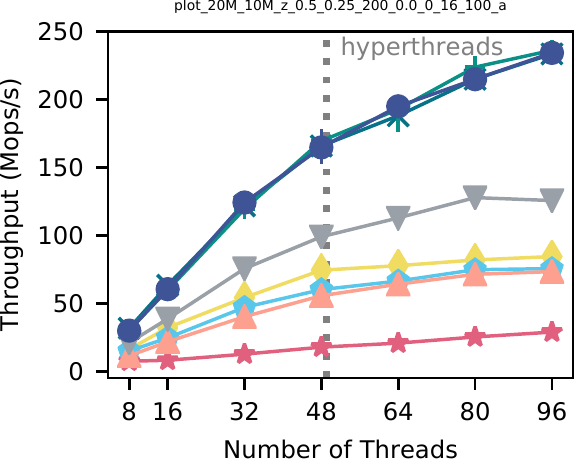}
    & 
    \includegraphics[scale=\plotscale, trim={{\lefttrim} 0 0 \toptrim}, clip]{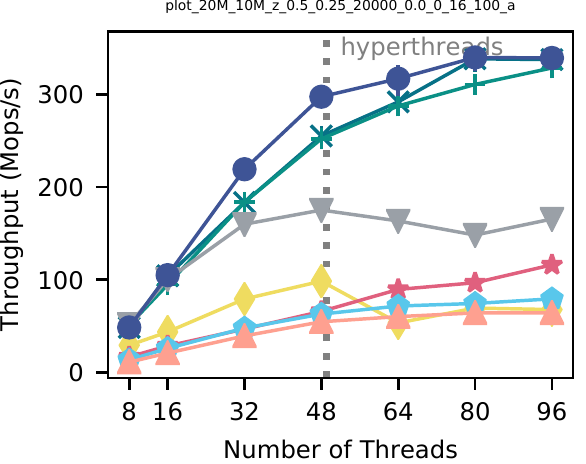} \\
    
    &
    \STAB{\rotatebox[origin=c]{90}{Update throughput}}
    &
    \includegraphics[scale=\plotscale, trim={0 0 0 \toptrim}, clip]{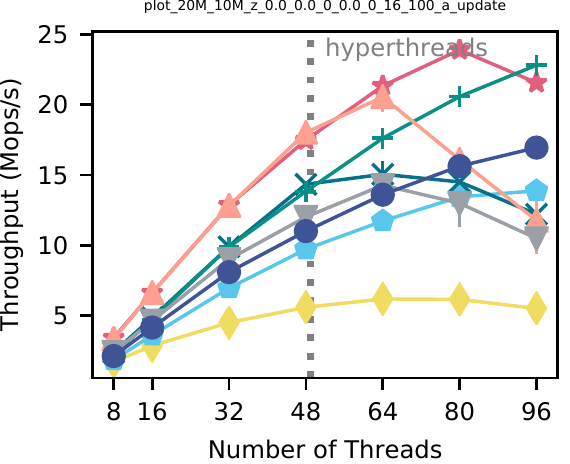}
    & 
    \includegraphics[scale=\plotscale, trim={{\lefttrim} 0 0 \toptrim}, clip]{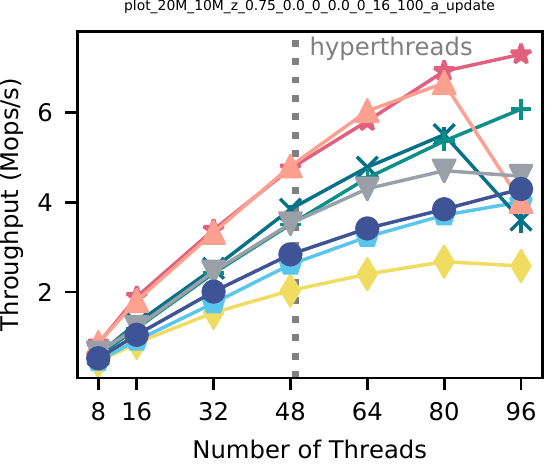} 
    & 
    \includegraphics[scale=\plotscale, trim={{\lefttrim} 0 0 \toptrim}, clip]{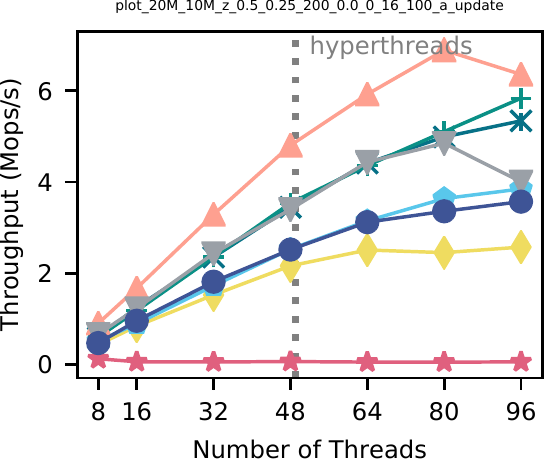}
    & 
    \includegraphics[scale=\plotscale, trim={{\lefttrim} 0 0 \toptrim}, clip]{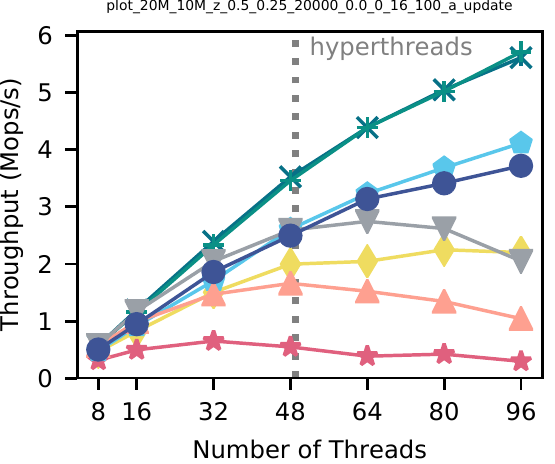} \\
    
   & & \multicolumn{4}{l}{\hspace{2cm}\includegraphics[scale=\plotscale, trim={0 0 2cm 0}, clip]{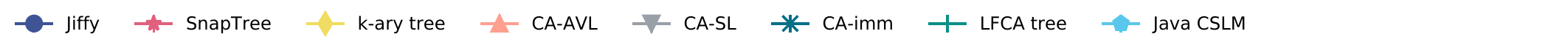}}\\[2pt]
    
   \hline\\[-5pt]
    
    \multirow{2}{*}{\STAB{\rotatebox[origin=c]{90}{10-op. batch updates}}}
    &
    \STAB{\rotatebox[origin=c]{90}{Total throughput}}
    &
    \includegraphics[scale=\plotscale, trim={0 0 0 \toptrim}, clip]{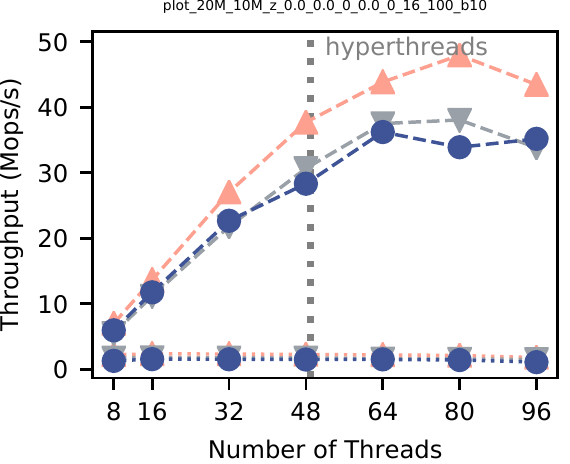}
    & 
    \includegraphics[scale=\plotscale, trim={{\lefttrim} 0 0 \toptrim}, clip]{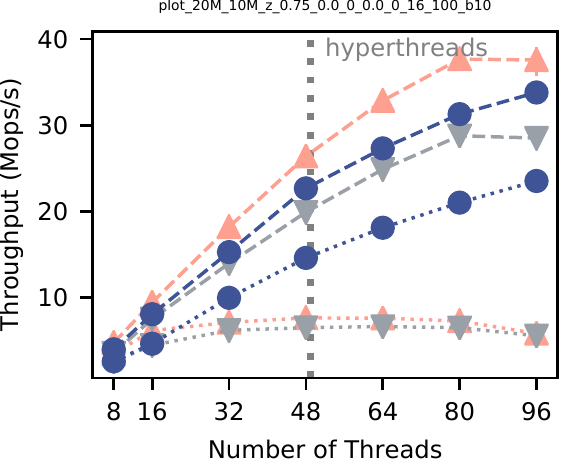} 
    & 
    \includegraphics[scale=\plotscale, trim={{\lefttrim} 0 0 \toptrim}, clip]{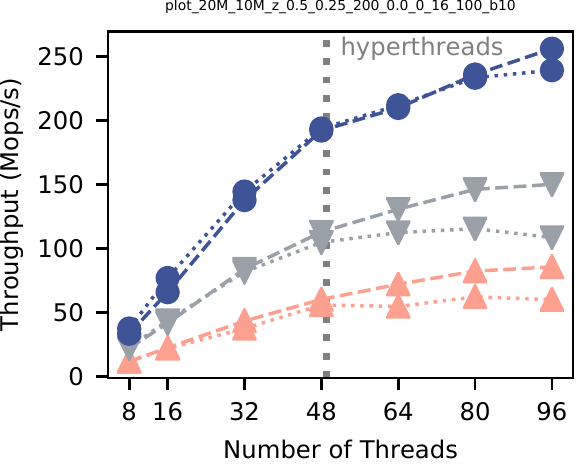}
    & 
    \includegraphics[scale=\plotscale, trim={{\lefttrim} 0 0 \toptrim}, clip]{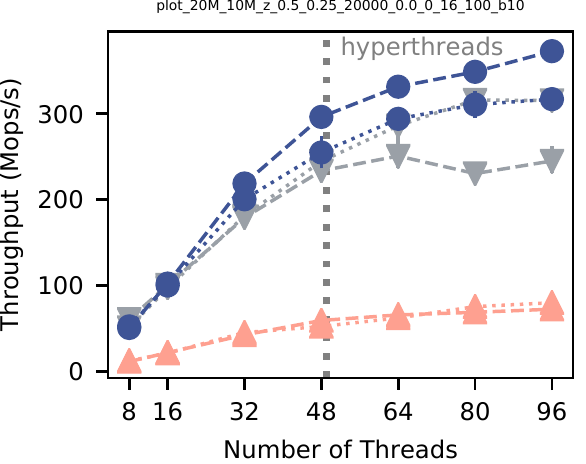} \\
    
    &
    \STAB{\rotatebox[origin=c]{90}{Update throughput}}
    &
    \includegraphics[scale=\plotscale, trim={0 0 0 \toptrim}, clip]{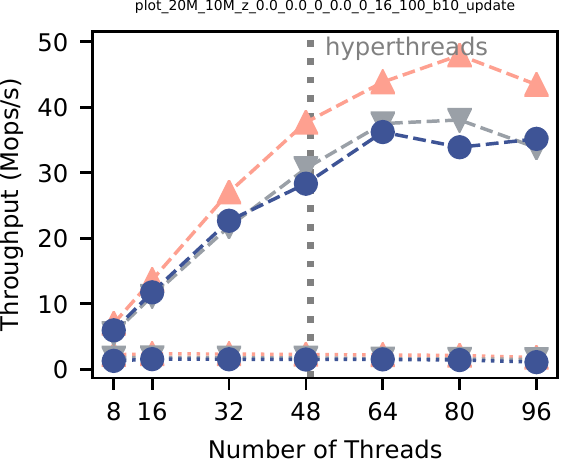}
    & 
    \includegraphics[scale=\plotscale, trim={{\lefttrim} 0 0 \toptrim}, clip]{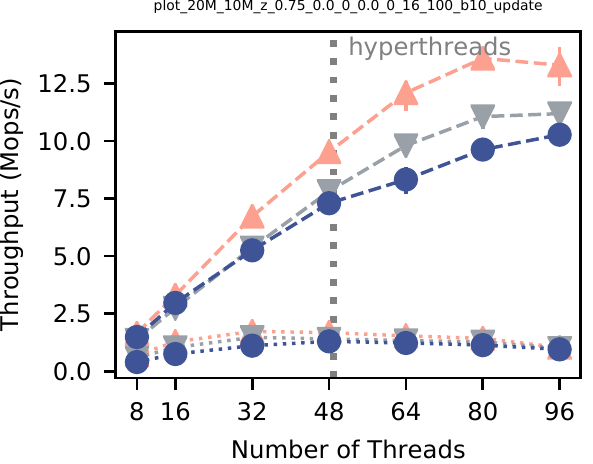} 
    & 
    \includegraphics[scale=\plotscale, trim={{\lefttrim} 0 0 \toptrim}, clip]{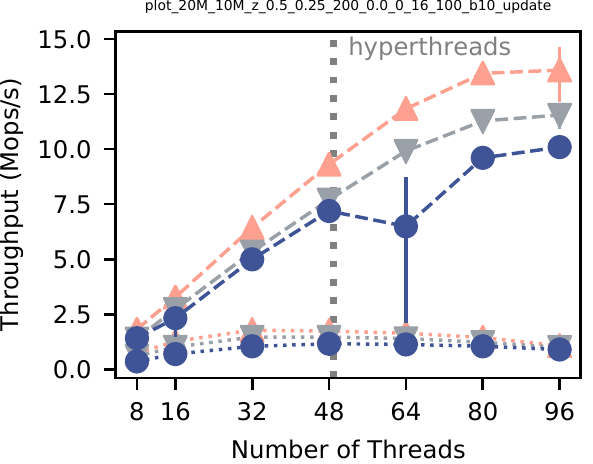}
    & 
    \includegraphics[scale=\plotscale, trim={{\lefttrim} 0 0 \toptrim}, clip]{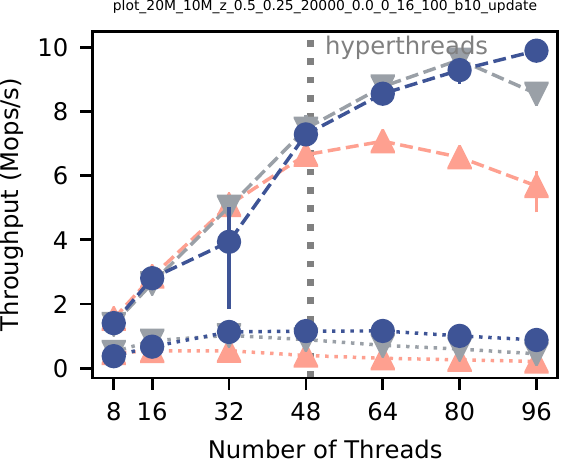} \\
    
%     & \multicolumn{4}{l}{\hspace{2cm}\includegraphics[scale=\plotscale, trim={0 0 2cm 0}, clip]{plots/plot_20M_10M_z_0_0_0_0_0_0_0_0_16_100_b10_legend_single_row.pdf}}\\
    
    \hline\\[-5pt]
    
    \multirow{2}{*}{\STAB{\rotatebox[origin=c]{90}{100-op. batch updates}}}
    &
    \STAB{\rotatebox[origin=c]{90}{Total throughput}}
    &
    \includegraphics[scale=\plotscale, trim={0 0 0 \toptrim}, clip]{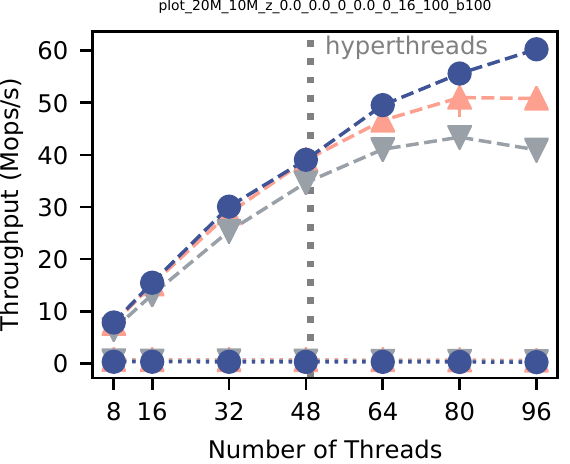}
    & 
    \includegraphics[scale=\plotscale, trim={{\lefttrim} 0 0 \toptrim}, clip]{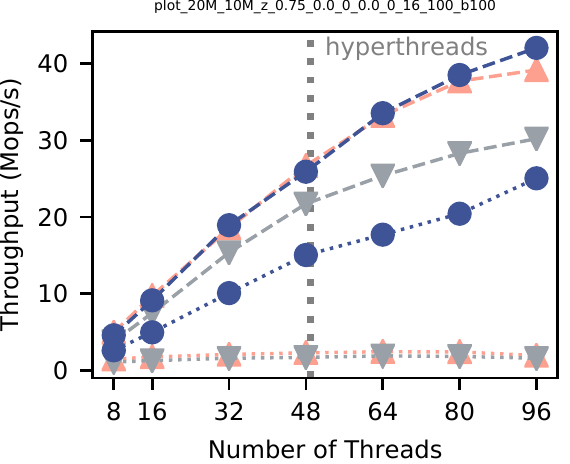} 
    & 
    \includegraphics[scale=\plotscale, trim={{\lefttrim} 0 0 \toptrim}, clip]{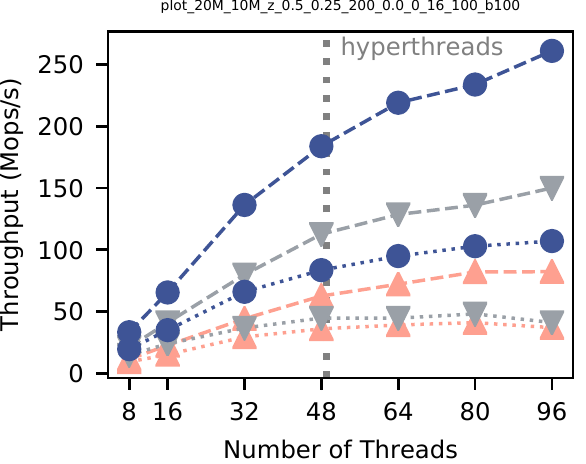}
    & 
    \includegraphics[scale=\plotscale, trim={{\lefttrim} 0 0 \toptrim}, clip]{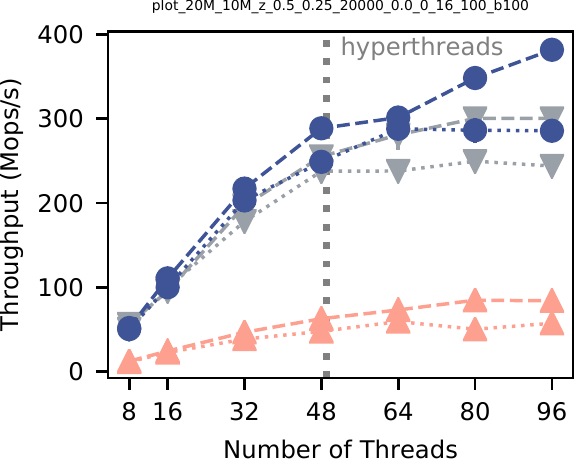} \\
    
    &
    \STAB{\rotatebox[origin=c]{90}{Update throughput}}
    &
    \includegraphics[scale=\plotscale, trim={0 0 0 \toptrim}, clip]{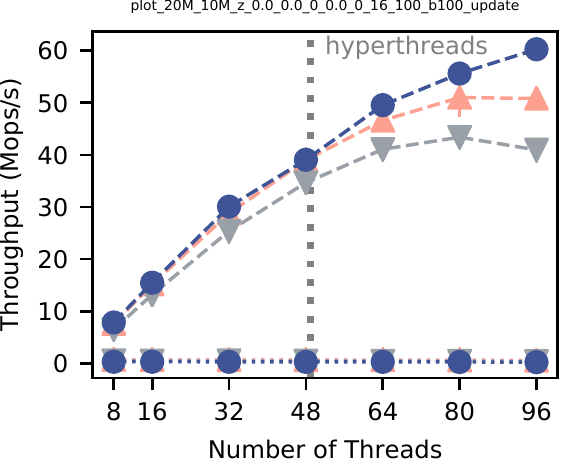}
    & 
    \includegraphics[scale=\plotscale, trim={{\lefttrim} 0 0 \toptrim}, clip]{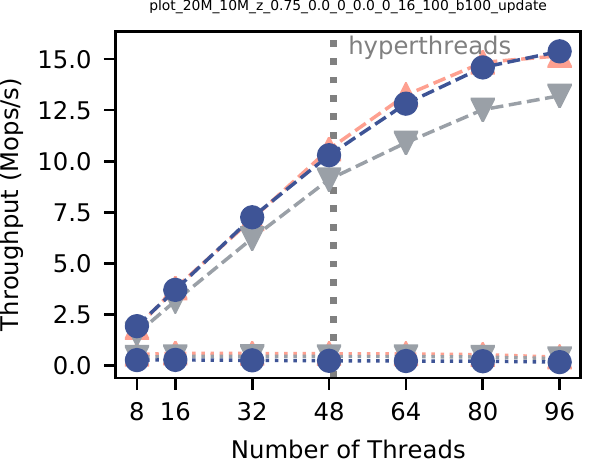} 
    & 
    \includegraphics[scale=\plotscale, trim={{\lefttrim} 0 0 \toptrim}, clip]{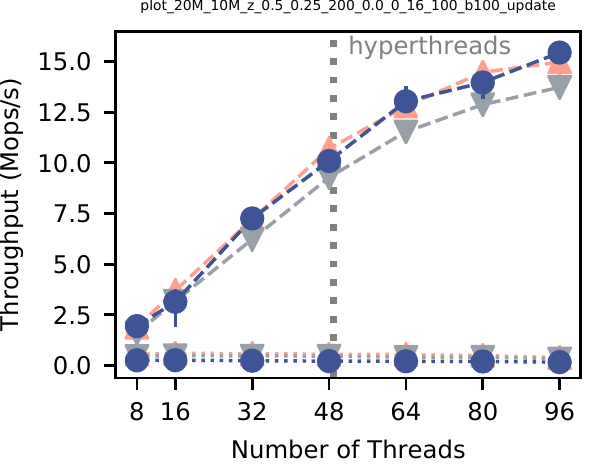}
    & 
    \includegraphics[scale=\plotscale, trim={{\lefttrim} 0 0 \toptrim}, clip]{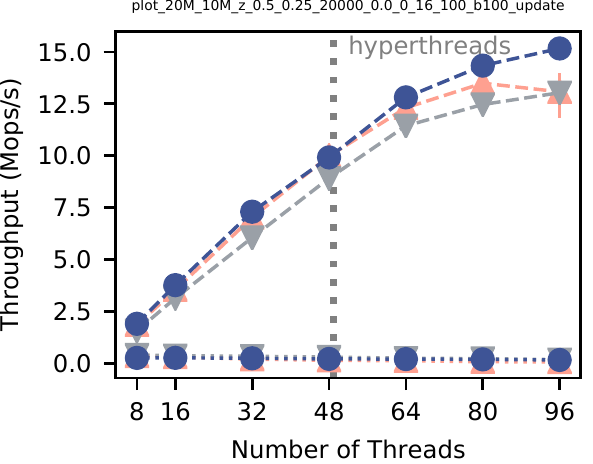} \\
    
    & & \multicolumn{4}{l}{\hspace{2cm}\includegraphics[scale=\plotscale, trim={0 0 2cm 0}, clip]{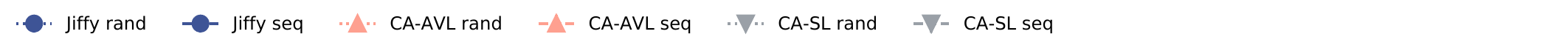}}\\
  \end{tabular}
  }
% \end{table}
\caption{Throughput scalability results (16~B keys, 100~B values, keys chosen with Zipfian distribution). }
\label{fig:a_throughput_z_16_100}
\end{figure*}
\egroup

\providecommand{\plotscale}{0.6}
\providecommand{\STAB}[1]{\vspace{0.5cm}#1}

\bgroup
\def\arraystretch{1.2}
\def\colwidth{1.65in}
\def\toptrim{0.15cm}
\def\lefttrim{0.35cm}
\begin{figure*}[t]
% \begin{table}
%   \caption{Throughput.}
    \scalebox{0.90}{
  \begin{tabular}{C{0.2in}C{0.2in}C{1.7in}C{\colwidth}C{\colwidth}C{\colwidth}}
%   \multicolumn{2}{c|}{Uniform distribution} & \multicolumn{2}{c}{Zipfian distribution (skewness: 0.99)}\\
%   \multicolumn{2}{c|}{Uniform distribution} & \multicolumn{2}{c}{Zipfian distribution}\\
%   \hline
    & & (a) 100\% threads: put/remove & (b) 25\% threads: put/remove & \multicolumn{2}{c}{25\% threads: put/remove, 50\% threads: get, 25\% threads: scan} \\
    & & & 75\% threads: get & (c) Short scans (100 ops) & (d) Long scans (10000 ops)\\ 
    
    \hline
%     (e) 100\% threads: put/remove & (f)
%     75\% threads: get & 
    
%     100\% threads: get\\
    \multirow{2}{*}{\STAB{\rotatebox[origin=c]{90}{Simple put/remove}}}
    & 
    \STAB{\rotatebox[origin=c]{90}{Total throughput}}
    &
    \includegraphics[scale=\plotscale, trim={0 0 0 \toptrim}, clip]{plots/plot_20M_10M_u_0_0_0_0_0_0_0_0_a.pdf}
    & 
    \includegraphics[scale=\plotscale, trim={{\lefttrim} 0 0 \toptrim}, clip]{plots/plot_20M_10M_u_0_75_0_0_0_0_0_0_a.pdf} 
    & 
    \includegraphics[scale=\plotscale, trim={{\lefttrim} 0 0 \toptrim}, clip]{plots/plot_20M_10M_u_0_5_0_25_200_0_0_0_a.pdf}
    & 
    \includegraphics[scale=\plotscale, trim={{\lefttrim} 0 0 \toptrim}, clip]{plots/plot_20M_10M_u_0_5_0_25_20000_0_0_0_a.pdf} \\
    
    &
    \STAB{\rotatebox[origin=c]{90}{Update throughput}}
    &
    \includegraphics[scale=\plotscale, trim={0 0 0 \toptrim}, clip]{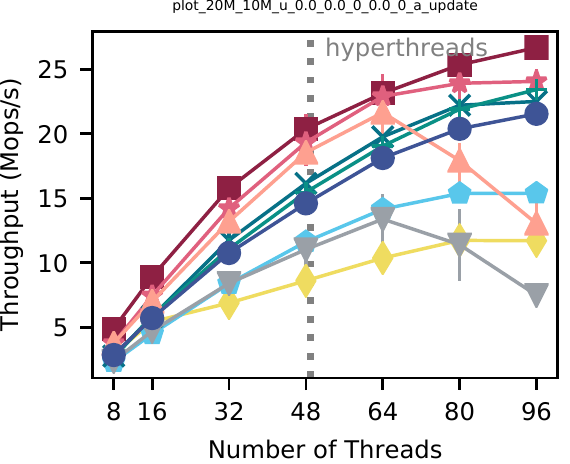}
    & 
    \includegraphics[scale=\plotscale, trim={{\lefttrim} 0 0 \toptrim}, clip]{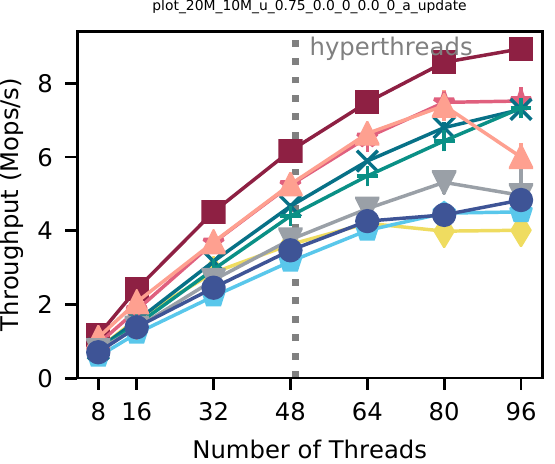} 
    & 
    \includegraphics[scale=\plotscale, trim={{\lefttrim} 0 0 \toptrim}, clip]{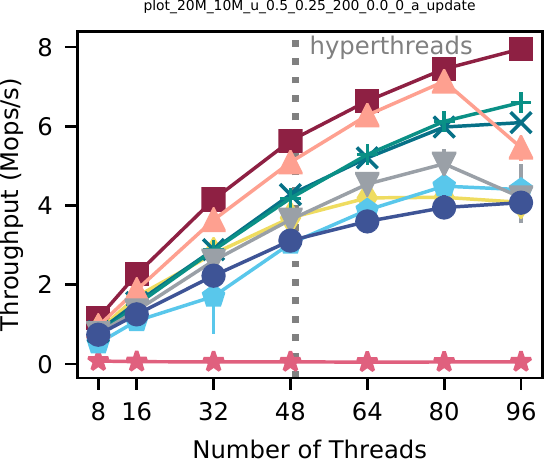}
    & 
    \includegraphics[scale=\plotscale, trim={{\lefttrim} 0 0 \toptrim}, clip]{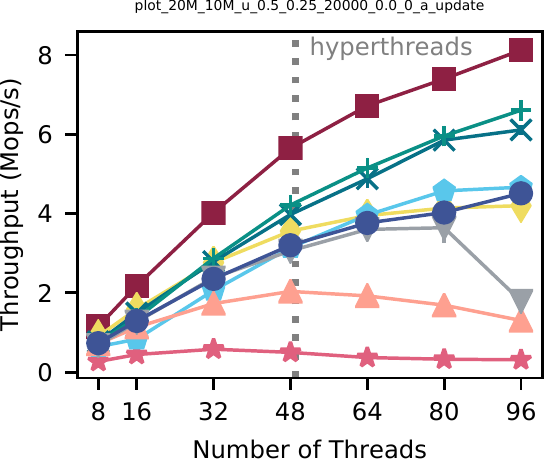} \\
    
   & & \multicolumn{4}{l}{\hspace{2cm}\includegraphics[scale=\plotscale, trim={0 0 2cm 0}, clip]{plots/plot_20M_10M_u_0_0_0_0_0_0_0_0_a_legend_single_row.pdf}}\\[2pt]
    
   \hline\\[-5pt]
    
    \multirow{2}{*}{\STAB{\rotatebox[origin=c]{90}{10-op. batch updates}}}
    &
    \STAB{\rotatebox[origin=c]{90}{Total throughput}}
    &
    \includegraphics[scale=\plotscale, trim={0 0 0 \toptrim}, clip]{plots/plot_20M_10M_u_0_0_0_0_0_0_0_0_b10.pdf}
    & 
    \includegraphics[scale=\plotscale, trim={{\lefttrim} 0 0 \toptrim}, clip]{plots/plot_20M_10M_u_0_75_0_0_0_0_0_0_b10.pdf} 
    & 
    \includegraphics[scale=\plotscale, trim={{\lefttrim} 0 0 \toptrim}, clip]{plots/plot_20M_10M_u_0_5_0_25_200_0_0_0_b10.pdf}
    & 
    \includegraphics[scale=\plotscale, trim={{\lefttrim} 0 0 \toptrim}, clip]{plots/plot_20M_10M_u_0_5_0_25_20000_0_0_0_b10.pdf} \\
    
    &
    \STAB{\rotatebox[origin=c]{90}{Update throughput}}
    &
    \includegraphics[scale=\plotscale, trim={0 0 0 \toptrim}, clip]{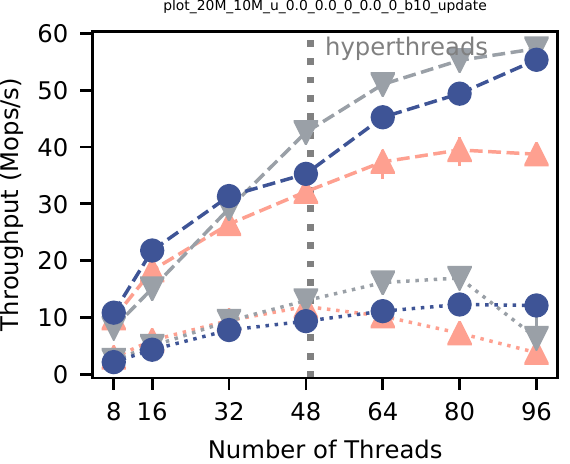}
    & 
    \includegraphics[scale=\plotscale, trim={{\lefttrim} 0 0 \toptrim}, clip]{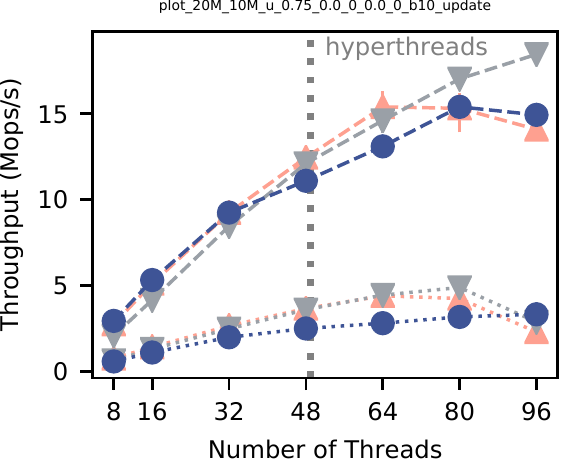} 
    & 
    \includegraphics[scale=\plotscale, trim={{\lefttrim} 0 0 \toptrim}, clip]{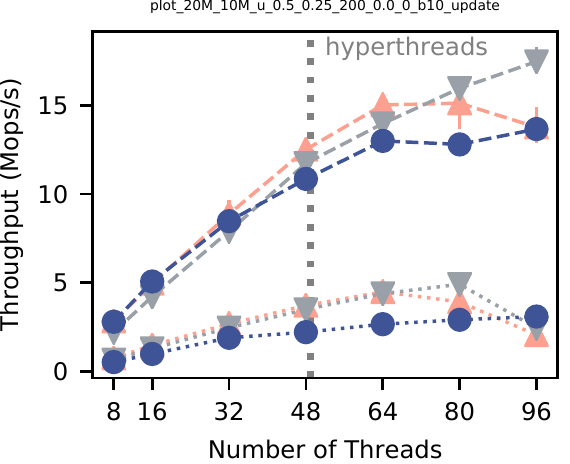}
    & 
    \includegraphics[scale=\plotscale, trim={{\lefttrim} 0 0 \toptrim}, clip]{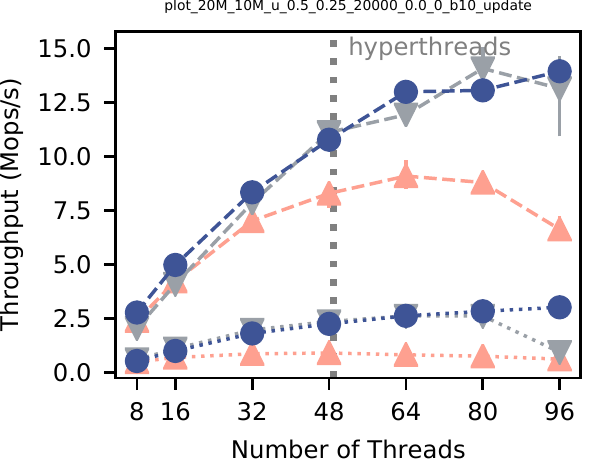} \\
    
%     & \multicolumn{4}{l}{\hspace{2cm}\includegraphics[scale=\plotscale, trim={0 0 2cm 0}, clip]{plots/plot_20M_10M_u_0_0_0_0_0_0_0_0_b10_legend_single_row.pdf}}\\
    
    \hline\\[-5pt]
    
    \multirow{2}{*}{\STAB{\rotatebox[origin=c]{90}{100-op. batch updates}}}
    &
    \STAB{\rotatebox[origin=c]{90}{Total throughput}}
    &
    \includegraphics[scale=\plotscale, trim={0 0 0 \toptrim}, clip]{plots/plot_20M_10M_u_0_0_0_0_0_0_0_0_b100.pdf}
    & 
    \includegraphics[scale=\plotscale, trim={{\lefttrim} 0 0 \toptrim}, clip]{plots/plot_20M_10M_u_0_75_0_0_0_0_0_0_b100.pdf} 
    & 
    \includegraphics[scale=\plotscale, trim={{\lefttrim} 0 0 \toptrim}, clip]{plots/plot_20M_10M_u_0_5_0_25_200_0_0_0_b100.pdf}
    & 
    \includegraphics[scale=\plotscale, trim={{\lefttrim} 0 0 \toptrim}, clip]{plots/plot_20M_10M_u_0_5_0_25_20000_0_0_0_b100.pdf} \\
    
    &
    \STAB{\rotatebox[origin=c]{90}{Update throughput}}
    &
    \includegraphics[scale=\plotscale, trim={0 0 0 \toptrim}, clip]{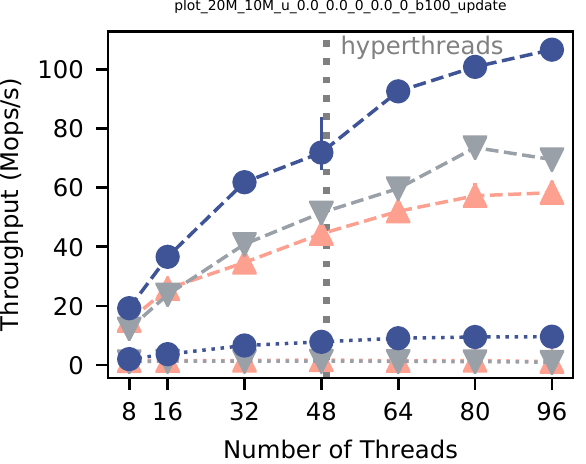}
    & 
    \includegraphics[scale=\plotscale, trim={{\lefttrim} 0 0 \toptrim}, clip]{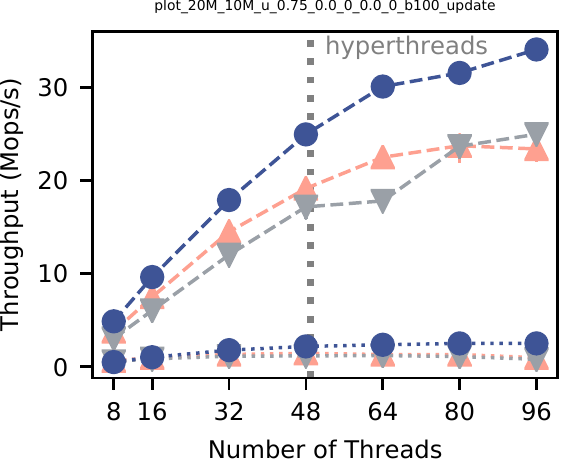} 
    & 
    \includegraphics[scale=\plotscale, trim={{\lefttrim} 0 0 \toptrim}, clip]{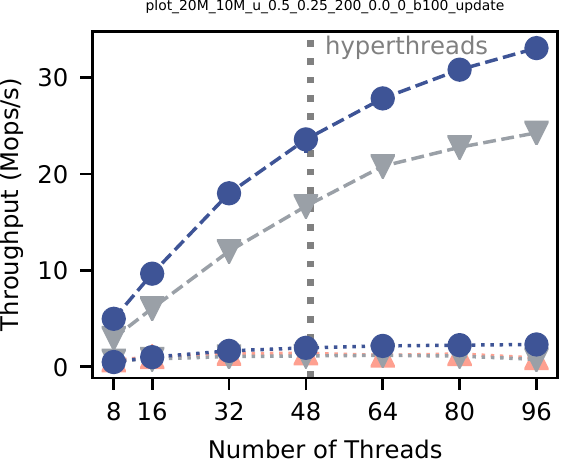}
    & 
    \includegraphics[scale=\plotscale, trim={{\lefttrim} 0 0 \toptrim}, clip]{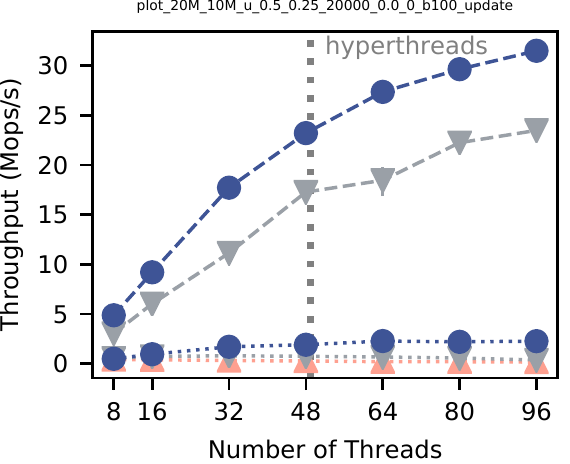} \\
    
    & & \multicolumn{4}{l}{\hspace{2cm}\includegraphics[scale=\plotscale, trim={0 0 2cm 0}, clip]{plots/plot_20M_10M_u_0_0_0_0_0_0_0_0_b10_legend_single_row.pdf}}\\
  \end{tabular}
  }
% \end{table}
\caption{Throughput scalability results (4~B keys, 4~B values, keys chosen with uniform distribution).}
\label{fig:a_throughput_u_4_4}
\end{figure*}
\egroup

\providecommand{\plotscale}{0.6}
\providecommand{\STAB}[1]{\vspace{0.5cm}#1}

\bgroup
\def\arraystretch{1.2}
\def\colwidth{1.65in}
\def\toptrim{0.15cm}
\def\lefttrim{0.35cm}
\begin{figure*}[t]
% \begin{table}
%   \caption{Throughput.}
    \scalebox{0.90}{
  \begin{tabular}{C{0.2in}C{0.2in}C{1.7in}C{\colwidth}C{\colwidth}C{\colwidth}}
%   \multicolumn{2}{c|}{Uniform distribution} & \multicolumn{2}{c}{Zipfian distribution (skewness: 0.99)}\\
%   \multicolumn{2}{c|}{Uniform distribution} & \multicolumn{2}{c}{Zipfian distribution}\\
%   \hline
    & & (a) 100\% threads: put/remove & (b) 25\% threads: put/remove & \multicolumn{2}{c}{25\% threads: put/remove, 50\% threads: get, 25\% threads: scan} \\
    & & & 75\% threads: get & (c) Short scans (100 ops) & (d) Long scans (10000 ops)\\ 
    
    \hline
%     (e) 100\% threads: put/remove & (f)
%     75\% threads: get & 
    
%     100\% threads: get\\
    \multirow{2}{*}{\STAB{\rotatebox[origin=c]{90}{Simple put/remove}}}
    & 
    \STAB{\rotatebox[origin=c]{90}{Total throughput}}
    &
    \includegraphics[scale=\plotscale, trim={0 0 0 \toptrim}, clip]{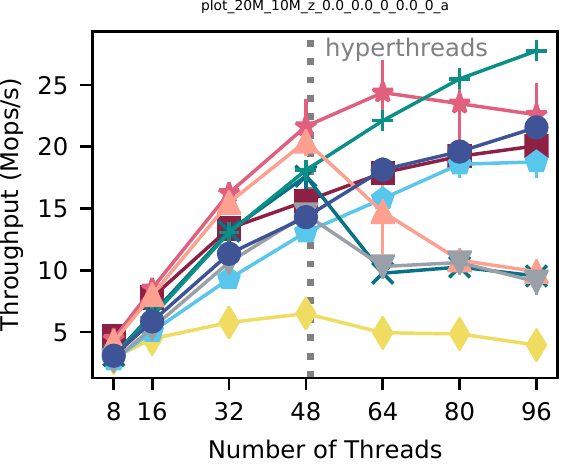}
    & 
    \includegraphics[scale=\plotscale, trim={{\lefttrim} 0 0 \toptrim}, clip]{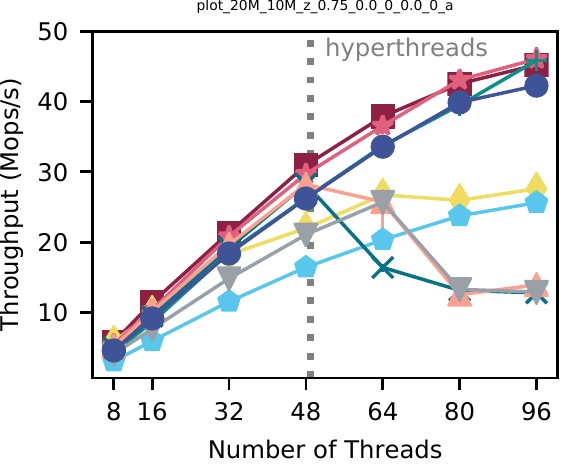} 
    & 
    \includegraphics[scale=\plotscale, trim={{\lefttrim} 0 0 \toptrim}, clip]{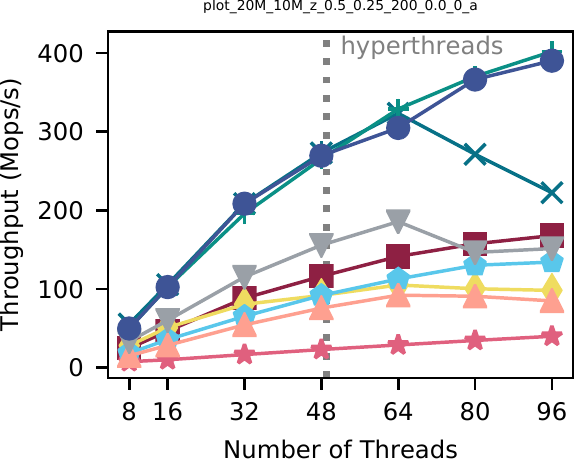}
    & 
    \includegraphics[scale=\plotscale, trim={{\lefttrim} 0 0 \toptrim}, clip]{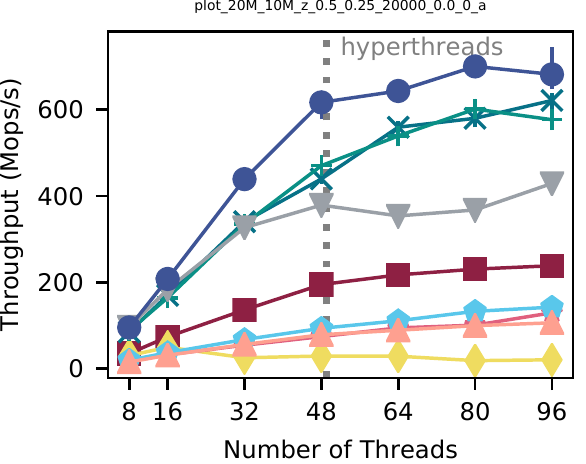} \\
    
    &
    \STAB{\rotatebox[origin=c]{90}{Update throughput}}
    &
    \includegraphics[scale=\plotscale, trim={0 0 0 \toptrim}, clip]{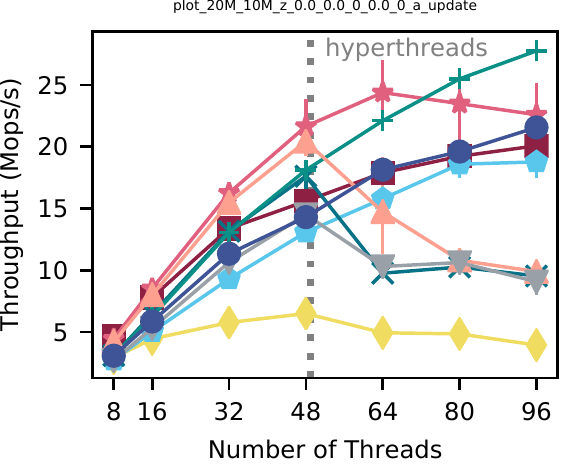}
    & 
    \includegraphics[scale=\plotscale, trim={{\lefttrim} 0 0 \toptrim}, clip]{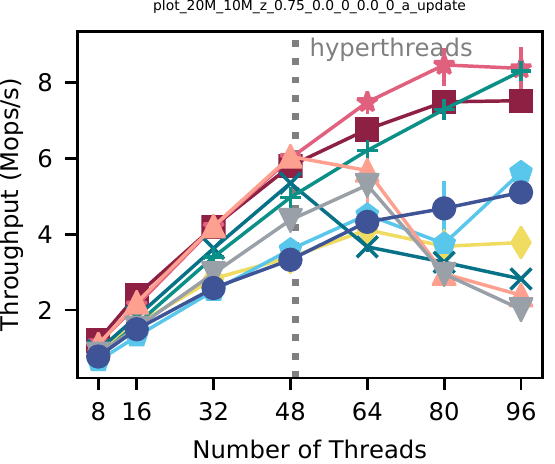} 
    & 
    \includegraphics[scale=\plotscale, trim={{\lefttrim} 0 0 \toptrim}, clip]{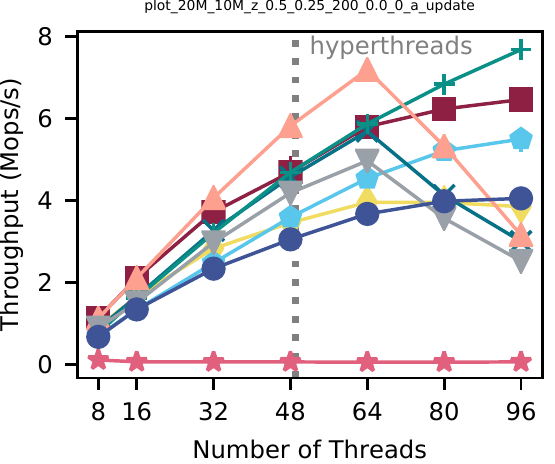}
    & 
    \includegraphics[scale=\plotscale, trim={{\lefttrim} 0 0 \toptrim}, clip]{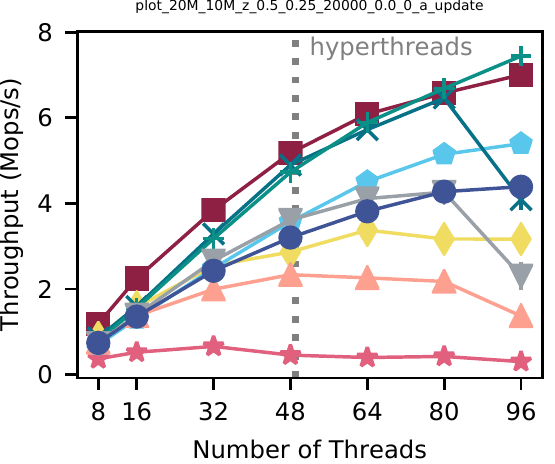} \\
    
   & & \multicolumn{4}{l}{\hspace{2cm}\includegraphics[scale=\plotscale, trim={0 0 2cm 0}, clip]{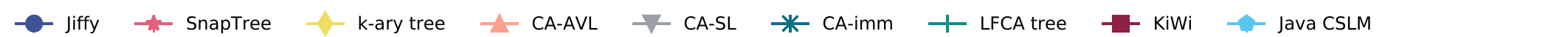}}\\[2pt]
    
   \hline\\[-5pt]
    
    \multirow{2}{*}{\STAB{\rotatebox[origin=c]{90}{10-op. batch updates}}}
    &
    \STAB{\rotatebox[origin=c]{90}{Total throughput}}
    &
    \includegraphics[scale=\plotscale, trim={0 0 0 \toptrim}, clip]{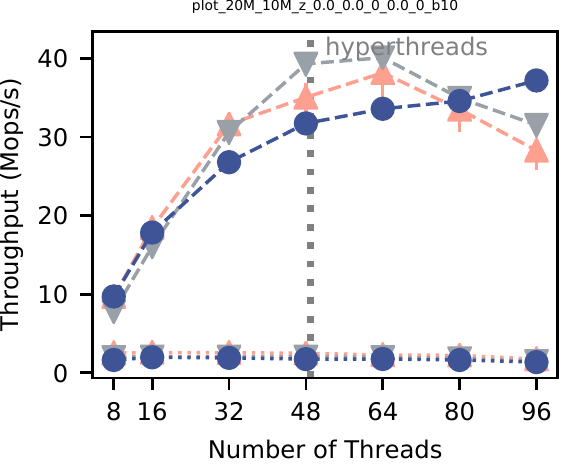}
    & 
    \includegraphics[scale=\plotscale, trim={{\lefttrim} 0 0 \toptrim}, clip]{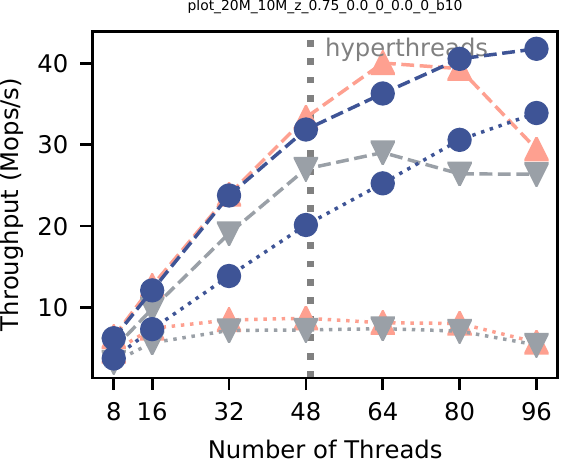} 
    & 
    \includegraphics[scale=\plotscale, trim={{\lefttrim} 0 0 \toptrim}, clip]{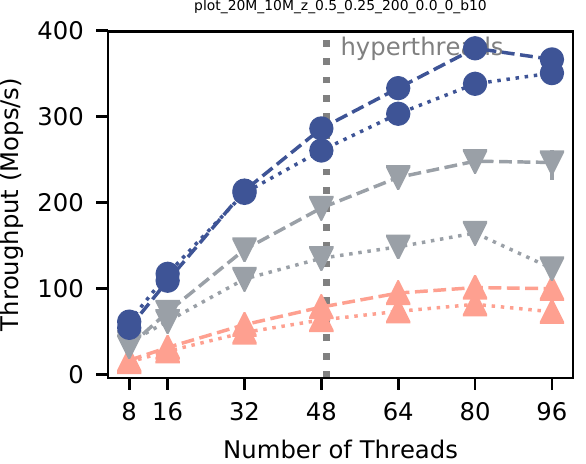}
    & 
    \includegraphics[scale=\plotscale, trim={{\lefttrim} 0 0 \toptrim}, clip]{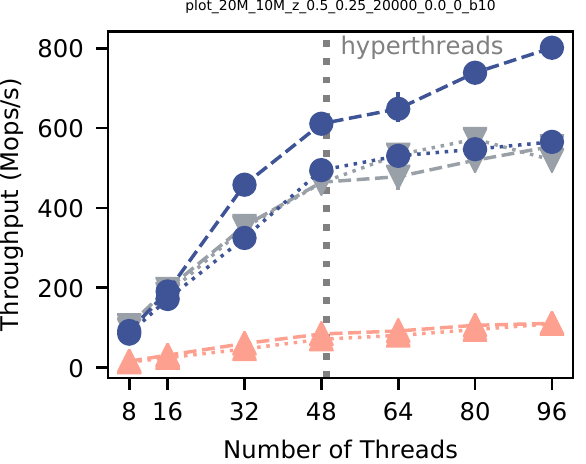} \\
    
    &
    \STAB{\rotatebox[origin=c]{90}{Update throughput}}
    &
    \includegraphics[scale=\plotscale, trim={0 0 0 \toptrim}, clip]{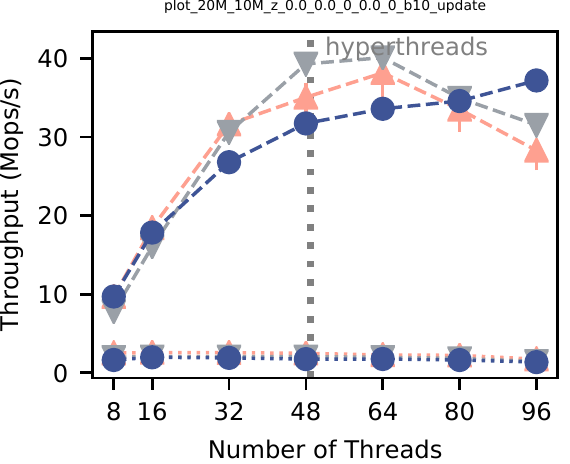}
    & 
    \includegraphics[scale=\plotscale, trim={{\lefttrim} 0 0 \toptrim}, clip]{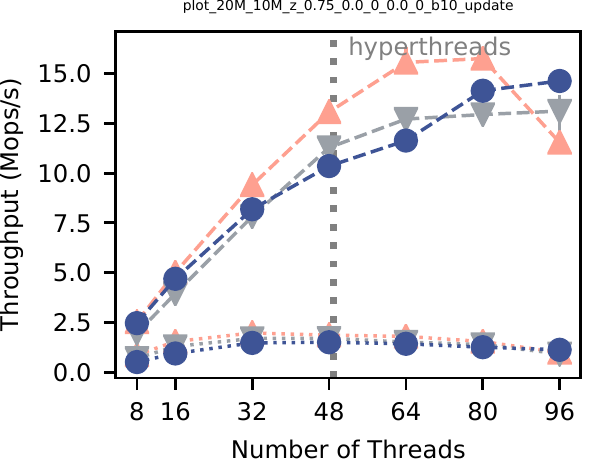} 
    & 
    \includegraphics[scale=\plotscale, trim={{\lefttrim} 0 0 \toptrim}, clip]{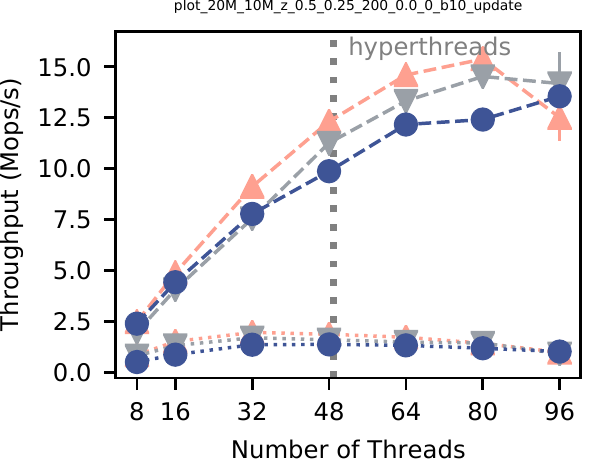}
    & 
    \includegraphics[scale=\plotscale, trim={{\lefttrim} 0 0 \toptrim}, clip]{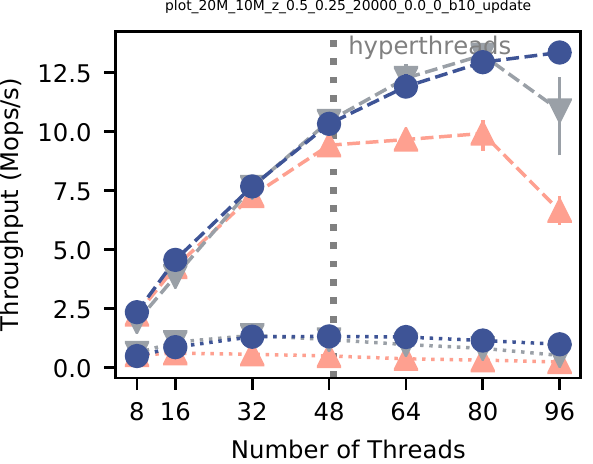} \\
    
%     & \multicolumn{4}{l}{\hspace{2cm}\includegraphics[scale=\plotscale, trim={0 0 2cm 0}, clip]{plots/plot_20M_10M_z_0_0_0_0_0_0_0_0_b10_legend_single_row.pdf}}\\
    
    \hline\\[-5pt]
    
    \multirow{2}{*}{\STAB{\rotatebox[origin=c]{90}{100-op. batch updates}}}
    &
    \STAB{\rotatebox[origin=c]{90}{Total throughput}}
    &
    \includegraphics[scale=\plotscale, trim={0 0 0 \toptrim}, clip]{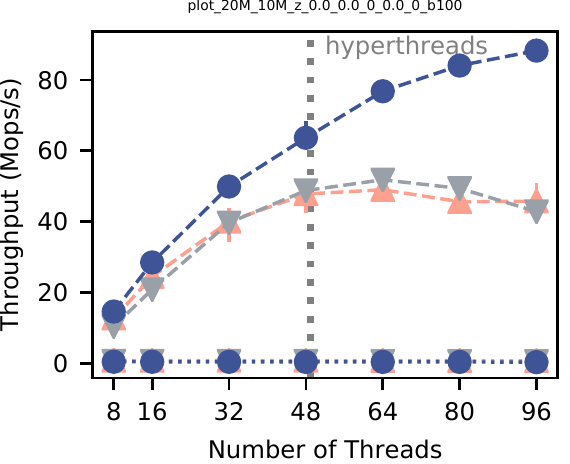}
    & 
    \includegraphics[scale=\plotscale, trim={{\lefttrim} 0 0 \toptrim}, clip]{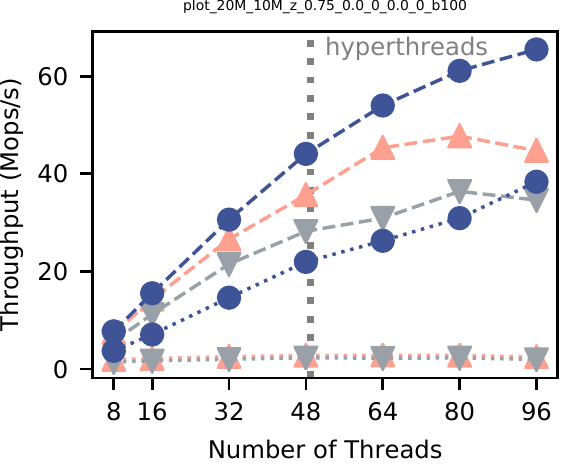} 
    & 
    \includegraphics[scale=\plotscale, trim={{\lefttrim} 0 0 \toptrim}, clip]{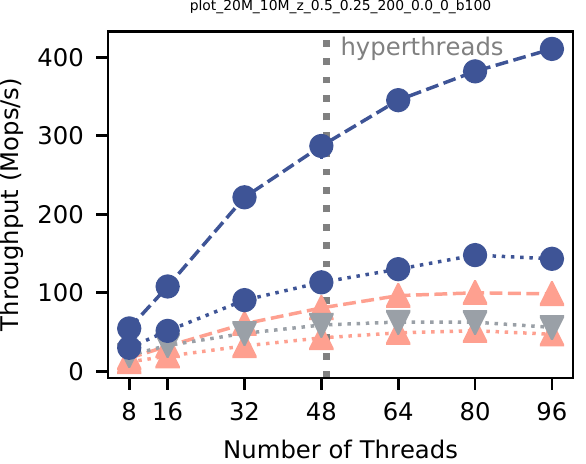}
    & 
    \includegraphics[scale=\plotscale, trim={{\lefttrim} 0 0 \toptrim}, clip]{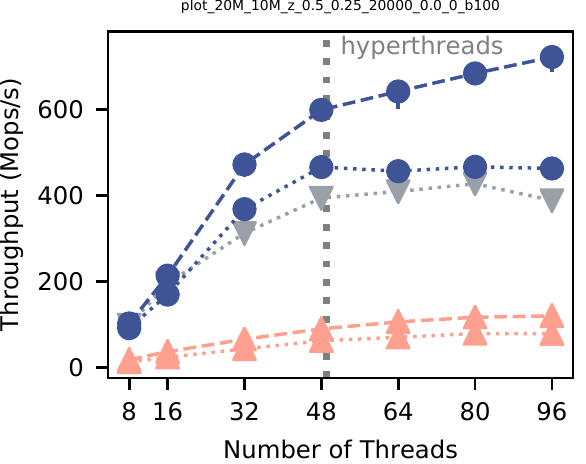} \\
    
    &
    \STAB{\rotatebox[origin=c]{90}{Update throughput}}
    &
    \includegraphics[scale=\plotscale, trim={0 0 0 \toptrim}, clip]{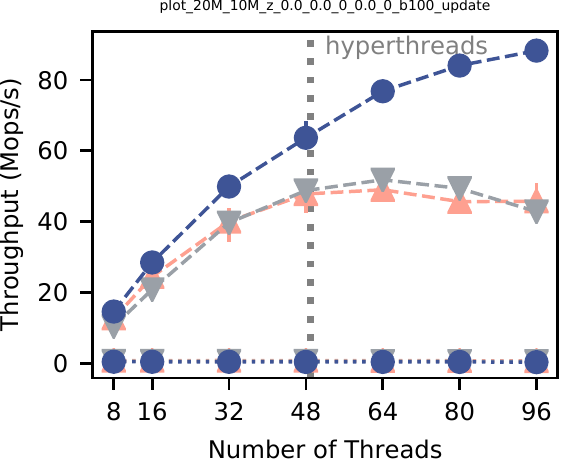}
    & 
    \includegraphics[scale=\plotscale, trim={{\lefttrim} 0 0 \toptrim}, clip]{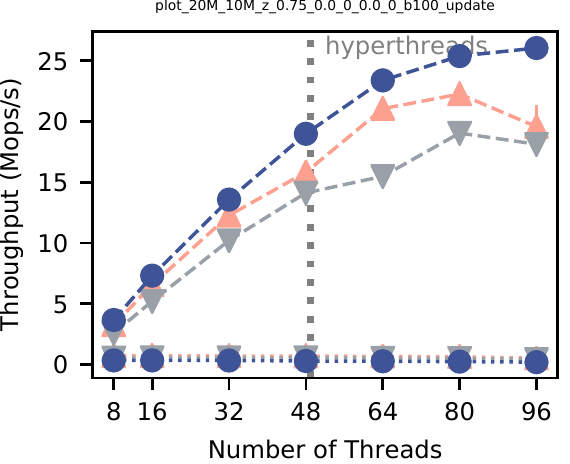} 
    & 
    \includegraphics[scale=\plotscale, trim={{\lefttrim} 0 0 \toptrim}, clip]{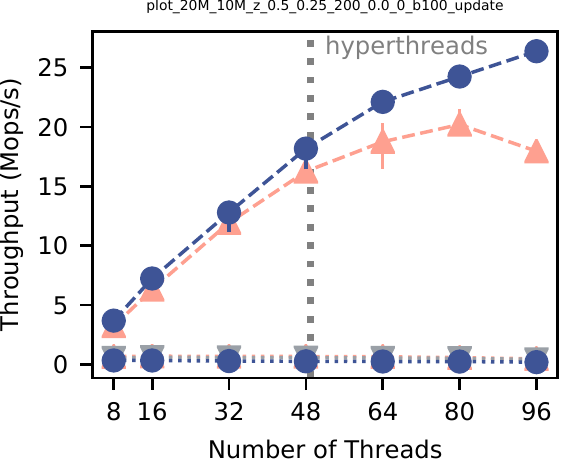}
    & 
    \includegraphics[scale=\plotscale, trim={{\lefttrim} 0 0 \toptrim}, clip]{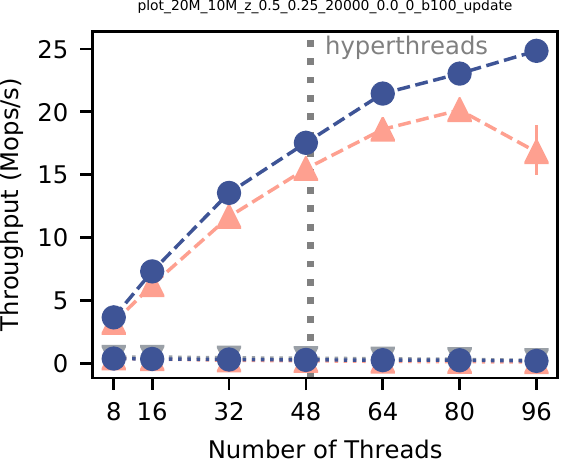} \\
    
    & & \multicolumn{4}{l}{\hspace{2cm}\includegraphics[scale=\plotscale, trim={0 0 2cm 0}, clip]{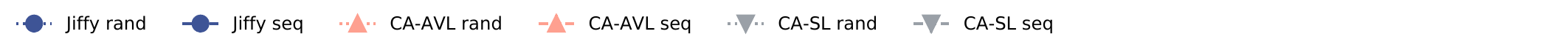}}\\
  \end{tabular}
  }
% \end{table}
\caption{Throughput scalability results (4~B keys, 4~B values, keys chosen with Zipfian distribution).}
\label{fig:a_throughput_z_4_4}
\end{figure*}
\egroup

}{}

\end{document}